\documentclass[amsmath,amssymb]{revtex4}

\usepackage{color}

\usepackage{graphicx}
\usepackage{dcolumn}
\usepackage{bm}
\input epsf

\pdfoutput=1

\def\sin{{\rm{sin}}}
\def\cos{{\rm{cos}}}

\usepackage{color} 

\begin{document}
\title{Evaporation of  ethanol-water droplet at different substrate temperatures and compositions}
\author{Pradeep Gurrala$^{\dagger}$, Pallavi Katre$^{\dagger\dagger}$, Saravanan Balusamy$^{\dagger}$, Sayak Banerjee$^{\dagger}$ and Kirti Chandra Sahu$^{\dagger\dagger}$\footnote{ksahu@iith.ac.in}}
\address{
$^{\dagger}$Department of Mechanical and Aerospace Engineering, Indian Institute of Technology Hyderabad, Sangareddy 502 285, Telangana, India \\
$^{\dagger\dagger}$Department of Chemical Engineering, Indian Institute of Technology Hyderabad, Kandi, Sangareddy 502 285, Telangana, India \\}

\date{\today}

\begin{abstract}
We experimentally investigate the evaporation dynamics of sessile droplets of a fixed volume (5$\mu$l) consisting of different compositions of ethanol-water binary mixture at different substrate temperatures $(T_s)$. The experiments are conducted on a cellulose-acetate substrate placed on a customised goniometer. The surface roughness studied by an atomic-force-microscopy (AFM) and the micro-scale images taken using a scanning-electron-microscope (SEM) show that the substrate considered in the present study is stable even at high temperatures. It is well known that in a binary mixture, the differential {rates} of evaporation of the individual {components} result in a complex evaporation process. We found that the complexity is even {more pronounced} at elevated temperatures. In order to compare the dynamics for different compositions and at different substrate temperatures, it is necessary to perform systematic experiments {at} a fixed condition. Such an attempt is made in the present study. At $T_s=25^\circ$C, we observe pinned-stage linear evaporation for pure droplets, but a binary (50\% ethanol + 50 \% water) droplet undergoes two distinct {evaporation stages: an early pinned stage and a later} receding stage. In the binary droplet, {the more volatile ethanol}, evaporates faster {leading} to a nonlinear trend in the evaporation process at the early stage. The phenomenon observed in the present study at $T_s=25^\circ$C is similar to that presented by previous researchers at room temperature. More interesting dynamics is {observed} in the evaporation process of a binary droplet at an elevated substrate temperature ($T_s=60^\circ$C). We found that the lifetime of the droplet exhibits a non-monotonic trend with the increase in ethanol concentration in the binary mixture, which {can be attributed to} the non-ideal behaviour of water-ethanol binary mixtures. Increasing $T_s$ decreases the lifetime of the (50\% ethanol + 50 \% water) binary droplet in a logarithmic scale. For this composition, at $T_s=60^\circ$C, we observed an early spreading stage, an intermediate pinned stage and a late receding stage of evaporation. Unlike $T_s=25^\circ$C, at the early times of the evaporation process, the contact angle of the droplet of pure water at $T_s=60^\circ$C  is greater than 90$^\circ$ (hydrophobic). {Late stage interfacial instability and even droplet break-up are observed for some (though not all) binary mixture compositions.} The evaporation dynamics for different compositions at $T_s=60^\circ$C exhibit a self-similar trend. It is also found that at $T_s=60^\circ$C the {normalised volumetric} evaporation rate is {early} constant for the entire evaporation process, indicating that the evaporation dynamics of a binary droplet {of a given composition} at $T_s=60^\circ$C is equivalent to that of another pure fluid with a higher volatility at room temperature. Finally, the evaporation rates of pure and binary droplets at different substrate temperatures are compared against a theoretical model developed for pure and binary mixture droplets. {The model predictions were found to be quite satisfactory for the steady evaporation phase of the droplet lifetimes.} 
\end{abstract}

\maketitle


\section{Introduction}
\label{sec:intro}

Evaporation of droplets is commonly encountered in atmospheric phenomena associated with clouds and raindrops, in biological systems, and also in industrial applications, such as combustion, ink-jet printing, hot-spot cooling, droplet-based microfluidics, coating technology, to name a few (see for instance \cite{brutin2015droplet,yau1996short,tripathi2015evaporating,chang2006evaporation}). The dynamics is due to the interplay between interfacial physics and phase change. One of the early {studies} on the evaporation of a spherical droplet in air was by \cite{langmuir1918evaporation}. He found that, for slow evaporation, the radius of a spherical droplet decreases as the square root of time. However, this finding was questioned for non-diffusive evaporation observed in the case of water droplets \citep{shahidzadeh2006evaporating}.  For a sessile droplet on a heated substrate, the contact line dynamics increases the complexity significantly. {A proper} understanding of the underlying physics of evaporating droplets plays an important role in wetting and surface characterization processes \citep{picknett1977evaporation}. Thus, several researchers have investigated the evaporation dynamics of droplets of pure fluids and binary mixtures placed on heated substrates. An extensive review of the literatures on recent advances on wetting and evaporation of sessile droplets can be found in \cite{brutin2018recent}. 

The objective of {the} present work is to experimentally study the evaporation dynamics of {sessile droplets} having of different compositions of {the} ethanol-water binary mixture and at different substrate temperatures. The results obtained from our experiments are compared against the theoretical models. Before reviewing the previous studies on evaporation of droplets of binary mixtures placed on a substrate, first we briefly review the literature on evaporation of a sessile droplet of pure fluids. 

Many researchers have investigated the evaporation dynamics of a sessile droplet of pure fluids on hydrophilic and hydrophobic surfaces (see e.g. \cite{erbil2012evaporation,dash2014droplet} and references therein). By conducting experiments, \cite{birdi1989study} studied the evaporation of a sessile droplet on a hydrophilic substrate and observed a constant evaporation rate with {a} pinned contact line. \cite{bourges1995influence} experimentally investigated the evaporation of sessile droplets of water and n-decane placed on different hydrophilic substrates and observed four distinct stages of evaporation depending on the roughness of the substrates. In the case of complete wetting, \cite{picknett1977evaporation} demonstrated {the existence of two} stages of evaporation, namely, {the constant-contact-angle stage and the constant-contact-line-area stage}. Later, it was observed that for partially wetting substrates, there are four stages in the droplet evaporation process, which are the early short and rapid spreading stage, {the} slow spreading stage, {the} constant-contact-angle stage and the final stage where both the contact angle and the radius decrease until the drop completely evaporates \citep{kovalchuk2014evaporation}. 

A sessile droplet placed on a heated substrate exhibits a temperature gradient in the vicinity of the solid surface and along the liquid-vapour interface, which in turn creates thermo-capillary convection inside the droplet \citep{girard2008effect,barash2009evaporation,ristenpart2007influence} and instability or hydrothermal waves (undulation) at the liquid-vapour interface for some liquids \citep{sefiane2008self}. \cite{saenz2017dynamics} investigated the evaporation of non-spherical droplets via numerical simulations and experiments. They demonstrated a universal scaling law for evaporation, which was valid even for droplets with complex shapes. \cite{ristenpart2007influence} investigated the effect of conductivity of {the} substrate on the evaporation dynamics of a droplet by conducting an asymptotic analysis. \cite{dash2014droplet,gatapova2014evaporation} have investigated the evaporation of water droplets on hydrophobic and superhydrophobic surfaces at different temperatures and compared their experimental results with the theoretical models. \cite{sobac2012} have  experimentally investigated an evaporating ethanol droplet at different temperatures on a hydrophilic and hydrophobic substrates. They have developed an evaporation model based on the quasi-steady, diffusion-driven assumptions and compared their experimental results with the theoretical predictions. \cite{carle2016,kelly2011evaporation} performed many experiments on evaporation of pure liquids droplets at different substrate temperatures and  developed empirical relationships by considering the combined influence of diffusion and convective transport.

Next we discuss the literature on the evaporation of droplets of binary mixtures when the substrate is at the ambient temperature. \cite{sefiane2003experimental} studied the evaporation of {sessile droplets} of ethanol-water {mixtures} of different compositions on a polytetrafluoroethylene (PTFE) substrate at room temperature and the dynamics was compared against that observed in {the} case of pure droplets. They found that the evaporation dynamics in case of pure fluids is very different from that of the binary mixtures. It is observed that {the volume of a droplet of pure water} or pure ethanol decreases monotonically. However, the evaporation for ethanol-water mixtures occurs {through} three distinct stages: the first and the last stages are mainly dominated by the evaporation of {the} more volatile component (ethanol) and {the} less volatile component (water), respectively whereas at the intermediate stage the volume of the drop remains almost constant, but the contact angle varies significantly. They also concluded that the dynamics contact angle largely depends on the concentration of ethanol {in the droplet}. 

In order to explain the different stages, \cite{christy2011flow,bennacer2014vortices} have studied the flow field in an evaporating ethanol-water droplet by varying the composition of the binary mixture at the room temperature using particle image velocimetry (PIV). They observed multiple vortices of random orientations at the early stage, which is the consequence of the concentration gradient resulting due to the evaporation of ethanol. At the intermediate stage a spike of outward flow was noticed, which deposits the remaining ethanol close to the apex of the drop leading to a solutal-thermocapillary flow inside the droplet. This is followed by the radial flow towards the contact line in the final stage. The final stage is similar to the one observed in case of pure water droplet. \cite{saenz2017dynamics} also investigated complex shaped droplets consisting of ethanol-water binary mixtures by conducting theoretical modelling and infrared thermography at different substrate temperatures and concentrations. However, most of the results presented in their study are at room temperature only. \cite{schofield2018lifetimes} theoretically studied the influence of the thermal resistance of the liquid and the substrate on the lifetime of an evaporating ethanol-water droplet. 

\cite{wang2008evaporation} and \cite{shi2009wetting} also observed a similar behaviour in {the} case of a sessile droplet of ethanol-water mixtures on a poly-methyl-methacrylate (PMMA) and PTFE surfaces, respectively. \cite{wang2008evaporation} also proposed a diffusion model to describe the evaporation stages. \cite{cheng2006evaporation} investigated the evaporation dynamics of {droplets} of ethanol-water mixtures on a gold surface and demonstrated the physical mechanism for different stages of the evaporation process. They reported that the change in the evaporation mode of binary droplets is dominated by the wetting hysteresis and the initial evaporation of the more volatile component (ethanol).

Later, \cite{sefiane2008wetting} extended their previous study on ethanol-water mixtures to {the investigation} of the evaporation dynamics of {water-methanol droplets} on a smooth polymer coated substrate. They observed that although the first stage was still dominated by the {evaporation of the more} volatile component (methanol), a small amount of methanol {remained in the solution} even after the first stage, which influenced the wetting behaviour after the first stage. Four distinct stages of the contact angle dynamics were observed. By conducting a theoretical analysis of the evaporative flux, the dynamics was explained using the antagonistic effects of the evaporation and the resulting flow field. By considering the slow evaporation of a binary mixture consisting of carbon diols (less volatile) and pure water (more volatile), \cite{karpitschka2017marangoni} conducted experiments and numerical simulations to study the Marangoni contraction of the droplet at $21^\circ$C. By conducting lubrication analyses, \cite{karapetsas2016evaporation,karapetsas2014thermocapillary,karapetsas2013effect} investigated the dynamics of a sessile drop under the influence of thermo-capillary forces and evaporation flux. In the case of binary mixtures of non-azeotropic, high carbon alcohol solutions, which exhibit parabolic surface tension-temperature dependences with well-defined minima, they observed super-spreading behaviour of the droplet.

As the above-mentioned review shows, the evaporation dynamics of pure droplets have been investigated extensively for both low and {high temperature substrate conditions}; however, to the best of our knowledge, the evaporation dynamics of a droplet consisting of a binary mixture have been studied for different compositions, but only at room temperature. Moreover, even at room temperature, the behaviour reported by the previous studies for different concentrations were also different as the dynamics depends on several factors, such as the property of the substrate (which have not been provided in most of the previous studies) and ambient conditions.  Therefore, the present work focuses on the evaporation of a sessile droplet of ethanol-water mixture at different substrate temperatures. As the effect of the substrate properties is expected to be even more pronounced at elevated temperatures, we have studied the properties of the cellulose-acetate tape and the PTFE substrates using the atomic force microscopy (AFM) and the scanning electron microscope (SEM) techniques. We found that the cellulose-acetate tape is stable at elevated temperatures. Thus the experiments are conducted on this substrate. The wetting and the spreading dynamics of droplets of ethanol-water mixtures are investigated at different temperatures and for different compositions. The temperature of the substrate is varied from $25^\circ$C to $60^\circ$C as the boiling temperature of pure ethanol is about $78^\circ$C \citep{lemmon2007nist}. The concentration of ethanol is varied from 0\% ethanol (pure water) to 100 \% ethanol (pure ethanol) on a volume basis. We observed that the evaporation behaviour is non-monotonic for high concentrations of ethanol at elevated temperatures. In order to understand the underlying physics, theoretical models have been developed for the evaporation of droplets of pure and binary mixtures of water and ethanol. 
It is found that the predictions from the theoretical models agree well with the experimental results. 

The rest of the paper is organised as follows. In Section \ref{sec:expt}, we provide a detailed discussion of the experimental set-up and the procedure followed in the present work. The results obtained from our experiments are presented in Section \ref{sec:dis}. The theoretical models for the evaporation of a sessile droplet in different situations are developed in Section \ref{sec:theory_drop}. The behaviours obtained from the theoretical models have been compared against the present experimental results in Section \ref{theory_expt_comp} for sessile droplets of pure and binary mixtures at the room and elevated temperatures. Finally, concluding remarks are given in Section \ref{sec:conc}.


\section{Experimental set-up and data processing technique}
\label{sec:expt}

\begin{figure}[h]
\centering
\includegraphics[width=0.95\textwidth]{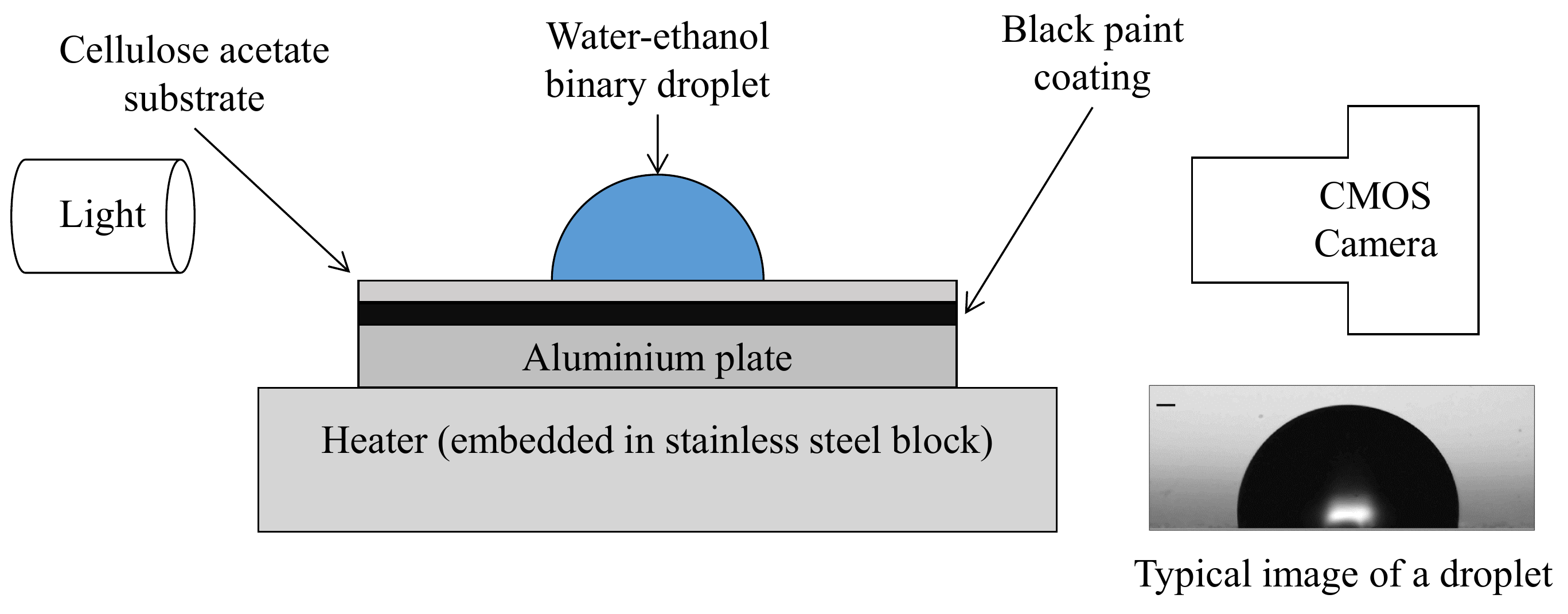}
\caption{Schematic diagram of the experimental set-up. It consists of a heater, a cellulose acetate substrate placed on a stainless steel plate, a light source and a camera. A typical image of an ethanol-water droplet recorded using a CMOS camera is also shown.}
\label{fig:geom}
\end{figure}

We investigate the evaporation dynamics of a sessile droplet of a binary mixture of water (W) and ethanol (E) on a heated substrate maintained at different temperatures. The experimental set-up consists of a heater, a motorised pump to create the droplet, cellulose acetate substrate placed on a multi-layered block, a light source and a complementary-metal-oxide-semiconductor (CMOS) camera {(make: Do3Think, model: DS-CBY501E-H)}. The schematic of the experimental set-up is shown in Figure \ref{fig:geom}. The substrate temperature is varied and maintained at a fixed temperature during each experiment using a proportional-integral-derivative (PID) controller. The entire setup is placed inside a big metallic box to minimise outside disturbances. This customised experimental set-up was fabricated by Holmarc Opto-Mechatronics Pvt. Ltd. The experiments are conducted in an air-conditioned room maintained at {a temperature of 22$^\circ$C and a relative humidity of} 36\%, which have been measured with the help of an HTC 288-ATH hygrometer. 

\begin{figure}[h]
	\centering
	\hspace{0.0cm}  (a) \hspace{6.0cm} (b) \\
	\includegraphics[width=0.425\textwidth]{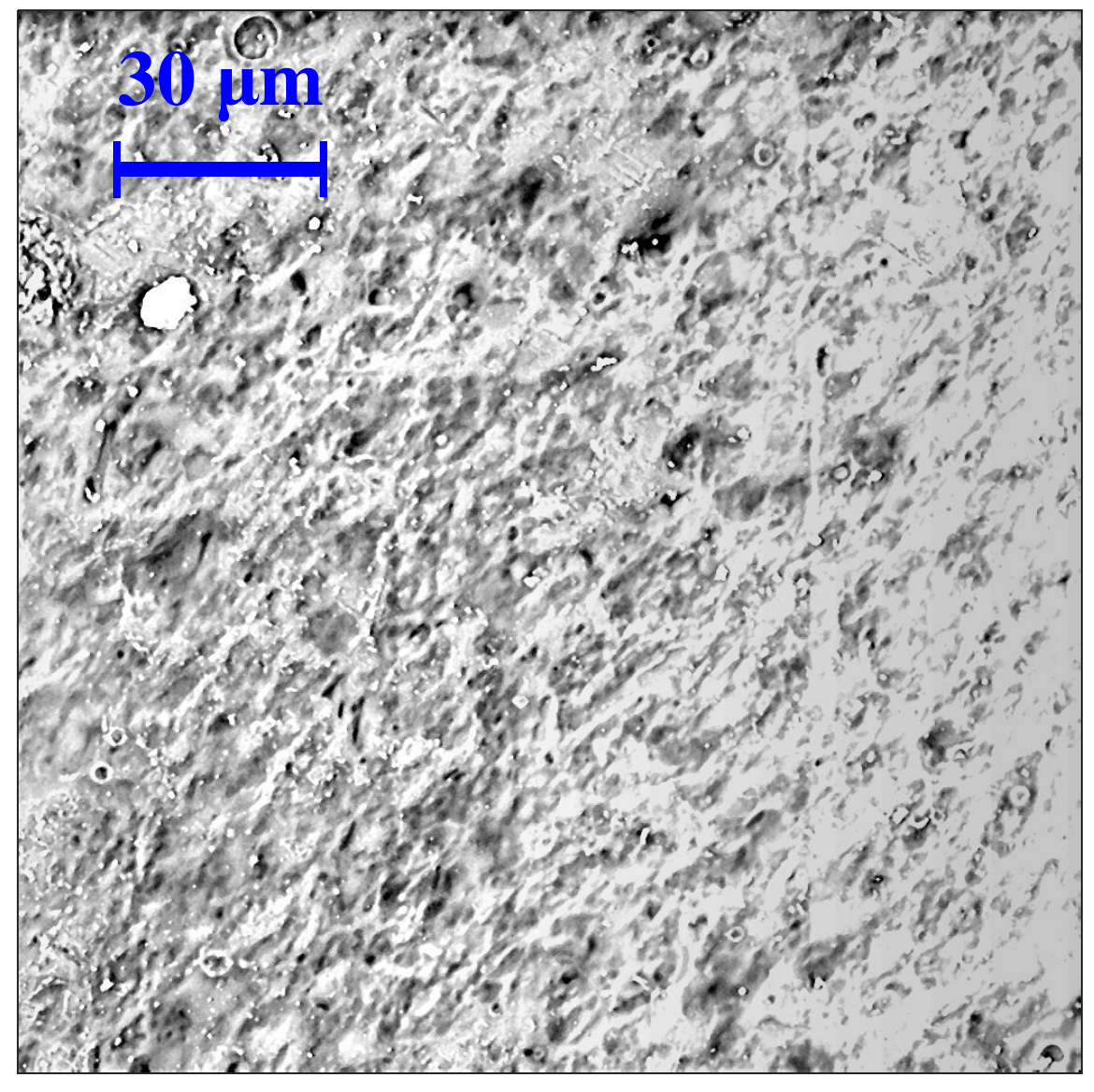} \hspace{2mm} \includegraphics[width=0.425\textwidth]{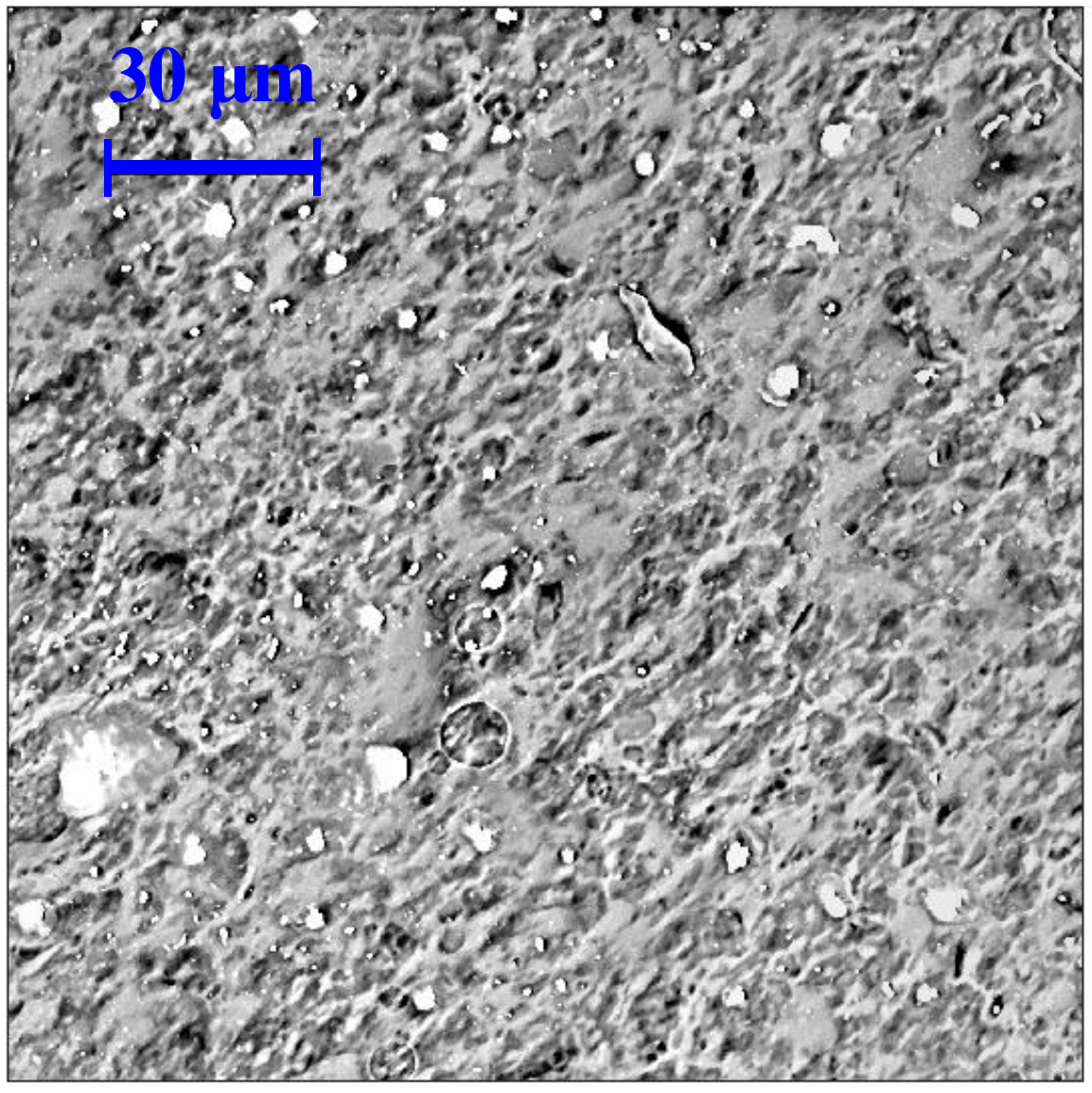} \\
	\hspace{0.0cm}  (c) \hspace{6.0cm} (d) \\
	\includegraphics[width=0.425\textwidth]{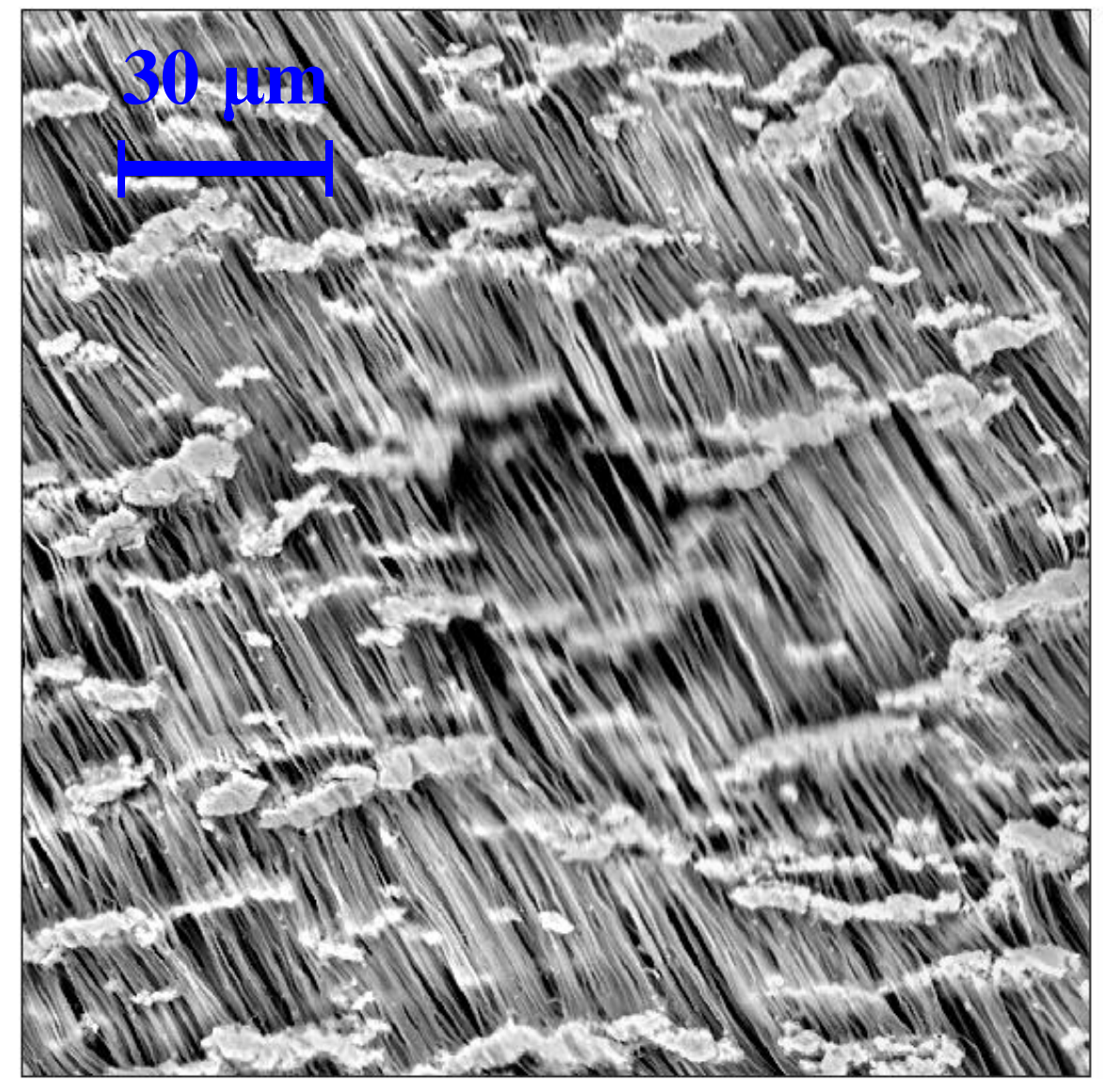} \hspace{2mm} \includegraphics[width=0.425\textwidth]{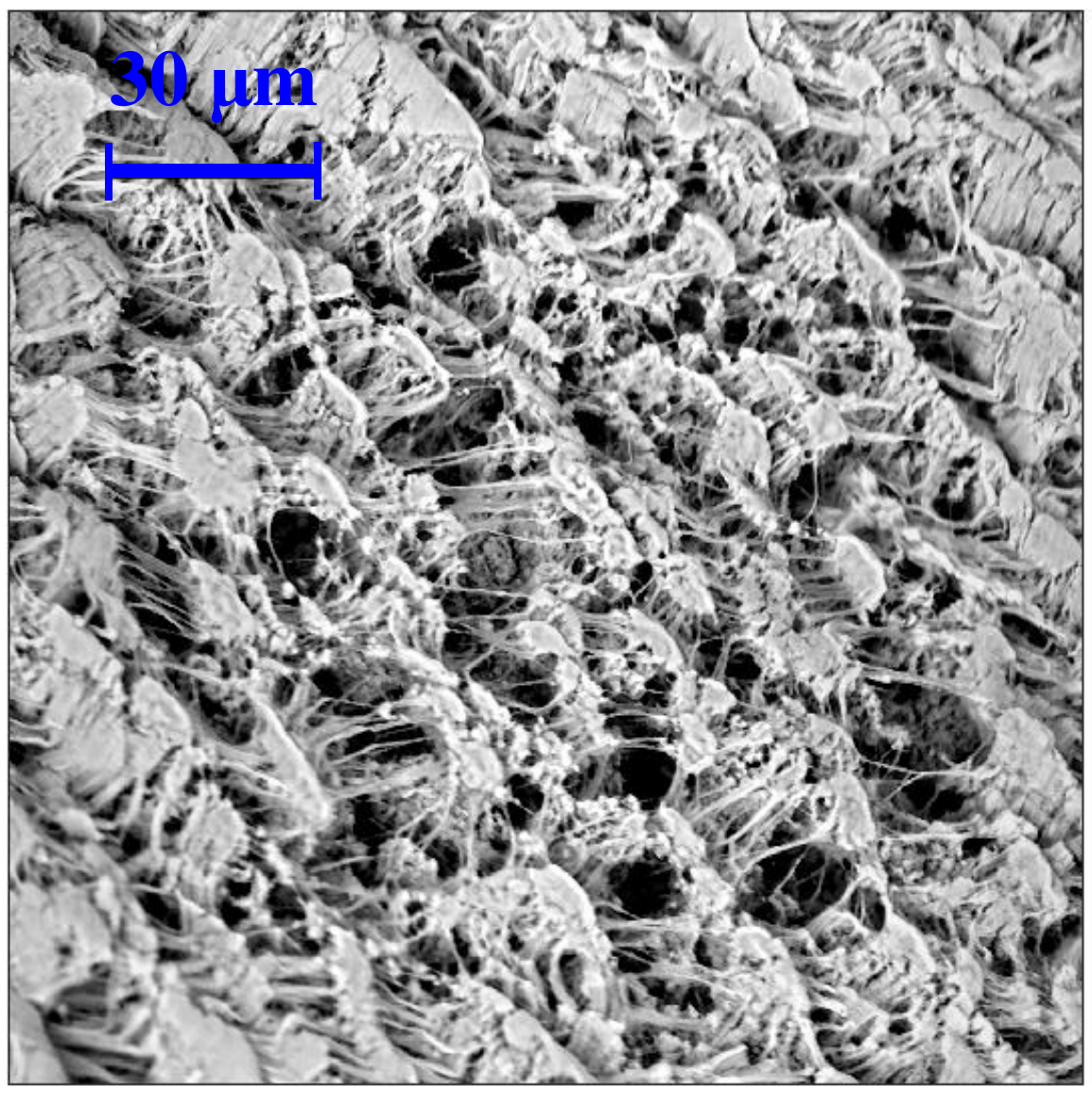} 
	\caption{The scanning electron microscope images of (a, b) cellulose acetate substrate at 25$^\circ$C and after using it at 60$^\circ$C, respectively; (c, d) polytetrafluoroethylene (PTFE) tape at 25$^\circ$C and after heating at 60$^\circ$C for about 10 minutes, respectively. The length of the scale bar is 30 microns.}
	\label{SEMimages}
\end{figure}

\begin{figure}[h]
	\centering
	\hspace{0.0cm}  (a) \hspace{6.0cm} (b) \\
	\includegraphics[width=0.48\textwidth]{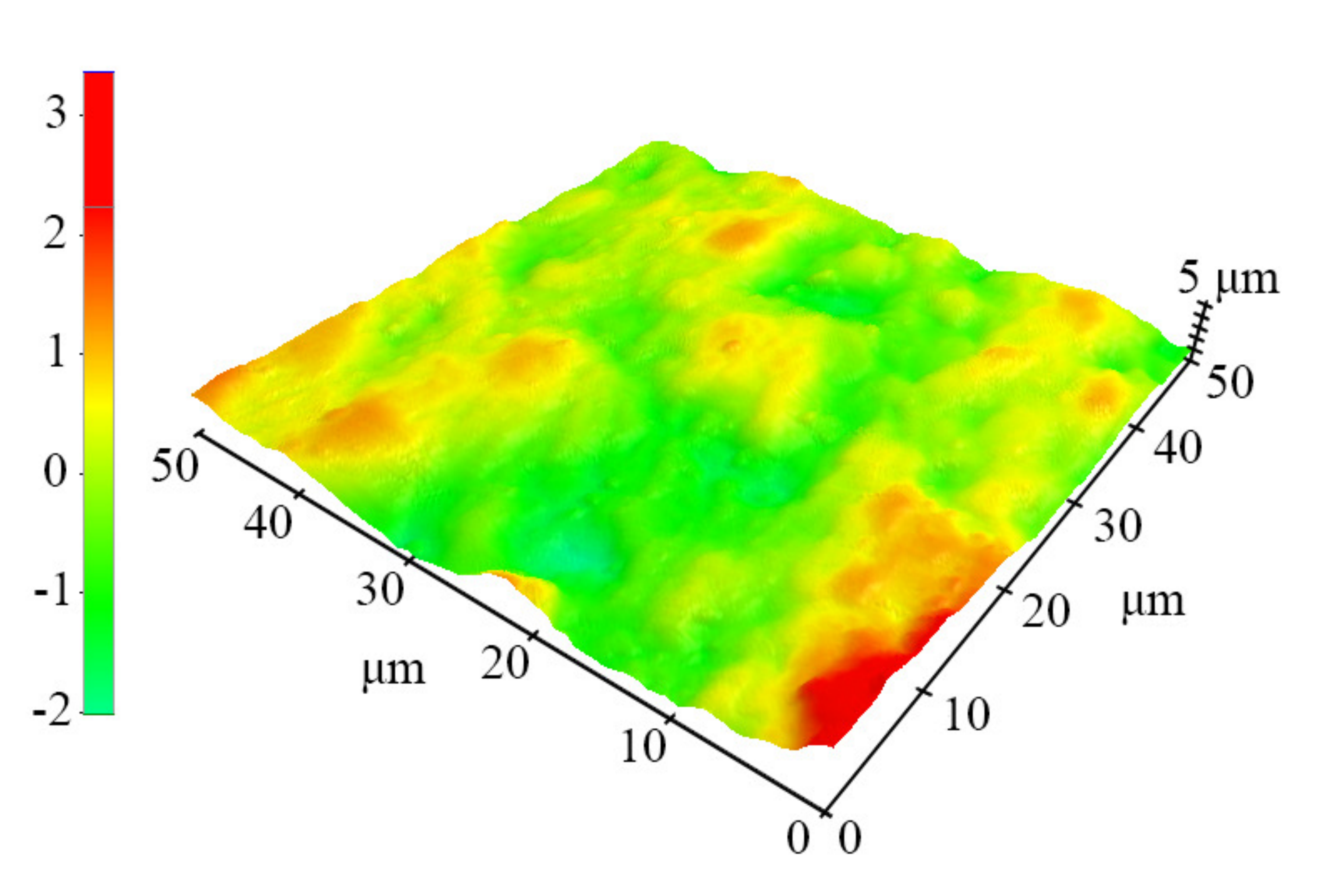} \hspace{0mm} \includegraphics[width=0.48\textwidth]{Figure3a.pdf} 
	\caption{The atomic force microscopy images of the cellulose acetate substrate at (a) 25$^\circ$C and (b) 60$^\circ$C. The colour bar indicates the surface roughness in microns.}
	\label{AFM}
\end{figure}

The multi-layered block contains a stainless steel plate of size 100 mm $\times$  80 mm $\times$ 15 mm with two PID regulated electrical heaters situated at its base (see Figure \ref{fig:geom}). The substrate is a cellulose-acetate tape of thickness 63 $\mu$m placed on an aluminium block of thickness 5 mm coated with a black paint, which is placed over the stainless steel plate. The scanning electron microscope {(make: Thermofisher Scientific, model: Phenom Prox)} images of the cellulose-acetate tape before and after experimentation at with 60$^\circ$C are shown in Figures \ref{SEMimages}(a) and (b), respectively. To check the effect of another substrate on the droplet evaporation dynamics, we have also considered polytetrafluoroethylene (PTFE) tape of thickness 75 $\mu$m. The SEM images of the PTFE tape at room temperature and after heating at 60 $^\circ$C for about 10 minutes are shown in Figures \ref{SEMimages}(c) and (d), respectively. It can be seen that the surface roughness of cellulose-acetate tape is uniform even after {experimentation at} 60$^\circ$C, but the surface property of PTFE tape changes when subjected to heating. Thus, we choose the cellulose-acetate tape as the substrate in the present study. Figures \ref{AFM}(a) and (b) show the atomic force microscopy {(make: Park System, model: NX10) images of the cellulose acetate substrate at room temperature and after heating at 60$^\circ$C, respectively.} The root-mean-square values of the roughness of the cellulose acetate substrate at 25$^\circ$C and 60$^\circ$C are 0.668 $\mu$m and 0.641 $\mu$m, respectively. The AFM images confirm that the surface roughness of the cellulose acetate substrate is not affected much by the high-temperature exposure.

The substrate temperature is set in the PID controller, and the heater is switched on to stabilise the set temperature. The temperature on the aluminium plate is measured using an infrared heat gun (which measures the surface temperature with an error of about $\pm 0.5^\circ$C), and the controller is adjusted to achieve the required substrate temperature. To achieve a steady state condition, we wait for an hour and then apply the cellulose-acetate tape on the stainless steel plate. Again we wait for 10 minutes before placing the droplet on the substrate. We observe a negligible temperature gradient on the surface owing to the small thickness of tape.

The binary solutions are prepared by varying the volume concentration of ethanol in purified deionised water (purity of 18:2 M$\Omega$) such that the total volume of the solution is 100 ml. {As water and ethanol are miscible at all concentrations, they are thoroughly mixed using a stirrer, and homogeneous solutions of different compositions are prepared.} An U-TeK chromatography syringe of size 100 $\mu$l with a needle of outer diameter 1.59 mm (supplied by Unitek Scientific Corporation) is connected with a motorised pump to create a droplet, which is gently placed on the substrate. The motorised pump controls the volume flow rate of the solution with an error $< 1\%$ at the desired flow rate. The droplet volume is kept constant at 5 $\mu$l for all the experiments conducted in the present study. The composition of the ethanol-water solution and the temperature of the substrate are varied, and the dynamics of the droplet is recorded using the CMOS camera at 10 frames-per-second (fps) with a spatial resolution of 1280 $\times$ 960 pixels. After each experiment, the syringe is cleaned with acetone and is allowed to dry, and the cellulose-acetate tape is also replaced. For each set of parameters, more than four repetitions of the experiment are conducted.

\begin{figure}[h]
	\centering
	\includegraphics[width=0.6\textwidth]{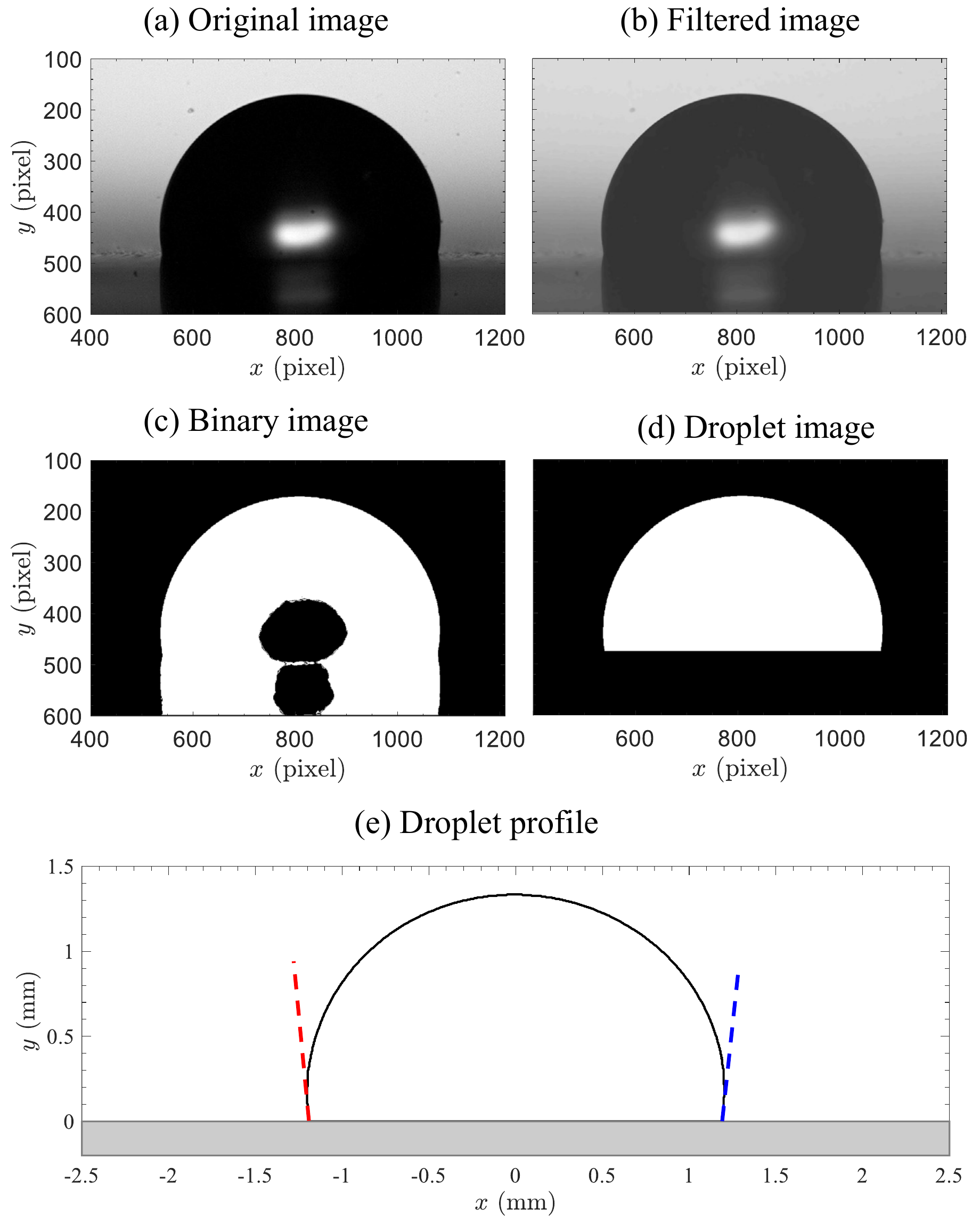}
	\caption{The image processing steps for a typical ethanol-water droplet recorded using the camera. (a) Typical original image of a droplet, (b) the corresponding filtered image, (c) the binary image processed using Matlab software, (d) the final re-constructed image of the droplet and (e) the droplet profile.}
	\label{fig:imageprocessing}
\end{figure}

The acquired droplet images are processed using the Matlab software to calculate the wetting diameter ($D$), the height ($h$), the right and left side contact angles ($\theta_r$ and $\theta_l$) {with respect to the camera field of view} and the volume of the droplet ($V$) as a function of time normalised by the lifetime of droplet, $t_e$ {(i.e. time taken by the droplet to evaporate completely). The time, $t$ is measured from the instant when the droplet reaches its initial equilibrium state.} Figure \ref{fig:imageprocessing}(a) shows the snapshot of the water droplet placed on the substrate,  where the substrate is situated around 500 pixels from the top and the dark image below is the reflection of the droplet. The reflection of the backlighting is also visible as a white spot in the middle of the droplet. To extract the droplet contour from the image sequences, we have developed an in-house image processing tool using the Matlab software. First, the random noises in the images are eliminated using median filtering technique. Then, the gradients are improved by sharpening the image using unsharp masking technique. Figure \ref{fig:imageprocessing}(b) shows an example of the filtered image, where the boundary of the droplet is much sharper than in the original image. Next, {the filtered} image is converted into a binary image using a suitable threshold value, which separates the droplet boundary from the background as can be seen in Figure \ref{fig:imageprocessing}(c). The final step is the filling of holes within the droplet boundary and removing the reflection part from the droplet image. The droplet contour is then traced from the image (Figure \ref{fig:imageprocessing}(d)) using the Matlab function, which is plotted in Figure \ref{fig:imageprocessing}(e) in world dimensions (scaling factor = 227.27 pixels/mm). The red and blue dotted lines are the tangents calculated at the left and right side of contact points, respectively. From the contour profile, all the geometrical information and the volume of the droplet are calculated by assuming it to be {of spherical-cap shape}. However, for high concentrations of ethanol, undulations are observed on the free surface of the droplet at later times, which limits the ability of the image processing tool to post-process the data at these later times (the end stage of evaporation).

\begin{figure}[h]
\centering
 \hspace{0.6cm}  (a) \hspace{6.0cm} (b) \\
\includegraphics[width=0.46\textwidth]{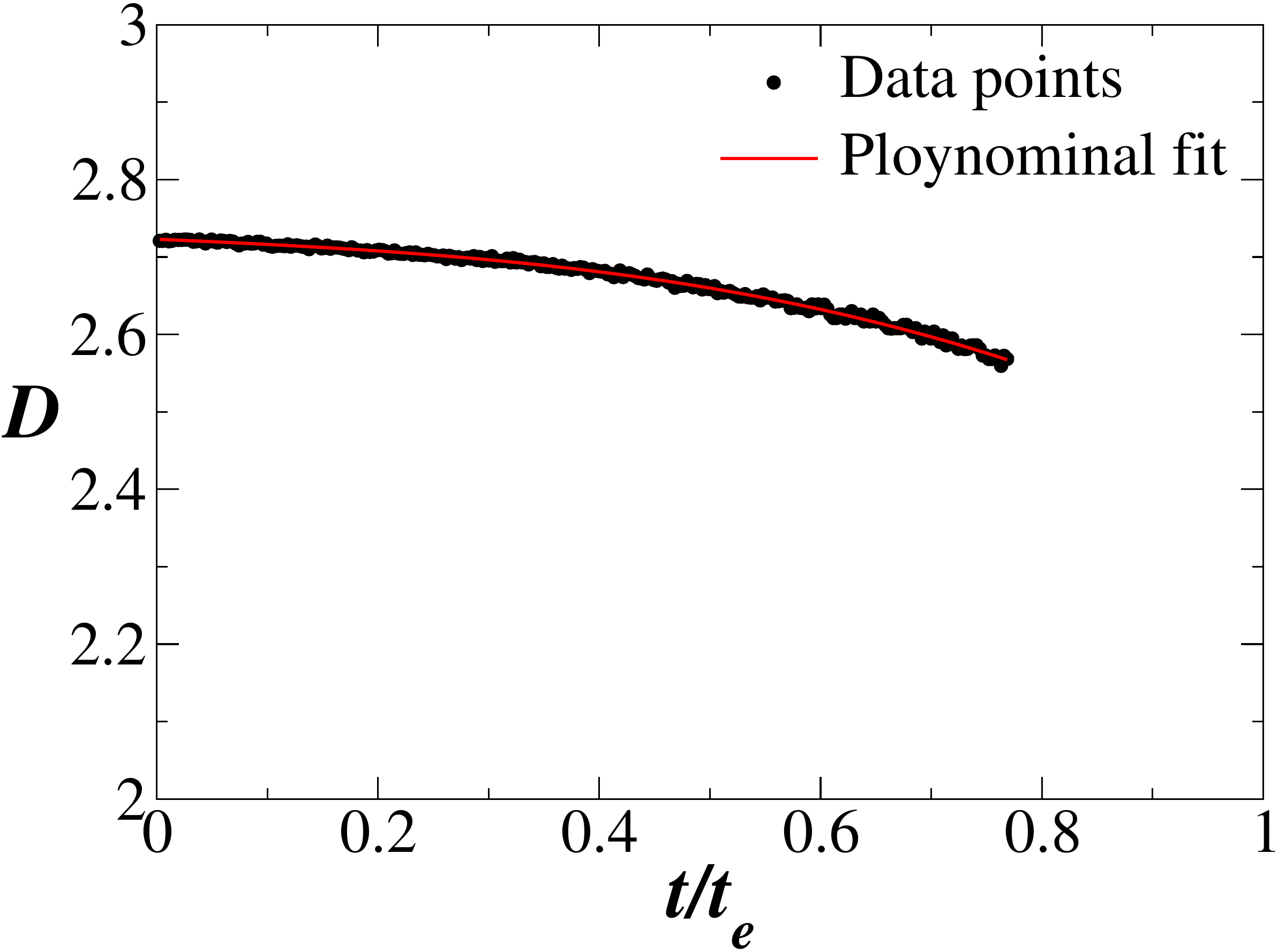} \hspace{2mm} \includegraphics[width=0.46\textwidth]{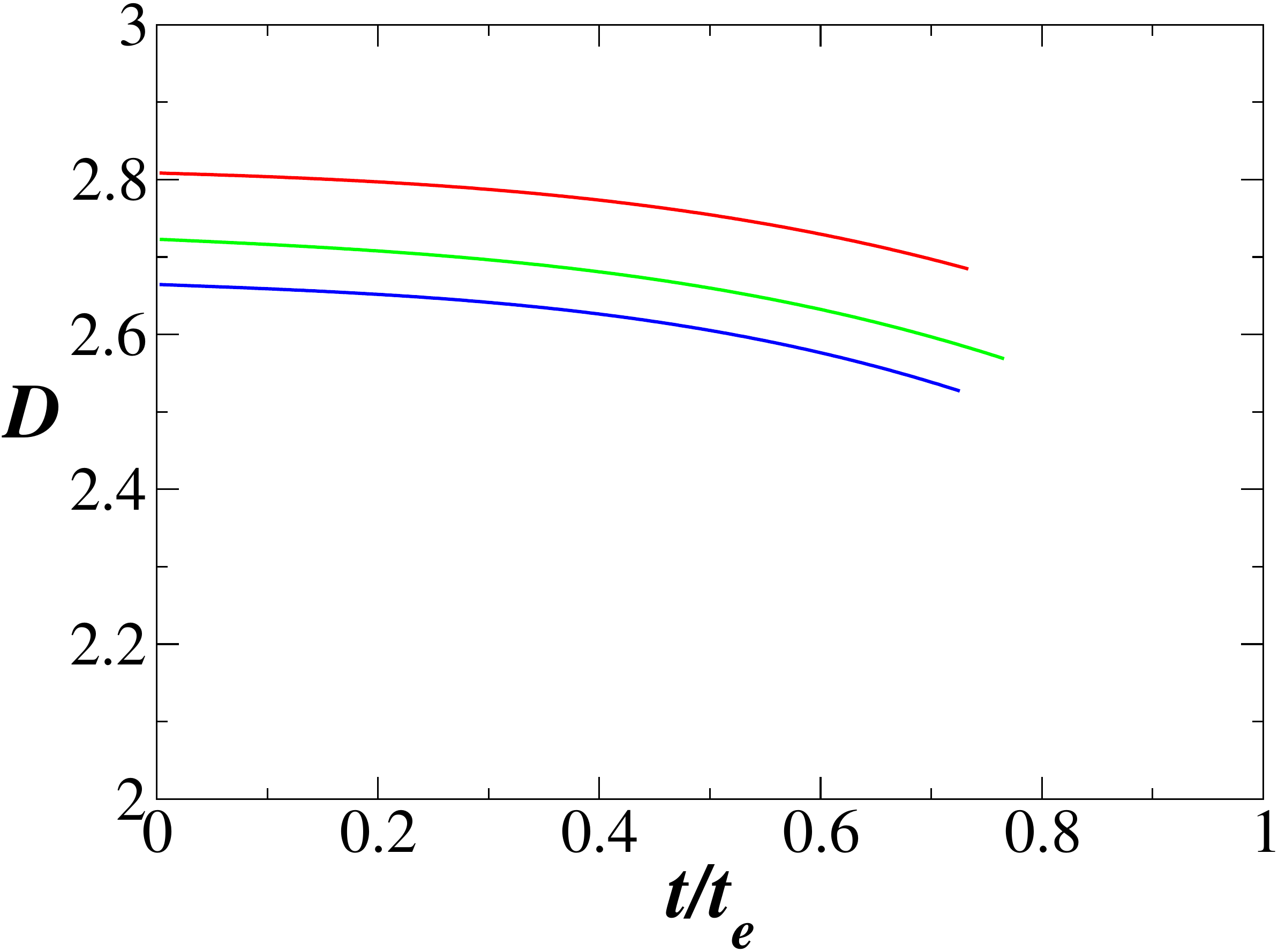} \\
 \hspace{0.6cm}  (c) \hspace{6.0cm} (d) \\
\includegraphics[width=0.46\textwidth]{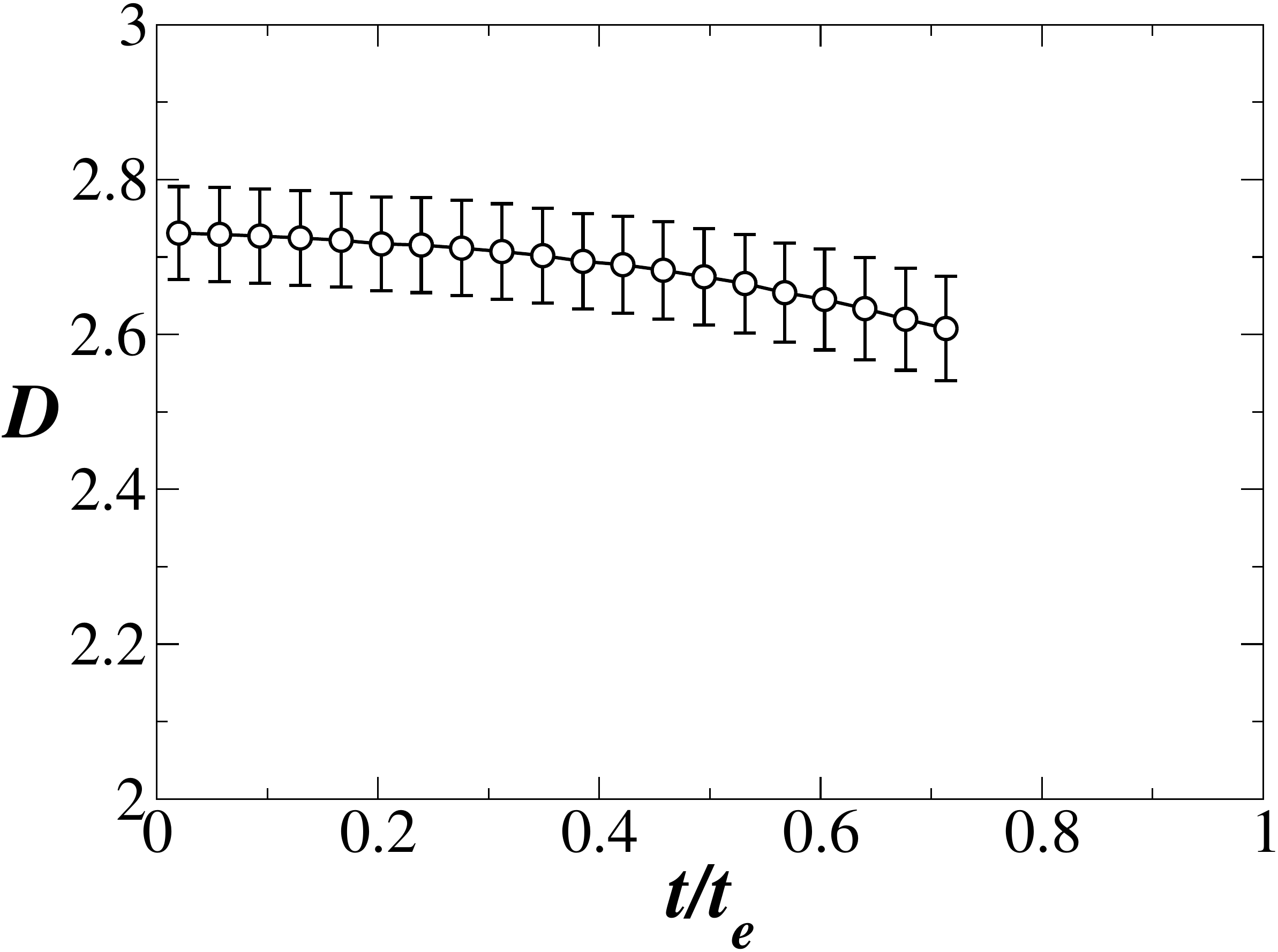} \hspace{2mm} \includegraphics[width=0.46\textwidth]{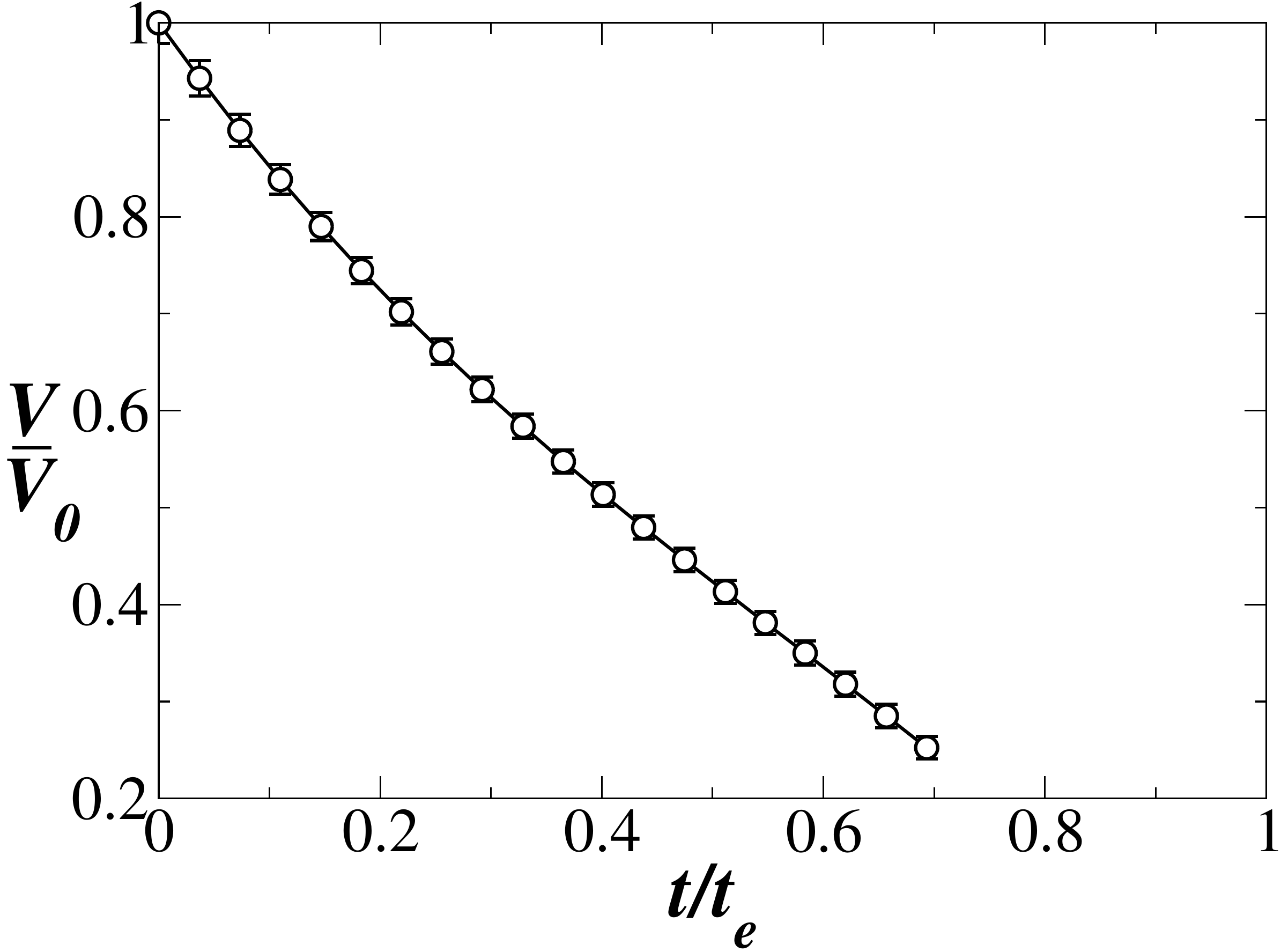} \\
\caption{The steps associated with the data processing for a typical droplet. (a) The typical data points for the diameter of the droplet, $D$ (mm) versus $t/t_e$ for one recording video and the polynomial fit, (b) the polynomial fits for three repetitions, (c) the resultant polynomial fit with an error bar and (d) the final $V/V_0$ versus $t/t_e$ plot. }
\label{fig:dataprocessing}
\end{figure}

An example of the evolution of the wetting diameter $(D)$ as a function of the normalised time scale $(t/t_e)$ is shown in Figure \ref{fig:dataprocessing}(a). The data points are gathered from one image sequence, and the red line shows the corresponding third-degree polynomial fit to the data points to reduce the high-frequency noise added into it by the digitization of the droplet boundary. Figure \ref{fig:dataprocessing}(b) shows the comparison of polynomial fit values of three repetitions for one test condition and the corresponding mean and standard deviation are plotted in Figure \ref{fig:dataprocessing}(c). Figure \ref{fig:dataprocessing}(d) shows the normalised volume $(V/V_0)$ evolution as a function of the normalised time with error bars. The maximum error in calculating the contact angles, the droplet height and its radius is found to be $< 6$\% for all the image sequences.

\section{Experimental results and discussion}
\label{sec:dis}

Based on the physics observed at different substrate temperatures and compositions of the binary mixture, we have presented our experimental results in three subsections: (i) the evaporation of a droplet of pure water (E 0\% + W 100\%), a binary mixture (E 50\% + W 50\%) and the pure ethanol (E 100\% + W 0\%) at $25^\circ$C substrate temperature; (ii) the evaporation of droplets of ethanol-water binary mixtures of different compositions at an elevated substrate temperature ($T_s=60^\circ$C); and (iii) the effect of varying temperature of the substrate on the evaporation of a droplet of (E 50\% + W 50\%) composition. {Here, (E $x$\% + W $y$\%), represents an ethanol-water binary solution containing $x$\% (volume based) of ethanol and $y$\% (volume based) of water.}

\subsection{Evaporation at a nearly ambient temperature}
\label{sec:exptdis_rt}
\begin{table}   
\begin{center}
\begin{tabular}{c|c|c}
Liquids & E (\%) \quad W (\%)   &$t_e$(s) at 25$^\circ$C \\
   &    & \\
Pure water & 0  \qquad 100 & 1488 $\pm$ 63 \\
Binary mixture  & 50  \qquad 50 & 1035 $\pm$ 13  \\
 Pure ethanol  &  100 \qquad  0 & 183 $\pm$ 3  \\
 \end{tabular}
\end{center}
\caption{Lifetime of the droplets, $t_e$(s) of different compositions of ethanol-water binary mixture at $T_s=25^\circ$C.} \label{T0}
\end{table}

In this section, we discuss the evaporation dynamics observed at the substrate temperature, $T_s=25^\circ$C and at 1 atmosphere pressure by considering droplets of pure water (E 0\% + W 100\%), a binary mixture (E 50\% + W 50\%) and pure ethanol (E 100\% + W 0\%). The observed droplet lifetimes $(t_e)$ are given in Table \ref{T0}. It can be seen that the droplet of pure ethanol (more volatile) evaporates about eight times faster than the droplet of  pure water (less volatile) of the same volume (5 $\mu$l) at $T_s=25^\circ$C. The evaporation time for the (E 50\% + W 50\%) binary droplet is only 30\% less than that of the pure water droplet, as more volatile ethanol component in the binary mixture evaporates at the short early stage leaving the less volatile water to dominate the long late stage evaporation process.

\begin{figure}[h]
\centering
\includegraphics[width=0.9\textwidth]{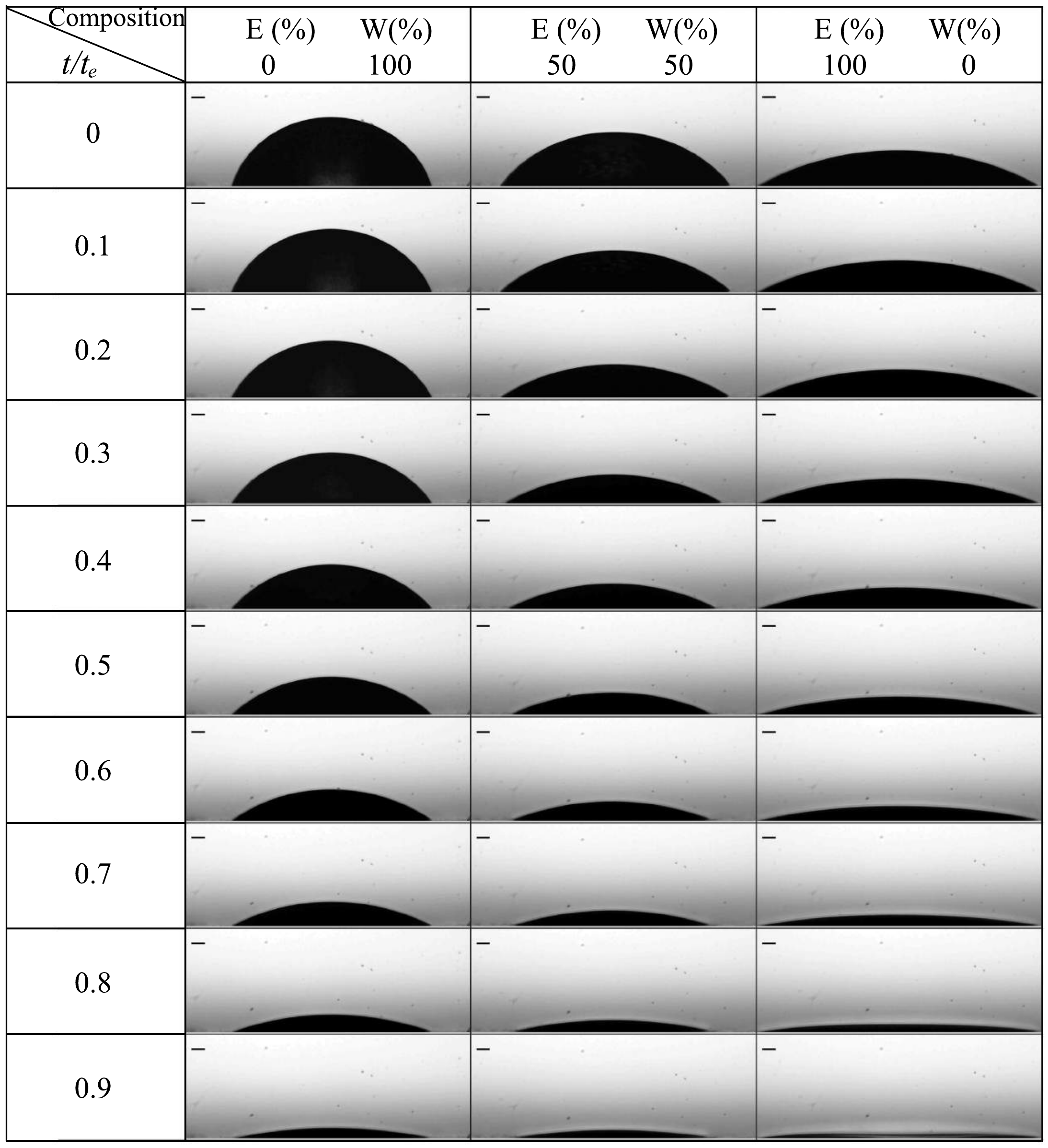}
\caption{Temporal evolution of droplet shape for pure water (E 0\% + W 100\%), (E 50\% + W 50\% solution) and pure ethanol (E 100\% + W 0\%) at 25$^\circ$C substrate temperature. The length of the scale bar shown in each panel is $200$ $\mu$m.}
\label{fig:fig3}
\end{figure}

The photographic images capturing the temporal evolution of the droplets of pure water, (E 50\% + W 50\%) binary mixture and pure ethanol at different times normalised by their corresponding lifetimes, $t_e$ are shown in Figure \ref{fig:fig3}. It can be seen in Figure \ref{fig:fig3} that the initial equilibrium wetting diameter of the droplet increases with the increase in the volume-fraction of ethanol in the ethanol-water binary mixture. For pure water (E 0\% + W 100\%) and pure ethanol (E 100\% + W 0\%), the droplet remains pinned for most of the duration of the evaporation process. However, for a droplet of the binary mixture with (E 50\% + W 50 \%), the wetting diameter of the droplet begins to recede for $t/t_e>0.2$. Close inspection of the pure ethanol droplet at $t/t_e=0.9$ also reveals the presence of surface undulations since the height of the droplet at the middle does not remain maximum at $t/t_e=0.9$. A similar plot was given by \cite{sterlyagov2018} for an ethanol-water droplet on a teflon substrate for different compositions at $24^\circ$C. The dynamics observed in the present study qualitatively agrees with that presented by \cite{sterlyagov2018}. 

\begin{figure}[h]
\centering
\includegraphics[width=0.98\textwidth]{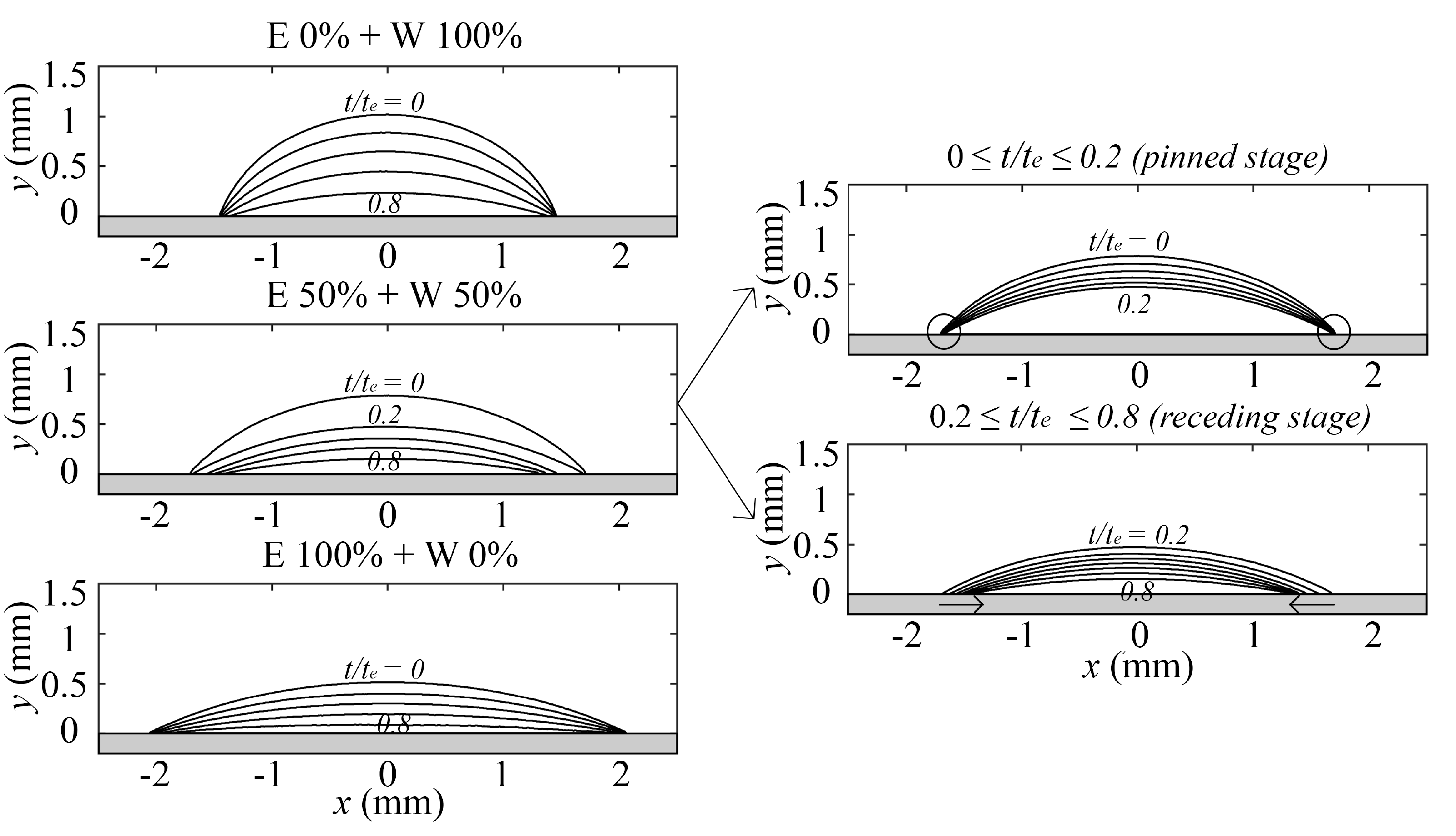}
\caption{Contours of the droplet for pure water (E 0\% + W 100\%), (E 50\% + W 50\%) solution and pure ethanol (E 100\% + W 0\%) at different normalised times, $t/t_e$. The contours are plotted at an interval of 0.2. The substrate temperature, $T_s$ is 25$^\circ$C. }
\label{fig:fig3b}
\end{figure}

The behavioural differences between the droplets of pure fluids and of the (E 50\% + W 50\%) binary mixture can be clearly seen in the contour diagrams presented in Figure \ref{fig:fig3b}. The pinned contact line behaviour is observed upto $t/t_e \cong 0.8$ in the case of pure water (E 0\% + W 100\%) and of pure ethanol (E 100\% + W 0\%) as is evident in the top and the bottom left panels of Figure \ref{fig:fig3b}. In contrast, for the droplet of (E 50\% + W 50\%) binary mixture, as shown in the middle left panel of Figure \ref{fig:fig3b}, the pinned phase is only upto $t/t_e \cong 0.2$ (also see the top right panel of Figure \ref{fig:fig3b}) and the droplet begins to recede for $t/t_e > 0.2$ (bottom right panel of Figure \ref{fig:fig3b}). 

\begin{figure}[h]
\centering
 \hspace{0.6cm}  (a) \hspace{6.0cm} (b) \\
\includegraphics[width=0.46\textwidth]{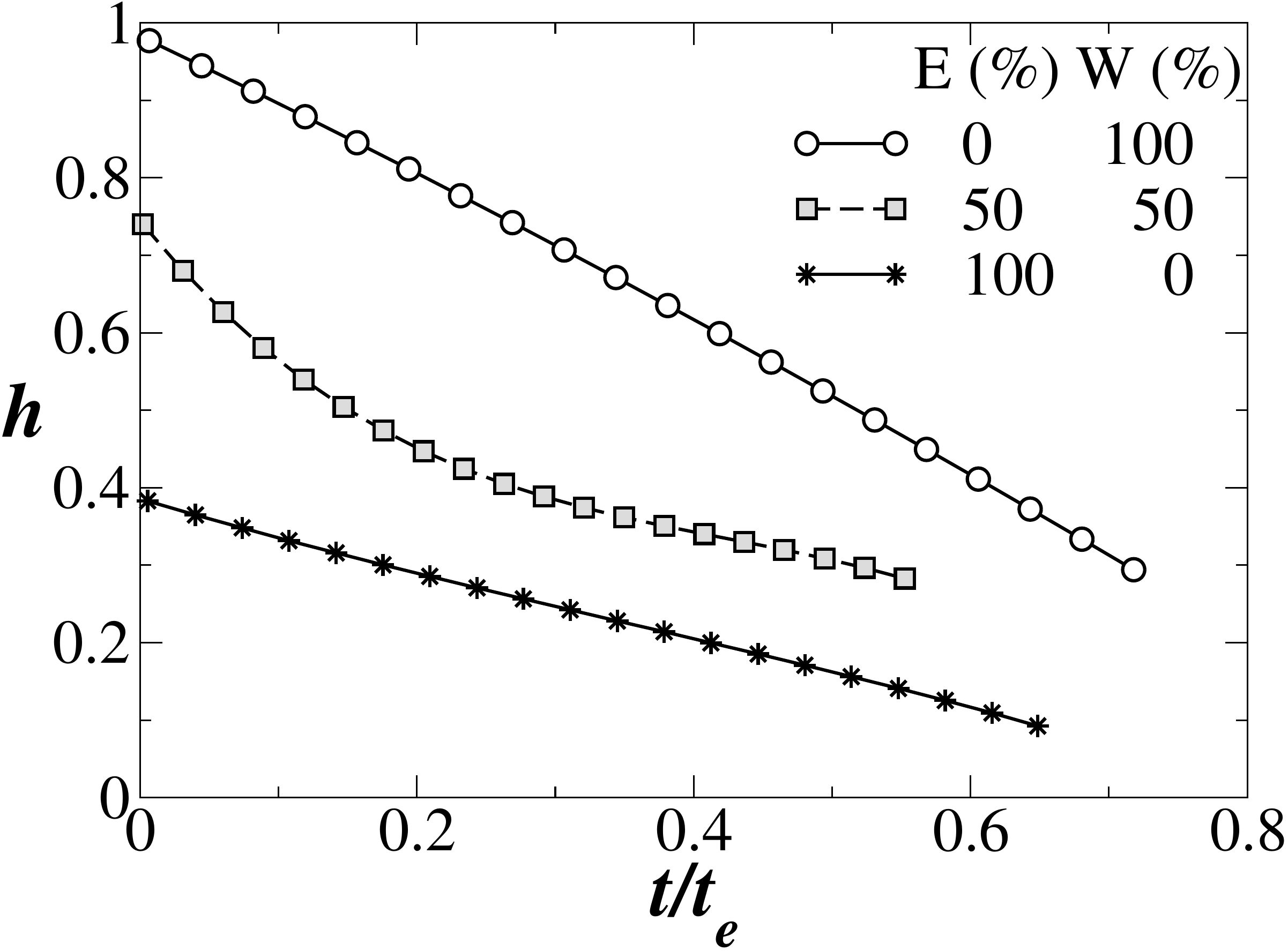} \hspace{2mm} \includegraphics[width=0.46\textwidth]{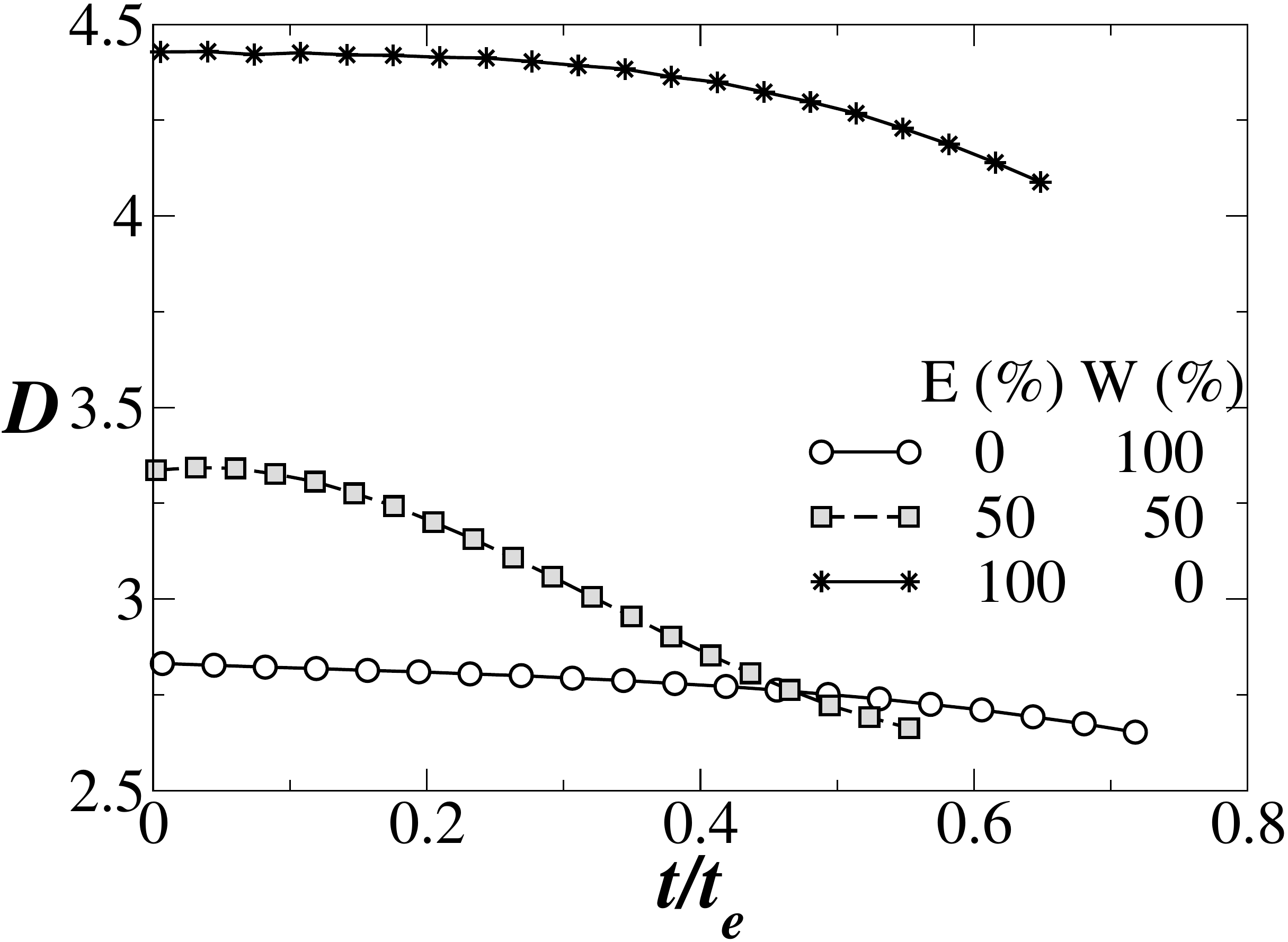} \\
 \hspace{0.6cm}  (c) \hspace{6.0cm} (d) \\
\includegraphics[width=0.46\textwidth]{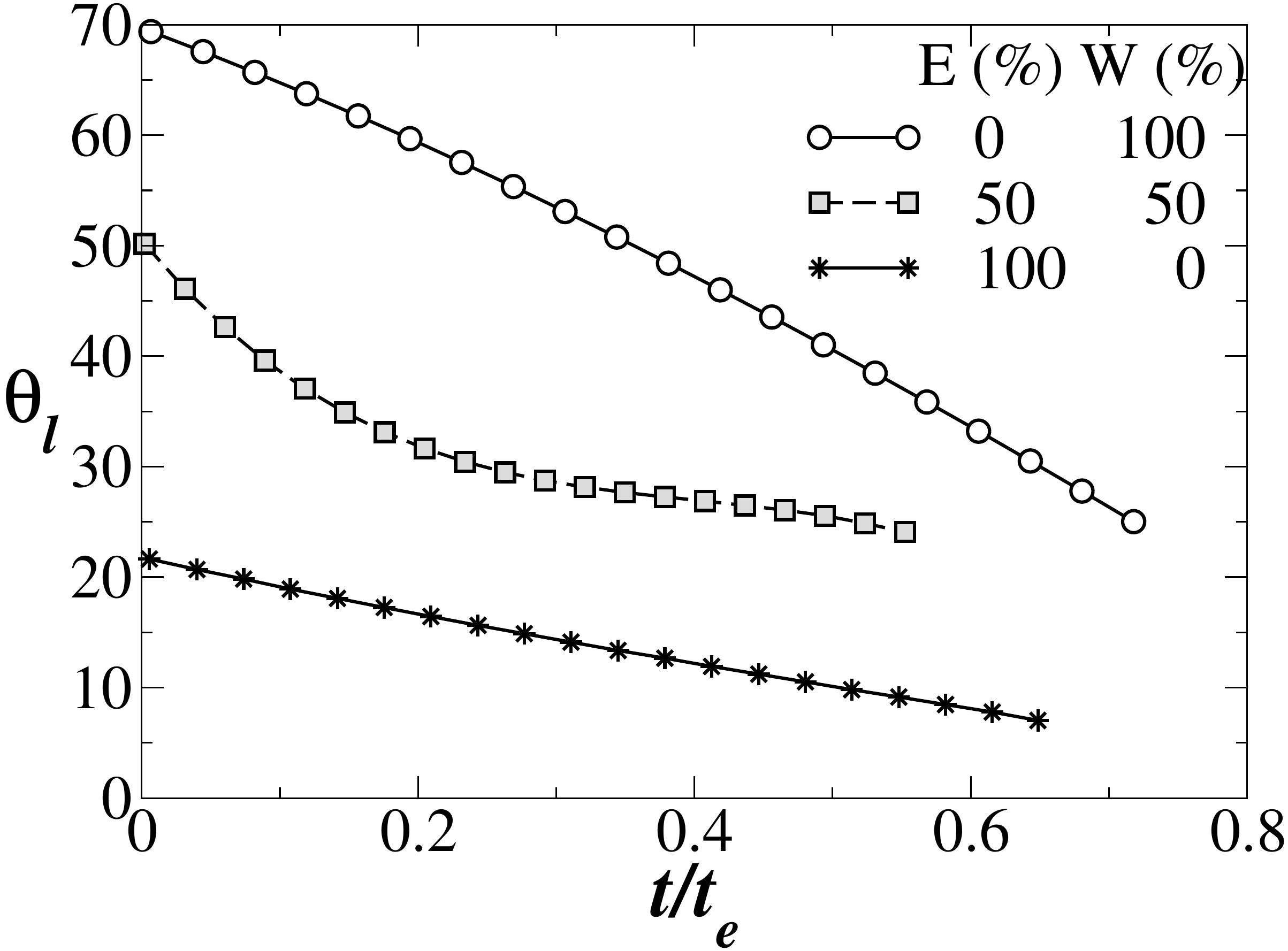} \hspace{2mm} \includegraphics[width=0.46\textwidth]{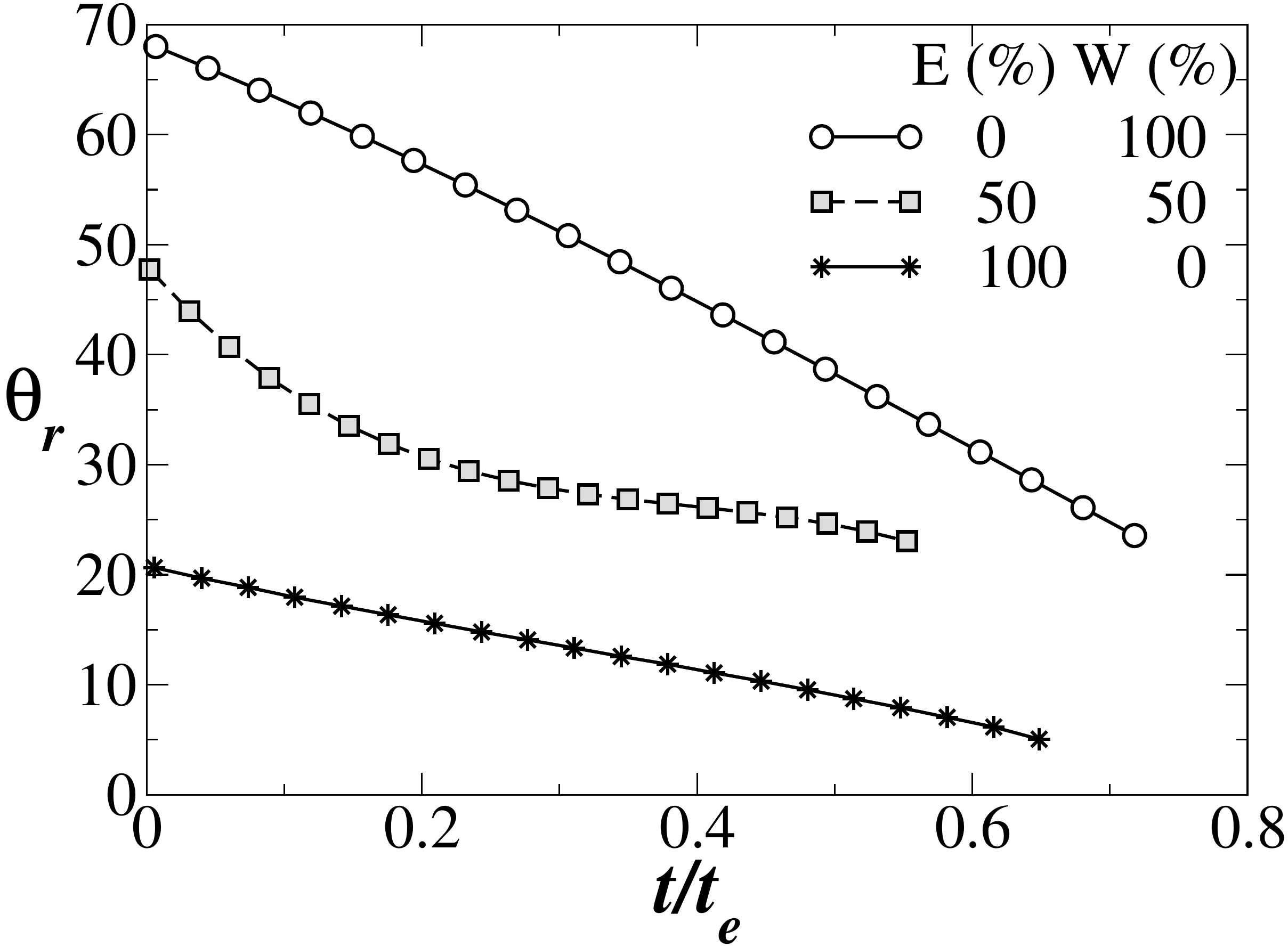} \\
 \hspace{0.6cm} (e) \\
\includegraphics[width=0.46\textwidth]{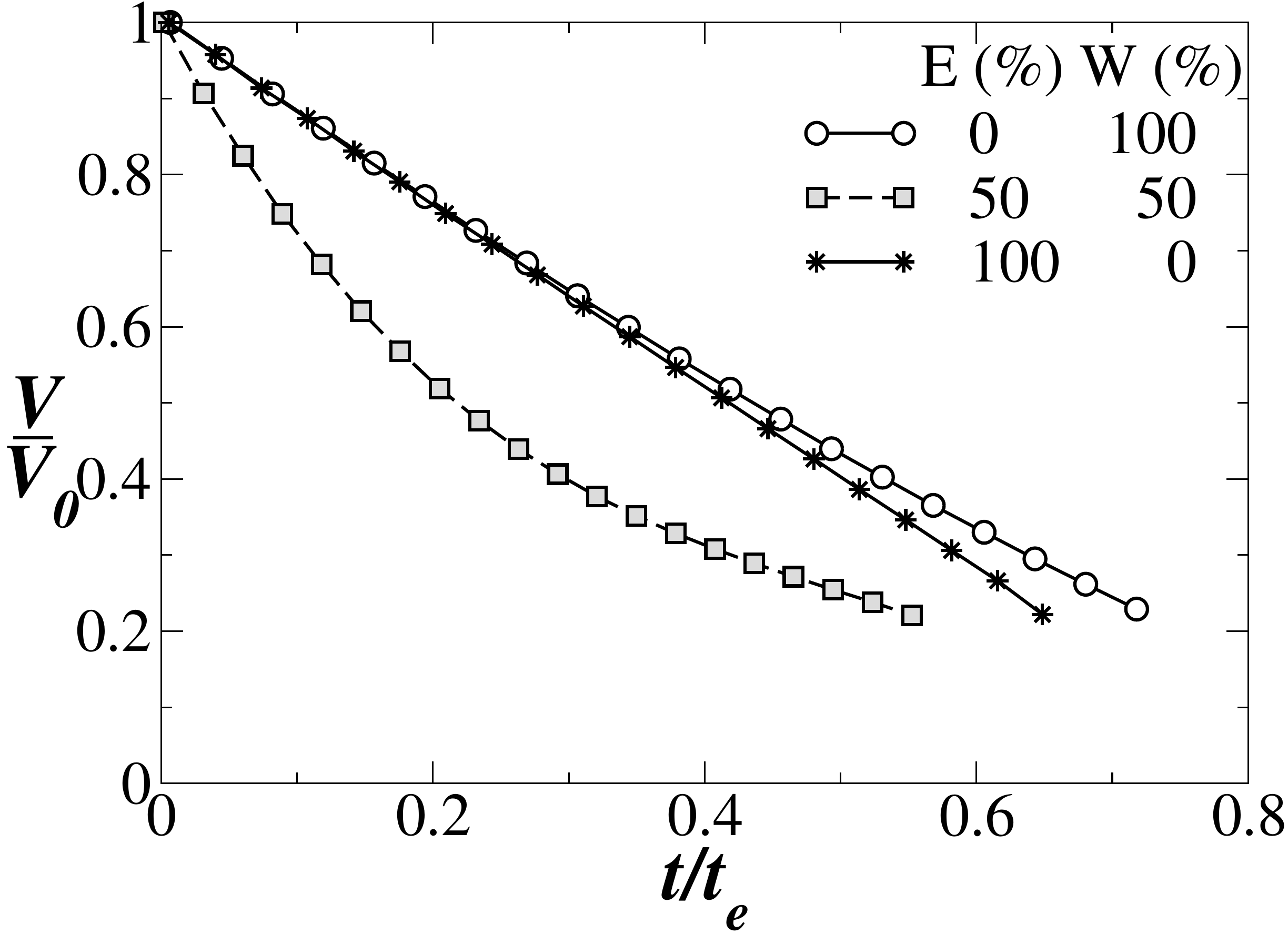}  \\
\caption{Variations of (a) the height ($h$) in mm, (b) the wetting diameter of the droplet ($D$) in mm, (c) the left contact angle ($\theta_l$) in degree, (d) the right contact angle ($\theta_r$) in degree and (e) the normalised volume with the initial volume of the droplet $\left ({V / V_0} \right)$ versus time normalised with the lifetime of the droplet $\left(t/t_e\right)$ for different ethanol (E) - water (W) compositions at $T_s=25^\circ$C. }
\label{fig:fig4}
\end{figure}

The non-monotonic behaviour of the binary droplet is further highlighted in Figures \ref{fig:fig4}(a)-(e), where the variations of the droplet height ($h$ in mm), the wetting diameter ($D$ in mm), the left contact angle ($\theta_l$), the right contact angle ($\theta_r$) (based on field of camera view) and of the droplet volume normalised  with initial the volume of the droplet $\left ({V / V_0} \right)$ are plotted against the normalised evaporation time $(t/t_e)$, and are shown for pure water, the (E 50\% + W 50\%) binary mixture and pure ethanol at 25$^\circ$C substrate temperature. Note that the left and right contact angles are defined based {on} the field of view. As expected due to the uniform evaporation process in the case of pure liquids, it can be seen in Figure \ref{fig:fig4}(a) that the droplet height, $h$ decreases mostly linearly for pure water and pure ethanol, whereas, for the droplet of (E 50\% + W 50\%) binary mixture, ethanol evaporates faster at the early time (in the pinned phase, $t/t_e \le 0.2$), leaving mostly water which evaporates at a slower rate. This is evident from the slope of $h$ versus $t/t_e$ curve for (E 50\% + W 50\%) binary mixture in Figure \ref{fig:fig4}(a), which is greater in the early pinned phase $(t/t_e < 0.2)$ than the later receding phase $(t/t_e>0.2)$. The pinned and receding phases for the pure fluids and the (E 50\% + W 50\%) binary mixture are also evident in Figure \ref{fig:fig4}(b), {where the droplet wetting radius ($D$) remains constant during the pinned phase and transitions into a monotonically decreasing curve during the receding phase.} 

The variations of the contact angles versus $t/t_e$ in Figures \ref{fig:fig4}(c) and (d) show that $\theta_l$ and $\theta_r$ decrease almost linearly for pure fluids. In contrast, the droplet of the (E 50\% + W 50\%) binary mixture enters a constant-contact-angle stage of evaporation for $t/t_e > 0.2$ (in the receding phase). The more complex evolution of the droplet shape for the binary mixture has an effect on the evaporation rates as well. It can be seen in Figure \ref{fig:fig4}(e) that in the case of pure liquids, the droplets exhibit a linearly decreasing volume trend with $t/t_e$. As the densities of pure fluids are constant at a given temperature, {the linear trend in the variations of $V / V_0$ against $t/t_e$ implies} that the evaporative mass flux is constant, at least upto $t/t_e = 0.8$ for the situation considered in the present study. In contrast, the trend is clearly nonlinear for the droplet of the (E 50\% + W 50\%) binary mixture, where we see a steep initial slope that flattens out at later times {as ethanol evaporates away and only water is left in the droplet}. 

\subsection{Evaporation at an elevated temperature}
\label{sec:exptdis_ht}
\begin{figure}[h]
\centering
\includegraphics[width=0.46\textwidth]{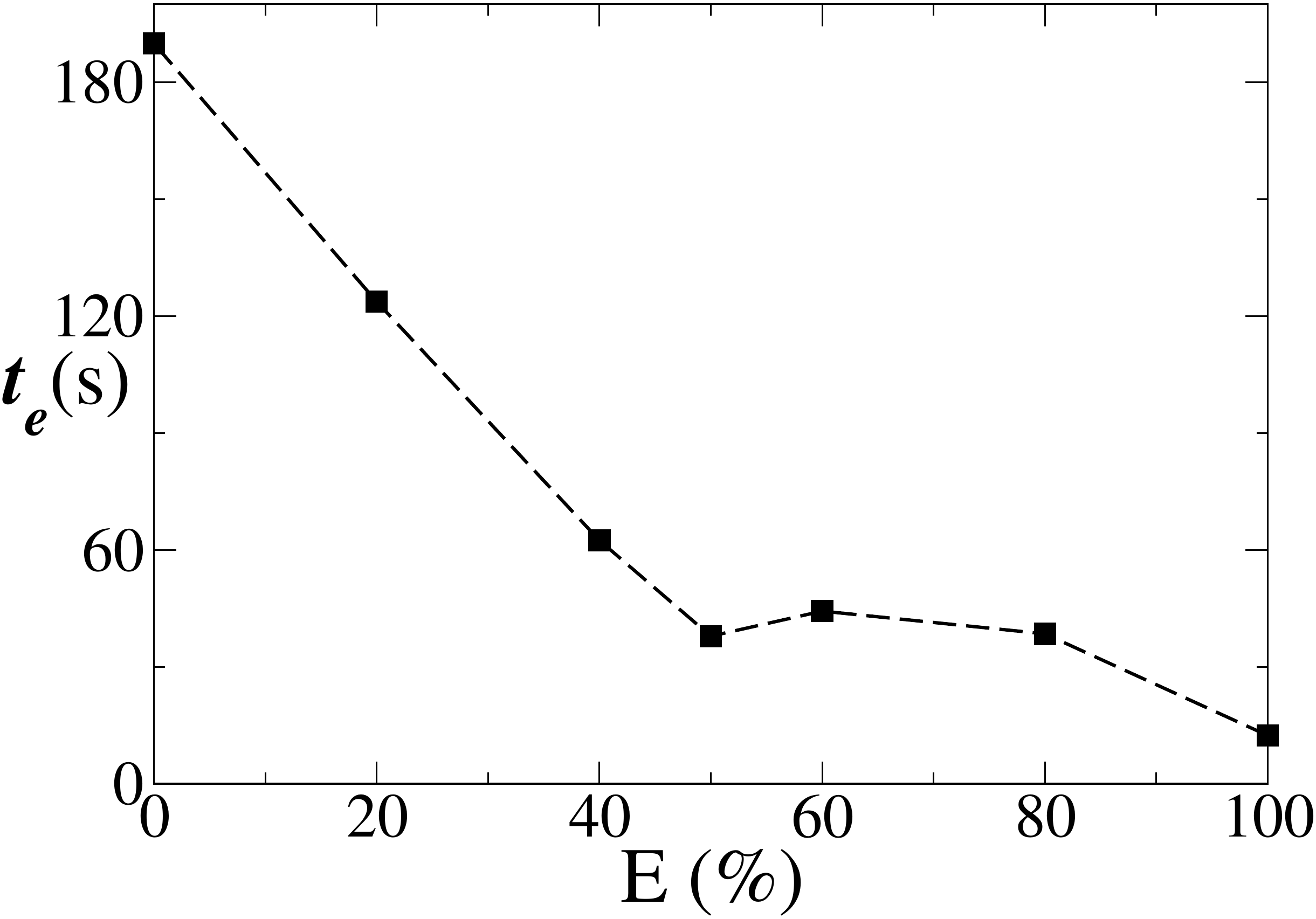} 
\caption{The variation of the lifetime time, $t_{e}$ (in s) of a droplet with the percentage concentration of ethanol (E) in the binary mixture at $T_s=60^\circ$C.}
\label{fig:te1}
\end{figure}

The experimental results for evaporating droplets (initial volume 5 $\mu$l) of pure water (E 0\% + W 100\%), binary mixture of varying compositions and pure ethanol (E 100\% + W 0\%) deposited on a heated substrate are discussed in this section. The substrate is maintained at $T_s=60^\circ$C and the experiments are conducted for seven compositions of ethanol-water binary mixtures. These compositions and the corresponding lifetimes of the droplet are presented in Table \ref{T2}. It can be seen that the droplet lifetimes, $t_e$ range from 12 seconds for a pure ethanol droplet to 190 seconds for a pure water droplet, which are much smaller than the corresponding values of $t_e$ at $T_s=25^\circ$C (Table \ref{T0}). The variation of $t_e$ with the percentage volumetric concentration of ethanol (E \%) in the binary mixture at $T_s=60^\circ$C is shown in Figure \ref{fig:te1}. It can be seen that there is a rapid linear decrease in the lifetime of the droplet with an increase in the concentration of ethanol till E = 50\%, after which the lifetime of the droplet is seen to vary only slightly with further increase in ethanol concentration till E=80\%. We believe that this behaviour can be attributed to the non-ideal vapour pressure phase diagram of water-ethanol binary mixtures \citep{bejan2016advanced} which is discussed further in Section \ref{sec:theory_drop_binary}.

\begin{table}   
\begin{center}
 \begin{tabular}{c|c}
 E (\%) \quad W (\%)   & $t_e$(s) at 60$^\circ$C \\
    &   \\
 0   \qquad 100 &   190 $\pm$ 5\\
 20  \qquad  80 &    124 $\pm$ 4 \\
 40  \qquad 60 &    62 $\pm$ 1 \\
 50  \qquad 50 &   38 $\pm$ 1 \\
 60 \qquad 40 &  44 $\pm$ 2 \\
 80  \qquad 20 & 38 $\pm$ 1 \\
 100 \qquad  0 &    12 $\pm$ 1 \\
\end{tabular}
\end{center}
\caption{Lifetime of the droplets, $t_e$(s) at $T_s=60^\circ$C for different compositions of ethanol-water binary mixture.} \label{T2}
\end{table}

\begin{figure}[h]
\centering
\includegraphics[width=0.9\textwidth]{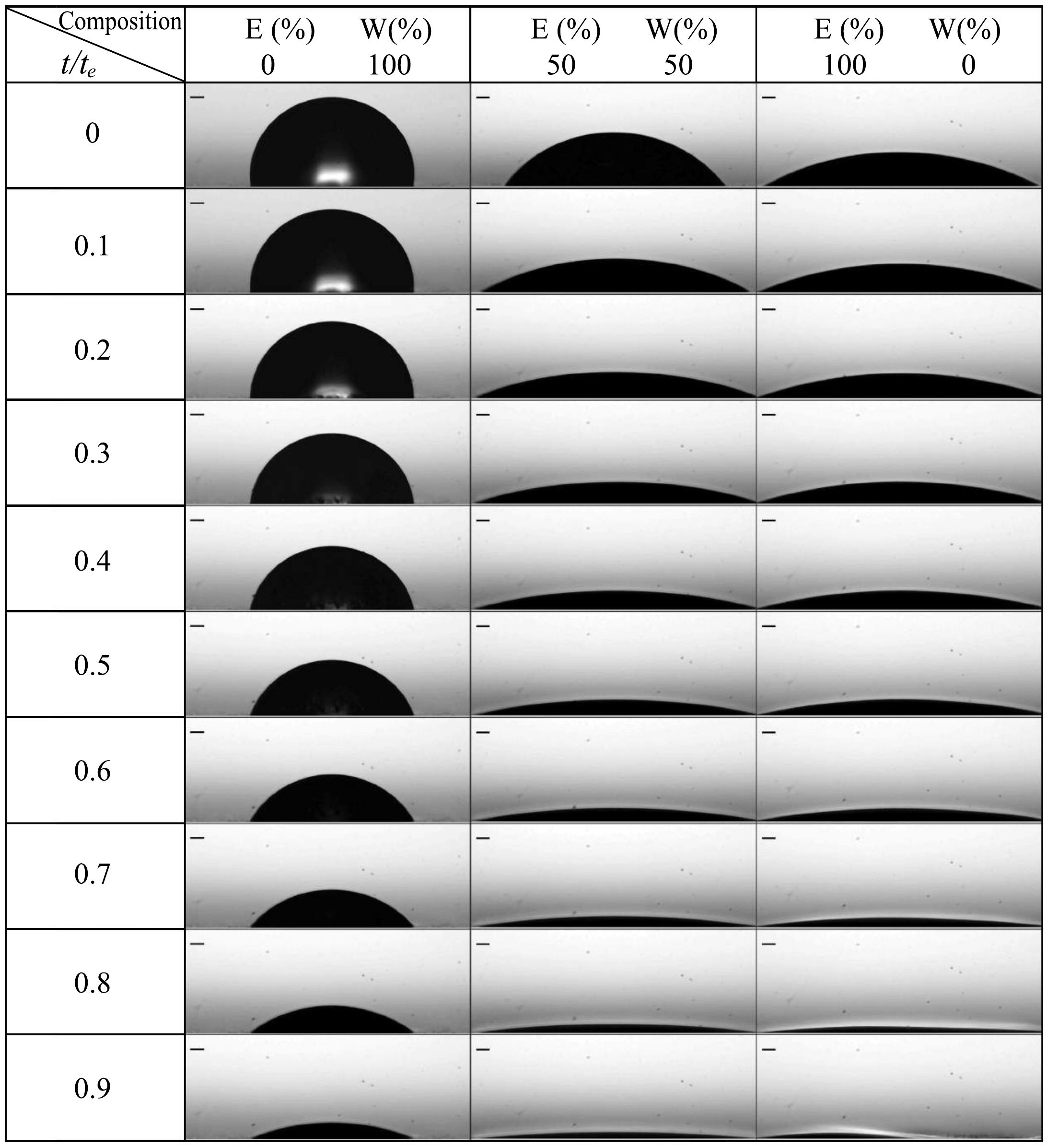}
\caption{Temporal evolution of droplet shape for pure water (E 0\% + W 100\%), (E 50\% + W 50\% solution) and pure ethanol (E 100\% + W 0\%) at 60$^\circ$C substrate temperature. The length of the scale bar shown in each panel is $200$ $\mu$m.}
\label{fig:fig5}
\end{figure}

The photographic images of the droplets of pure water (E 0\% + W 100\%), a binary mixture (E 50\% + W 50\%) and pure ethanol (E 100\% + W 0\%) are presented at different values of $t/t_e$ at $T_s=60^\circ$C. It can be seen in Figure \ref{fig:fig5} that, like in the case of $T_s=25^\circ$C (Figure \ref{fig:fig3}), the droplets of pure water and pure ethanol remain pinned for the majority of their steady lifetimes. However, for a (E 50\% + W 50\%) binary droplet, we see that the droplet has an initial spreading phase after its deposition on the heated substrate that lasts upto 20\% of its total evaporation time, after which it transitions into a pinned evaporation stage. At $T_s=60^\circ$C, a significantly higher undulation (interfacial instability driven by the Marangoni convection) is observed in Figure \ref{fig:fig5} for the pure ethanol droplet  near the end stages of the evaporation ($t/t_e \ge 0.7$) as compared to that at $T_s=25^\circ$C (Figure \ref{fig:fig3}). The late stage evaporation dynamics even becomes asymmetrical, as is evident at $t/t_e = 0.9$ in Figure \ref{fig:fig5}. 

\begin{figure}[h]
\centering
\includegraphics[width=0.95\textwidth]{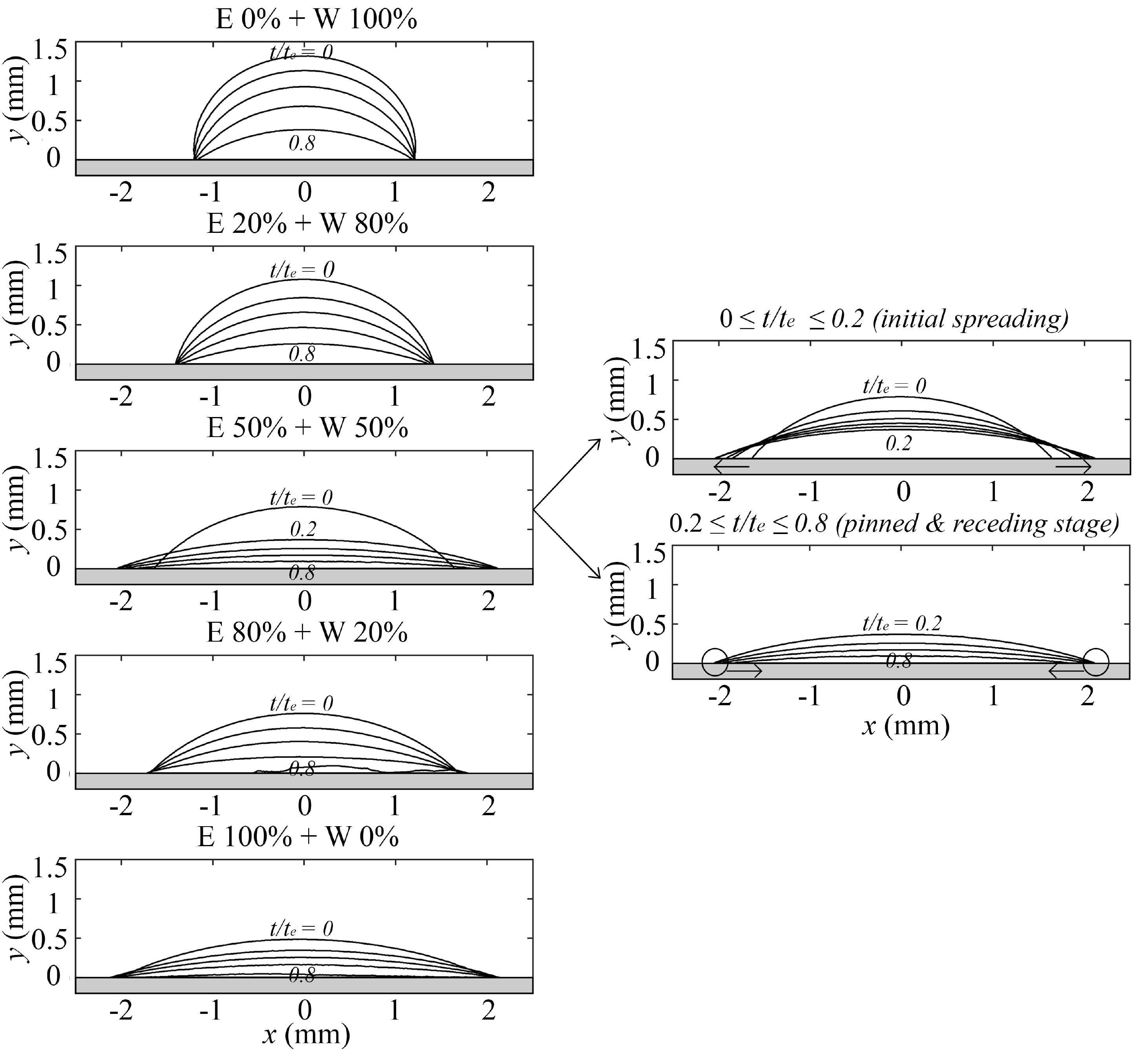}
\vspace{-0.2cm}
\caption{Comparison of the droplet spreading behaviour for different compositions. The contours are plotted in an interval of 0.2. The substrate temperature is $T_s=60^\circ$C. }
\label{fig:fig5b}
\end{figure}

\begin{figure}[h]
\centering
(a) \hspace{6.0cm} (b) \\
\includegraphics[width=0.95\textwidth]{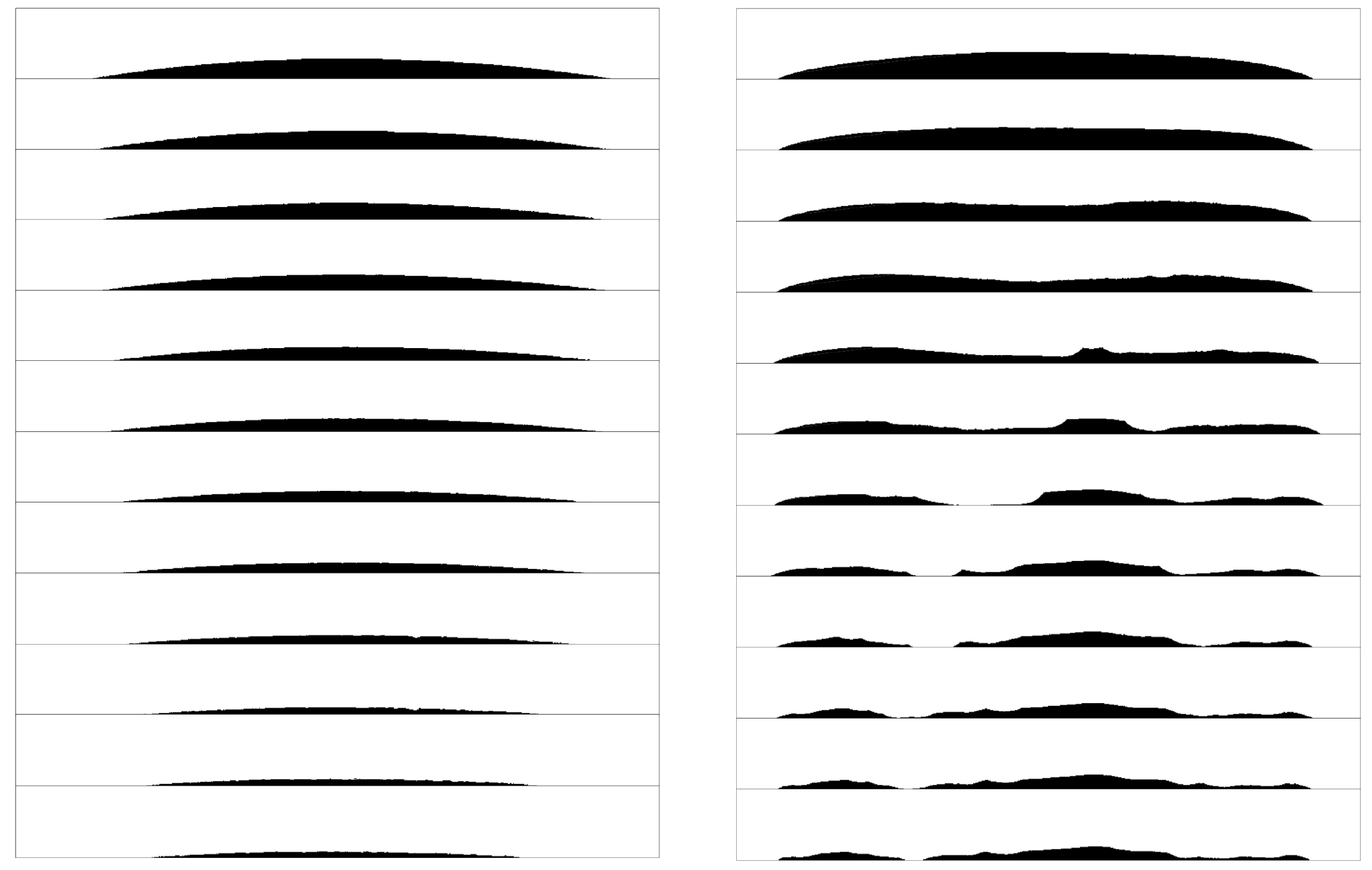} 
\caption{Comparison of late time behaviour of  evaporation of binary droplets of (a) (E 60\% + W 40\%) and (b) (E 80\% + W 20\%) at $T_s=60^\circ$C. The time evolution is from the top to bottom.}
\label{new}
\end{figure}

A detailed comparison of temporal contour evolutions of the droplets of pure water (E 0\% + W 100\%), binary mixtures of compositions (E 20\% + W 80\%), (E 50\% + W 50\%) and (E 80\% + W 20\%), and pure ethanol (E 100\% + W 0\%) at $T_s=60^\circ$C is presented in Figure \ref{fig:fig5b}. In contrast to the evaporation dynamics of the droplet of pure water (E 0\% + W 100\%) at $T_s=25^\circ$C (Figure \ref{fig:fig3b}), it can be seen in the top left panel of Figure \ref{fig:fig5b} that the contact angle at the early stage ($t/t_e \le 0.4$) is greater than $90^\circ$. As the droplet evaporates, the contact line remains pinned, but the contact angle decreases gradually. For droplets {of} (E 20\% + W 80\%),  (E 80\% + W 20\%) and pure ethanol (E 100\% + W 0\%), the contact angle is always less than $90^\circ$, and the pinned contact line behaviour is observed for most of the evaporation process. The droplet interface undulations for higher ethanol compositions ((E 80\% + W 20\%) and pure ethanol) are also evident in the bottom left two panels of Figure \ref{fig:fig5b}. {The late time behaviour of droplets of (E 60\% + W 40\%) and (E 80\% + W 20\%) at $T_s=60^\circ$C is shown in Figures \ref{new}(a) and (b), respectively. It can be seen in Figure \ref{new}(a) that the evaporation of a (E 60\% + W 40\%) droplet is smooth even at the later stages. In contrast, a droplet of (E 80\% + W 20\%), at the late stages, undergoes undulations at the liquid-vapour interface (Figure \ref{new}(b)). The undulations invariably begin as a deviation from the spherical cap profile such that there appears a pinching constriction at the middle of the droplet. The undulation waves progressively intensify and at the end of the evaporation stage, there is a clear break-up of the droplet into several satellite droplets.}

The early-time dynamics of a (E 50\% + W 50\%) droplet is also interesting. In this case, the initial spreading of the droplet is more pronounced, which is not prominent for other compositions in which one component is present in a greater proportion than the other component. {A} droplet of (E 50\% + W 50\%) composition exhibits three distinct stages at $T_s=60^\circ$C as highlighted in the right panels of Figure \ref{fig:fig5b}. They are, the early spreading stage $(t/t_e < 0.2)$, the intermediate pinned stage $(0.2 \le t/t_e \le 0.6)$ and the late receding stage $(t/t_e > 0.6)$. In the early spreading stage, the wetting diameter of the droplet increases and reaches to a value comparable to that for a pure ethanol droplet (bottom left panel of Figure \ref{fig:fig5b}). Subsequently, upto $t/t_e \approx 0.6$, the (E 50\% + W 50\%) droplet remains pinned and finally undergoes a slow contact line recession till the completion of the evaporation process. It is to be noted that the above behaviour was reproducible and was observed in all experimental runs (6 times). Thus, the complex behaviour of binary mixtures, particularly those where both component concentrations are initially comparable, is underscored by the above observations.

\begin{figure}[h]
\centering
\hspace{0.6cm}  (a) \hspace{6.0cm} (b) \\
\includegraphics[width=0.46\textwidth]{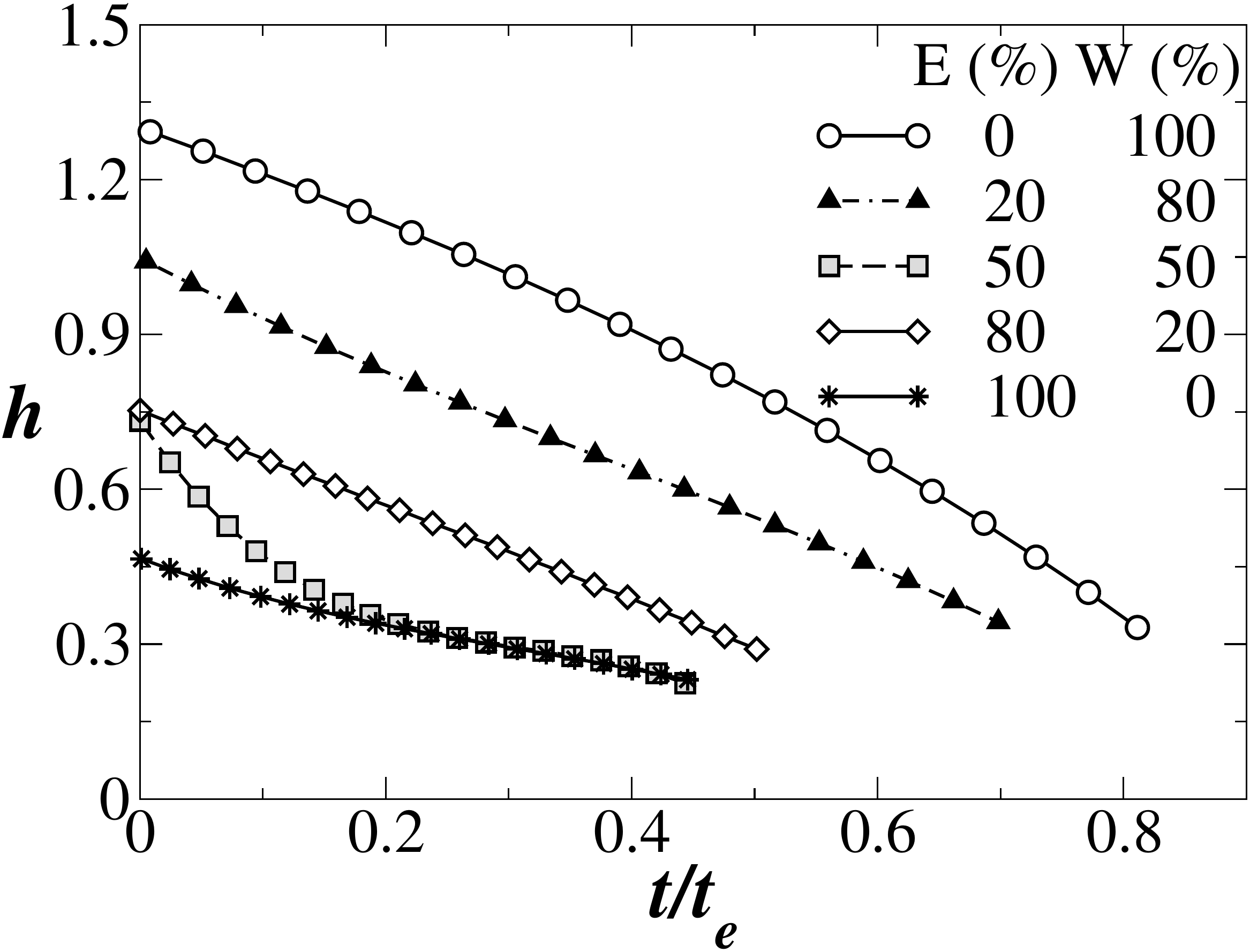} \hspace{2mm} \includegraphics[width=0.46\textwidth]{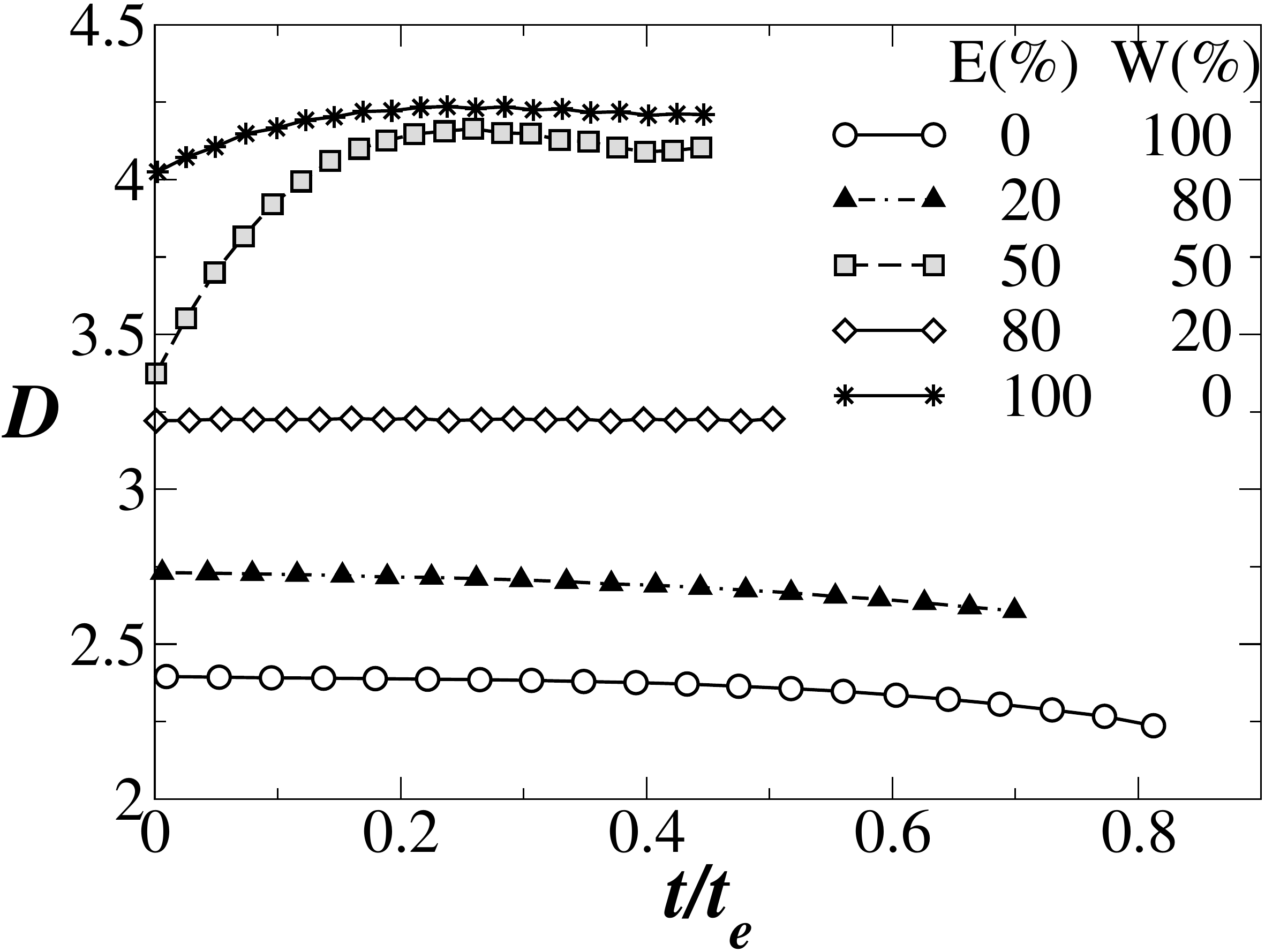} \\
\hspace{0.6cm}  (c) \hspace{6.0cm} (d) \\
\includegraphics[width=0.46\textwidth]{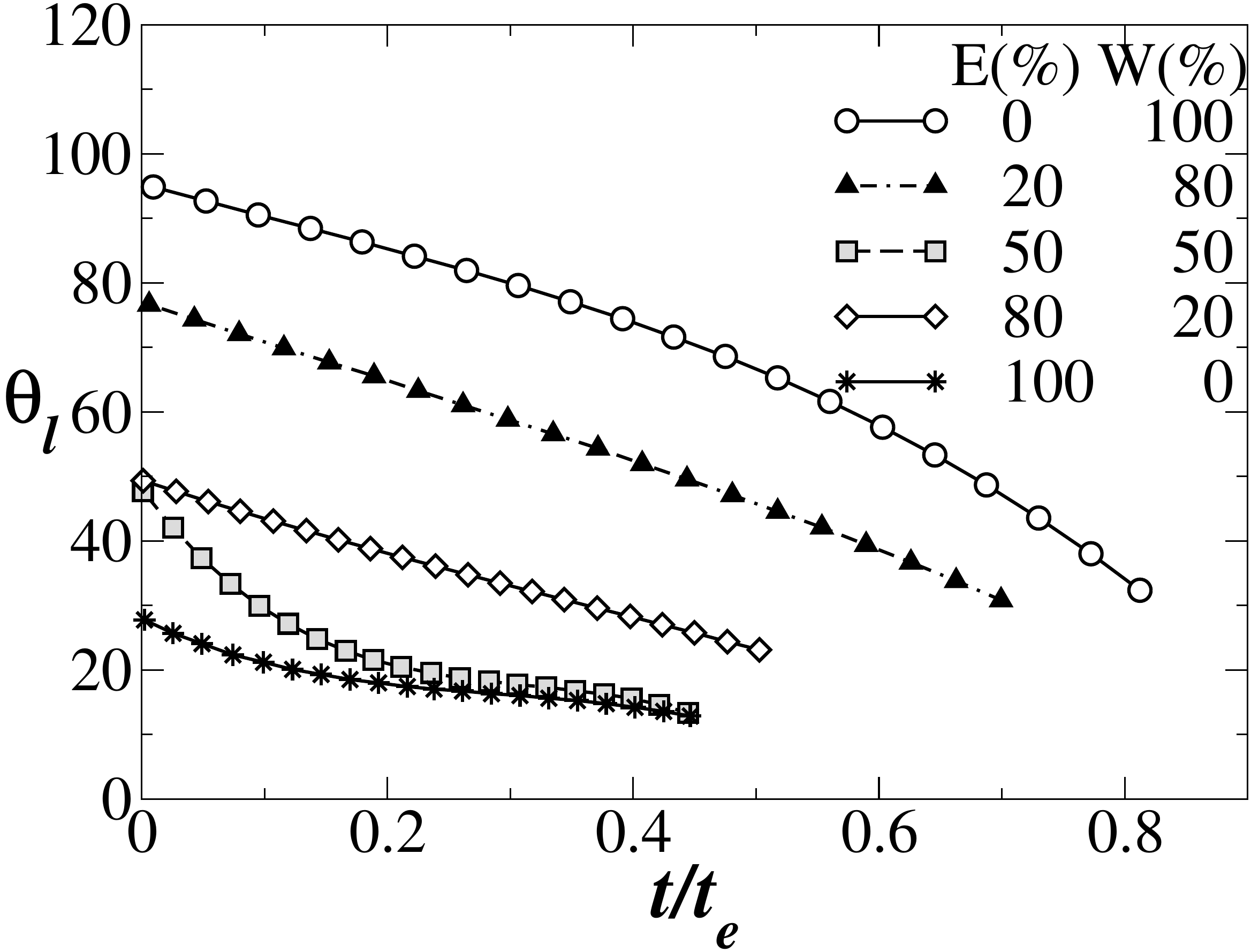} \hspace{2mm} \includegraphics[width=0.46\textwidth]{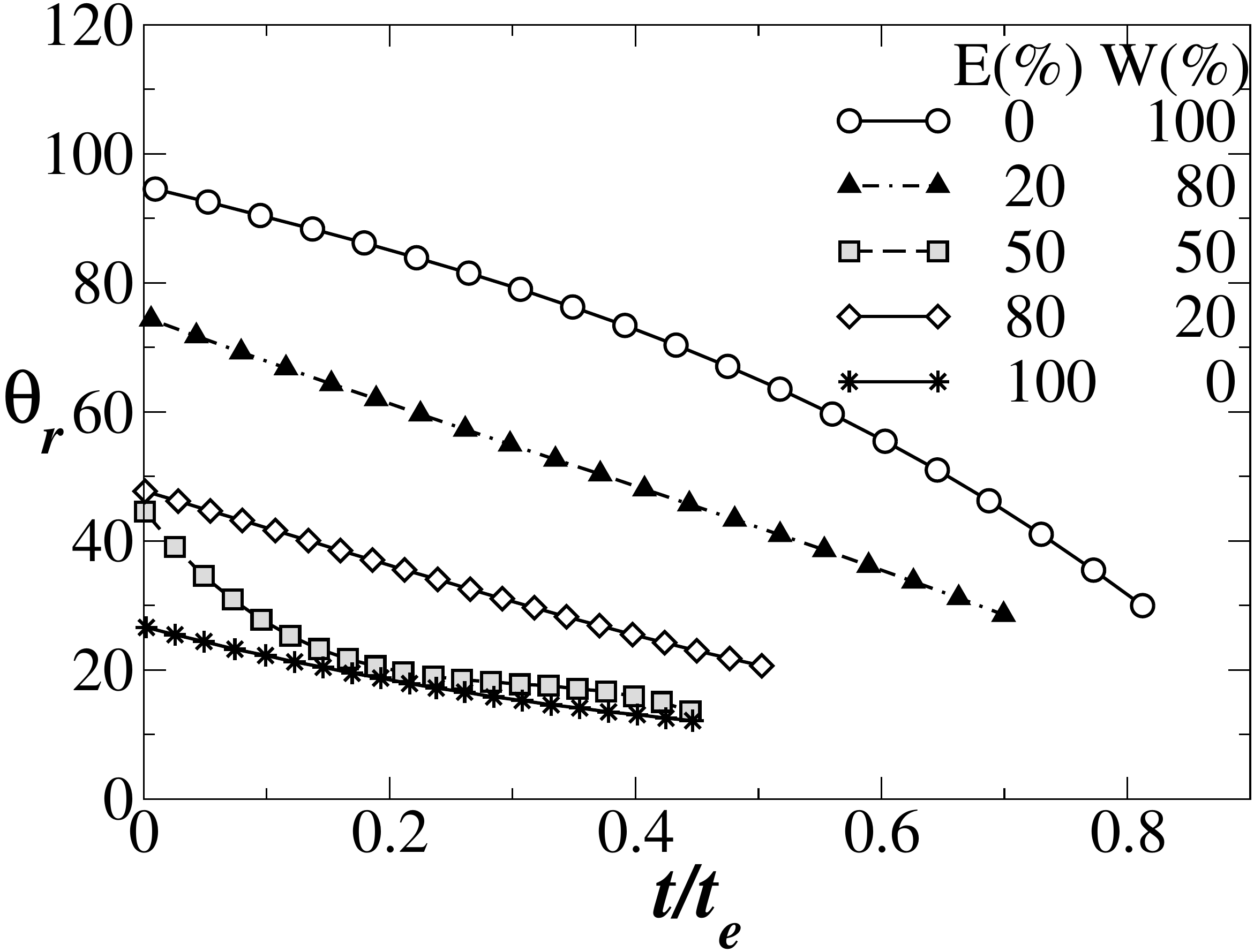} \\
\hspace{0.6cm} (e) \\
\includegraphics[width=0.46\textwidth]{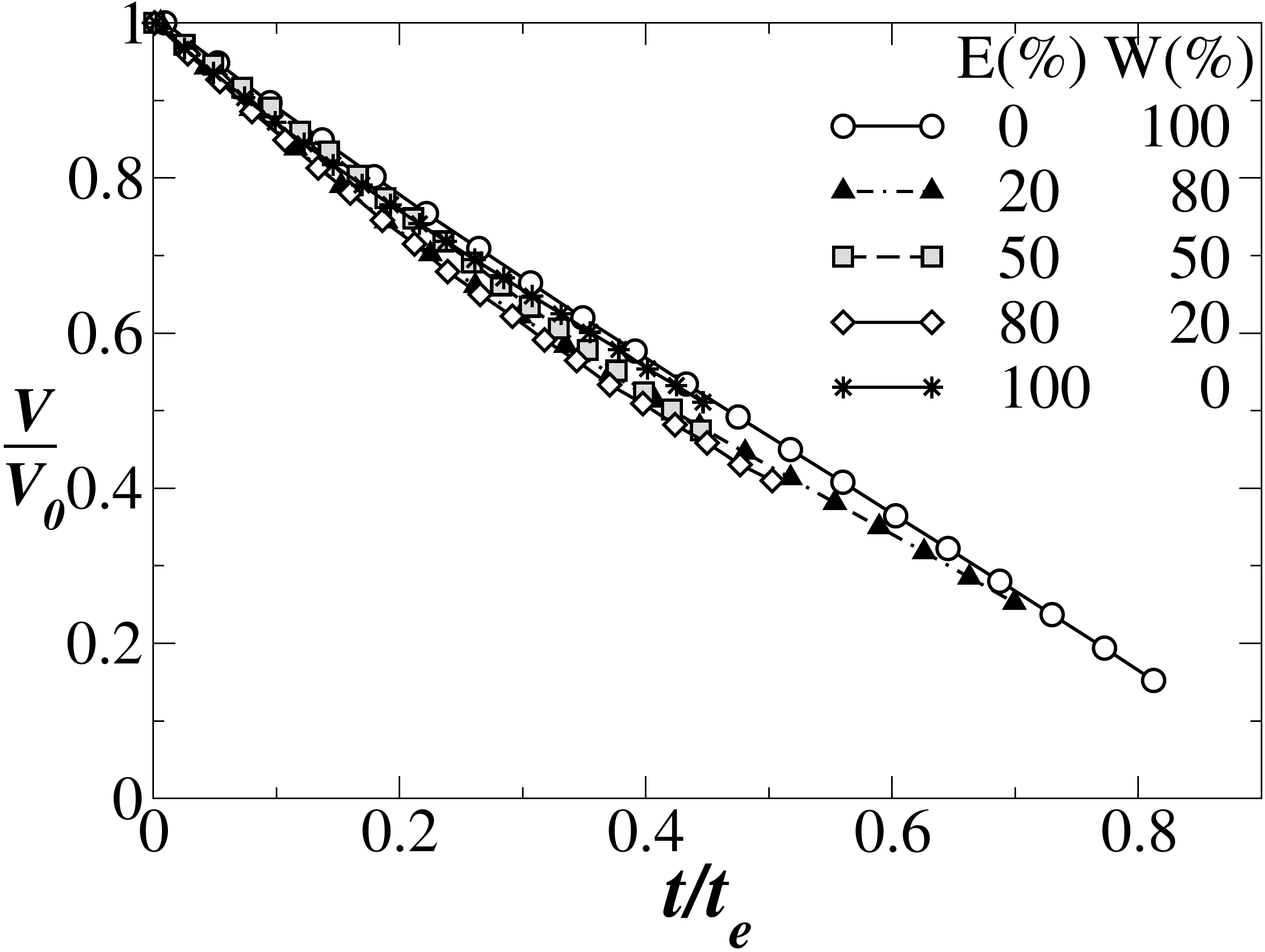}  \\
\caption{Variations of (a) the height ($h$) in mm, (b) the wetting diameter of the droplet ($D$) in mm, (c) the left contact angle ($\theta_l$) in degree, (d) the right contact angle ($\theta_r$) in degree and (e) the normalised volume with initial volume of the droplet $\left ({V / V_0} \right)$ versus time normalized with the lifetime of the droplet $\left(t/t_e\right)$ for different ethanol (E) - water (W) concentration at $T_s=60^\circ$C. }
\label{fig:fig6}
\end{figure}

The plots for the droplet height ($h$), the droplet wetting diameter ($D$), the left and right contact angles ($\theta_l$ and $\theta_r$) and the normalised volume of the droplet ($V/V_0$) against the normalised evaporation time $(t/t_e)$ for different compositions of ethanol and water mixture at $T_s=60^\circ$C are presented in Figure \ref{fig:fig6}. It can be seen in Figure \ref{fig:fig6}(a) that the droplet height decreases almost linearly with the increase in $t/t_e$ for all the compositions, except in case of the droplet of (E 50\% + W 50\%) binary mixture. In Figure \ref{fig:fig6}(b), no monotonic trend on the variations of $D$ versus $t/t_e$ is observed. For the droplets of (E 50\% + W 50\%) and (E 100\% + W 0\%), the droplet wetting diameter $(D)$ increases with the increase in $t/t_e$ (spreading stage) and reaches a plateau designating the intermediate pinned stage ($0.2 \le t/t_e \le 0.6$). A modest initial spreading is observed for the droplet of pure ethanol (E 100\% + W 0\%), whereas a significantly high contact line spreading is observed in case of a droplet of (E 50\% + W 50\%) binary mixture. At the later times, the droplet undergoes a receding stage of evaporation as explained in Figure \ref{fig:fig5b}. For other compositions, the wetting diameter, $D$ remains almost constant (no spreading), while the height of the droplet, $h$ decreases continuously. Inspection of Figures \ref{fig:fig6}(c) and (d) reveals that for pure water (E 0\% + W 100\%), $\theta_l \approx 96^\circ$ and $\theta_r \approx 95.3$ at $t/t_e=0$. This initial hydrophobic behaviour of the droplet on the substrate at $T_s=60^\circ$C is visually evident in Figure \ref{fig:fig5b}. The droplets of other compositions exhibit hydrophilic behaviour during the entire evaporation process. It can also be seen in Figures \ref{fig:fig6}(c) and (d) that the contact angle dynamics of the droplets of (W 50\% +  E 50\%) mixture and pure ethanol becomes nonlinear; the nonlinearity is more for the droplet of (W 50\% + E 50\%) mixture. Note that due to the increase in interfacial undulations at the late stages for droplets with high ethanol percentages, the post-processing method discussed in Section \ref{sec:expt} becomes difficult. Thus in Figures \ref{fig:fig6}(a)-(e), we present the results till the time the undulation is minimum.

Another interesting feature is observed in Figure \ref{fig:fig6}(e), where the normalised volume $(V/V_0)$ against the normalised time $(t/t_e)$ plots for all the compositions considered are seen to nearly collapse into a single monotonically decreasing line. This suggests that despite the complexities just alluded to, the global evaporation flux rates for all the cases have an inherent self-similar nature that may perhaps be modelled through simple analytical methods. Such an attempt is made in Sections \ref{sec:theory_drop_pure} and \ref{sec:theory_drop_binary} for pure and binary droplets, respectively.

\subsection{Effect of substrate temperature on the evaporation of (E 50\% + W 50\%) droplet}
\label{sec:expttempt}

\begin{figure}[h]
\centering
\includegraphics[width=0.46\textwidth]{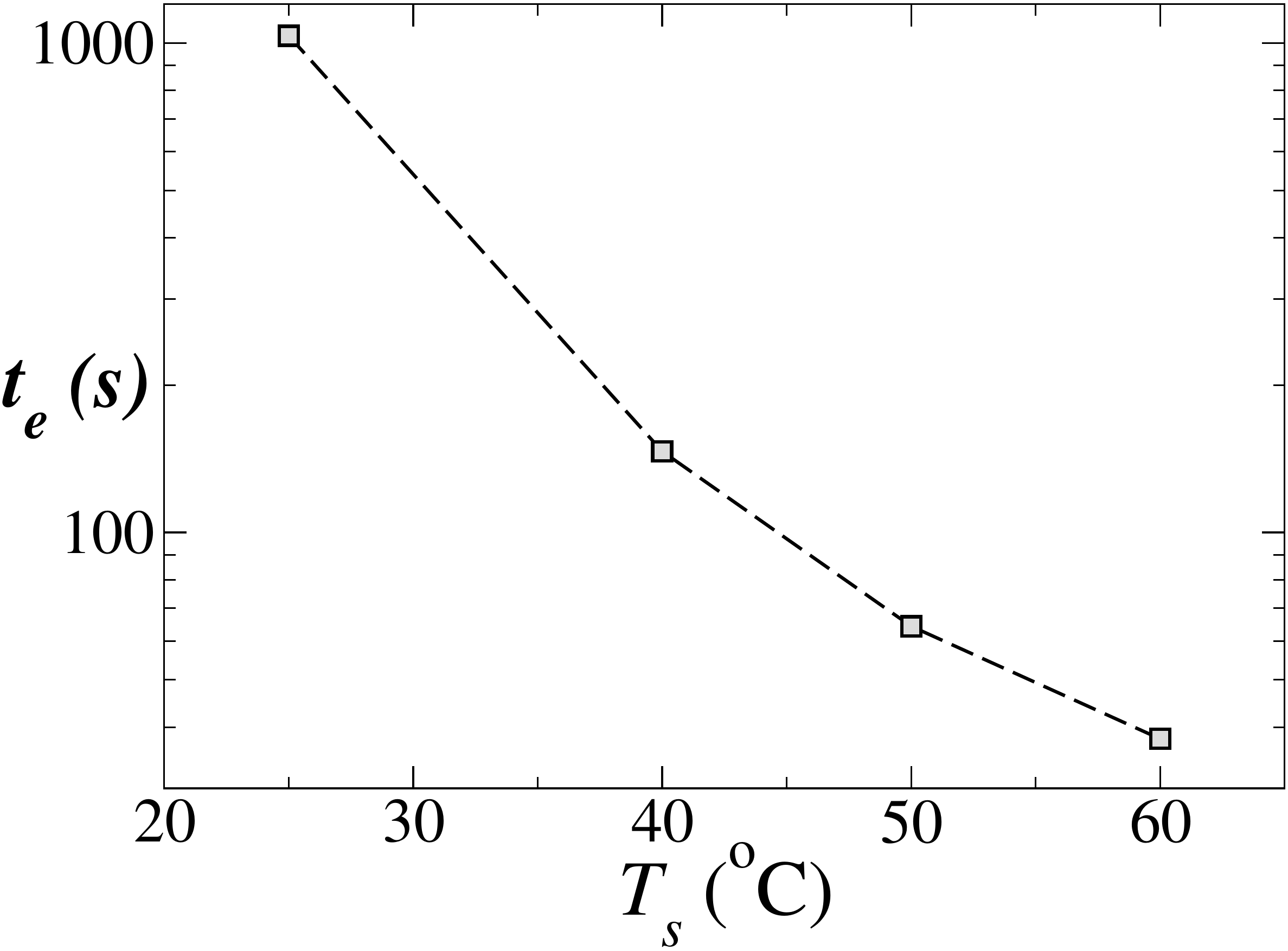} 
\caption{The variation of the lifetime time, $t_{e}$ (in s) of a droplet of (E 50\% + W 50\%) binary mixture with temperature of the substrate, $T_s$.}
\label{fig:te2}
\end{figure}

Next we investigate the effect of varying the substrate temperature, $T_s$ on the evaporation dynamics of a (E 50\% + W 50\%) binary droplet. In order to compare the lifetimes of the (E 50\% + W 50\%) binary droplet of initial volume 5 $\mu$l, in Table \ref{T3}, the values of the total evaporation time, $t_e$ for pure water (E 0\% + W 100\%), a binary mixture of (E 50\% + W 50\%) and pure ethanol (E 100\% + W 0\%) at different substrate temperatures are presented. We found that, as expected, $t_e$ decreases with the increase in the temperature of the substrate, $T_s$ (Table \ref{T3} and Figure \ref{fig:te2}). Also it can be seen in Table \ref{T3} that for all the values of $T_s$ considered, increasing the ethanol concentration in the binary mixture decreases the lifetime of the droplet. 

\begin{table}
  \begin{center}
    \begin{tabular}{c|c|c|c}
    &$t_e$(s) for & $t_e$(s) for & $t_e$(s) for \\
     $T_s$ ($^\circ$C) & E 0\%  + W 100\% & E 50\%  + W 50\% & E 100\%  + W 0\%\\
   & &  &  \\
      25                & 1488 $\pm$ 63          & 1035 $\pm$ 13           &  183 $\pm$ 3 \\
      40   & $-$ & 147 $\pm$ 6 &   $-$   \\
      50 & $-$ & 64 $\pm$ 2 & $-$  \\
      60 & 190 $\pm$ 5 & 38 $\pm$ 1 & 12 $\pm$ 1\\
    \end{tabular}
  \end{center}
  \caption{Lifetime of the droplets, $t_e$(s) of pure water (E 0\% + W 100\%), binary mixture of (E 50\% + W 50\%) and pure ethanol (E 100\% + W 0\%) at different temperatures of the substrate.}  \label{T3}
\end{table}

\begin{figure}[h]
\centering
\includegraphics[width=0.95\textwidth]{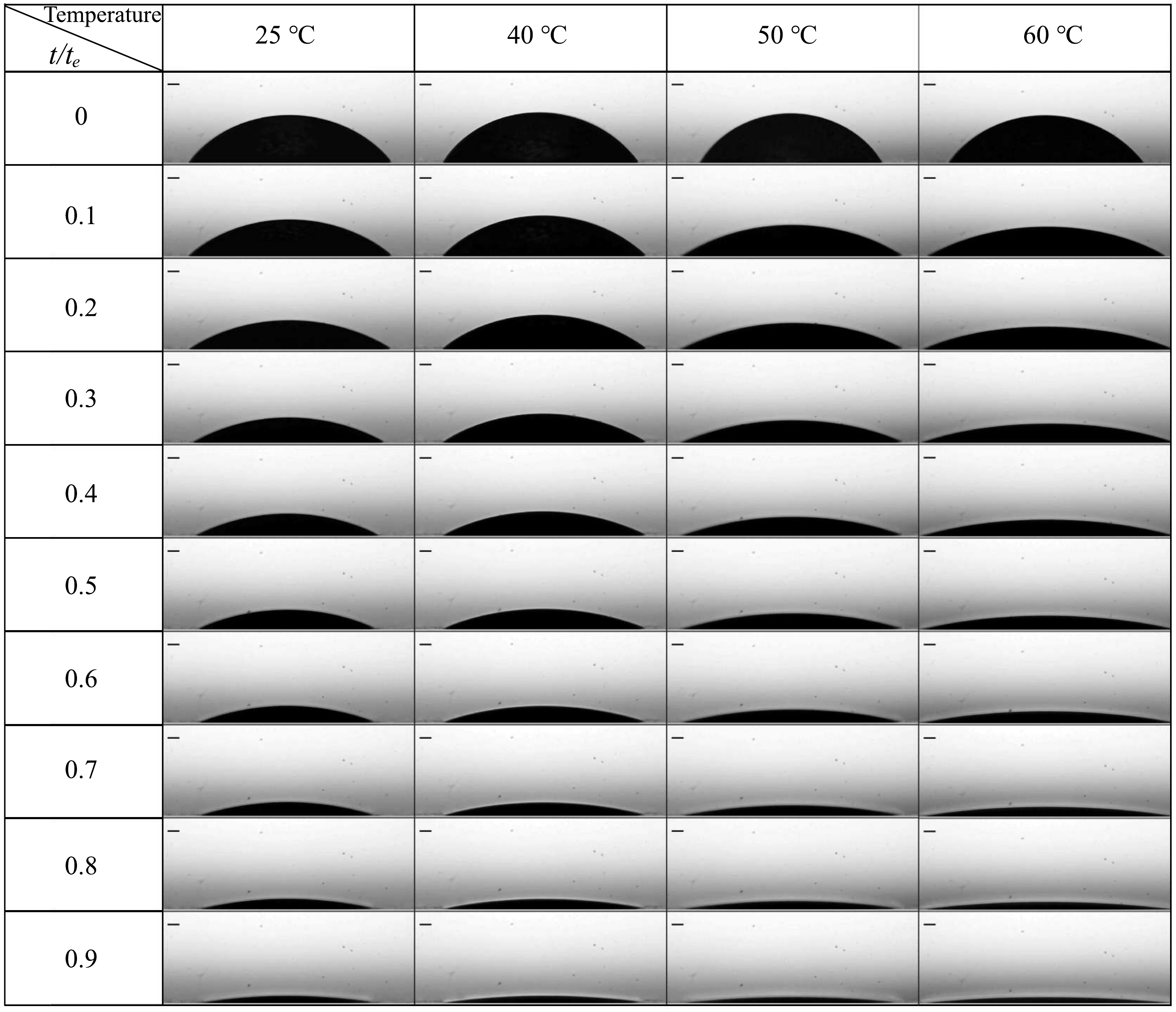}
\caption{Temporal evolutions of a (E 50\% + W 50\%) droplet at different substrate temperatures. The length of the scale bar shown in each panel is $200$ $\mu$m.}
\label{fig:fig7a}
\end{figure}

\begin{figure}[h]
\centering
\includegraphics[width=0.95\textwidth]{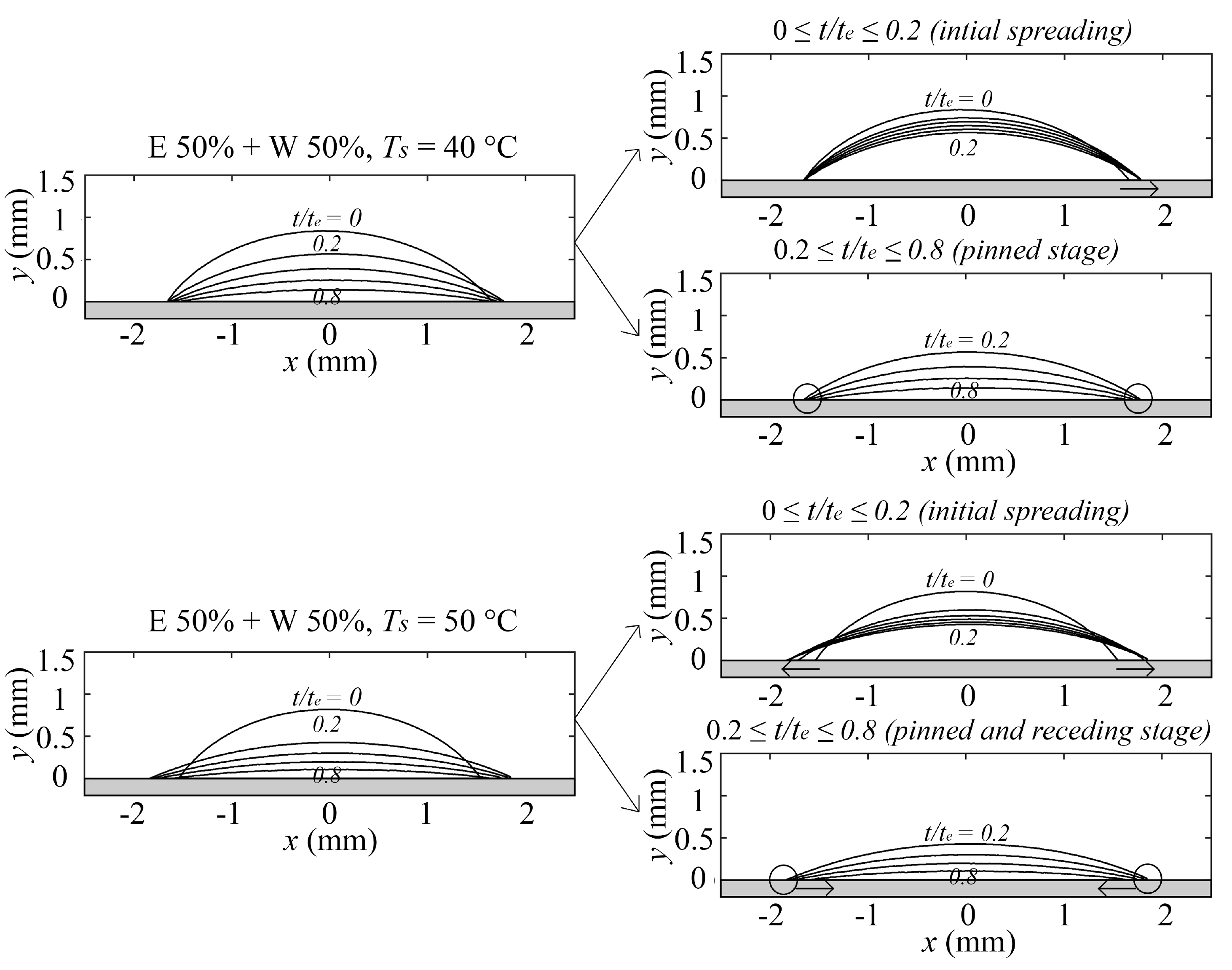}
\caption{Comparison of the droplet spreading behaviour at different substrate temperatures for (E 50\% + W 50\%) composition.}
\label{fig:fig7b}
\end{figure}

The temporal evolutions of the photographic images of the droplet of (E 50\% + W 50\%) binary mixture are shown in Figure \ref{fig:fig7a} at $T_s = 25^\circ$C, $40^\circ$C, $50^\circ$C and $60^\circ$C. The results for $T_s=25^\circ$C and $60^\circ$C are already discussed in Sections \ref{sec:exptdis_rt} and \ref{sec:exptdis_ht}, respectively, and are presented here only for the comparison purpose. 
The contours of the droplets at $T_s=40^\circ$C and $T_s=50^\circ$C, obtained by post-processing the droplet images shown in Figure \ref{fig:fig7a}, are presented in Figure \ref{fig:fig7b}. At $T_s=40^\circ$C, it can be seen that during $0 \le t/t_e \le 0.2$, the droplet of (E 50\% + W 50\%) binary mixture spreads a little; the spreading is observed only in the right contact line, while the left contact line is pinned (top right panel in Figure \ref{fig:fig7b}). In contrast, at $T_s=50^\circ$C, the droplet undergoes a more dominant spreading at both the left and right contact lines during $0 \le t/t_e \le 0.2$ (third right panel in Figure \ref{fig:fig7b}) between $0.2 \le t/t_e \le 0.6$. After the initial spreading stage, the droplet undergoes a pinned stage (second and fourth right panels of Figure \ref{fig:fig7b}). At this stage, the height of the droplet decreases due to evaporation. For $t/t_e>0.8$, the droplet contact line undergoes a slow recession till the end of evaporation.

\begin{figure}[h]
\centering
 \hspace{0.6cm}  (a) \hspace{6.0cm} (b) \\
\includegraphics[width=0.46\textwidth]{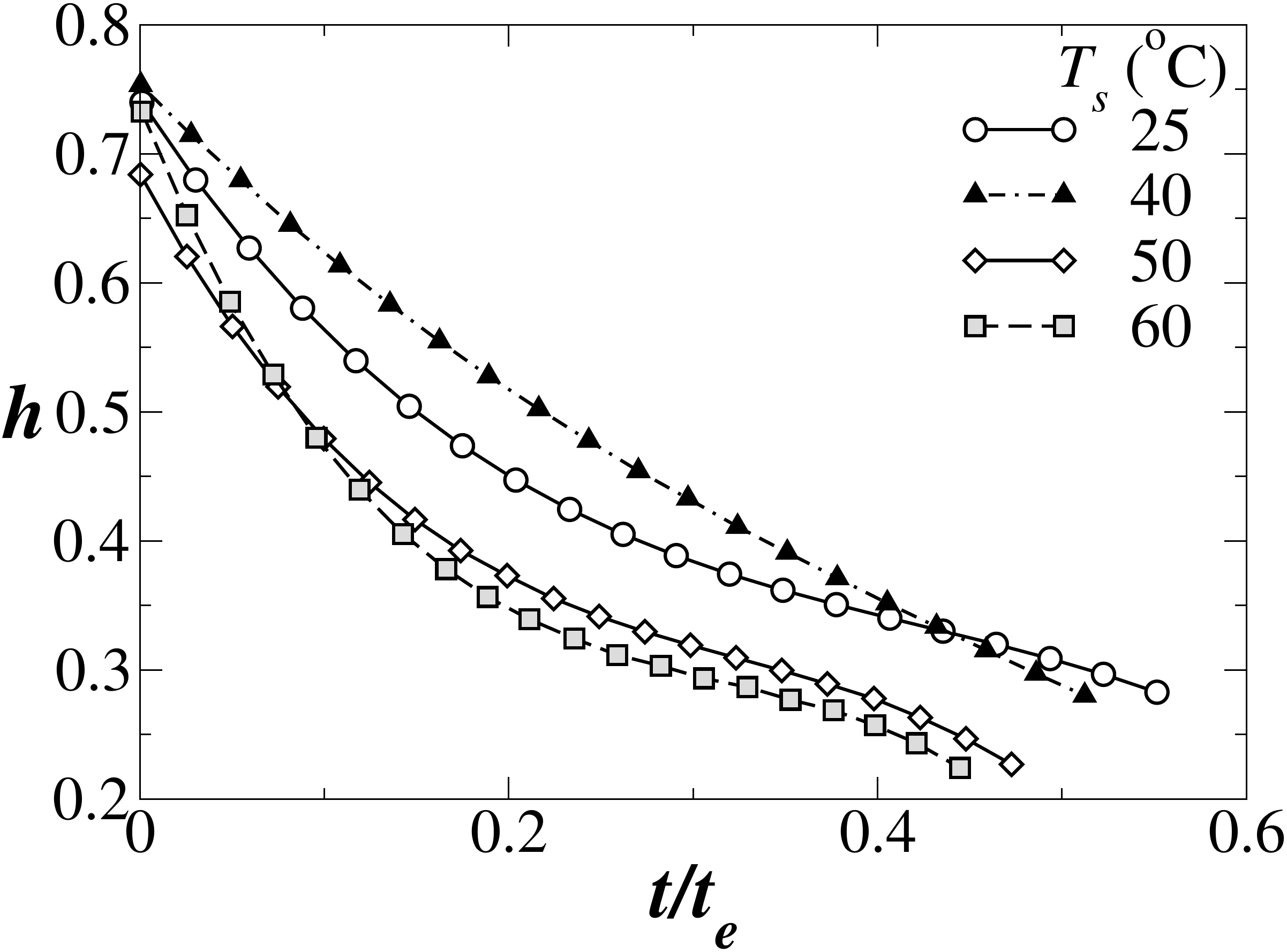} \hspace{2mm} \includegraphics[width=0.46\textwidth]{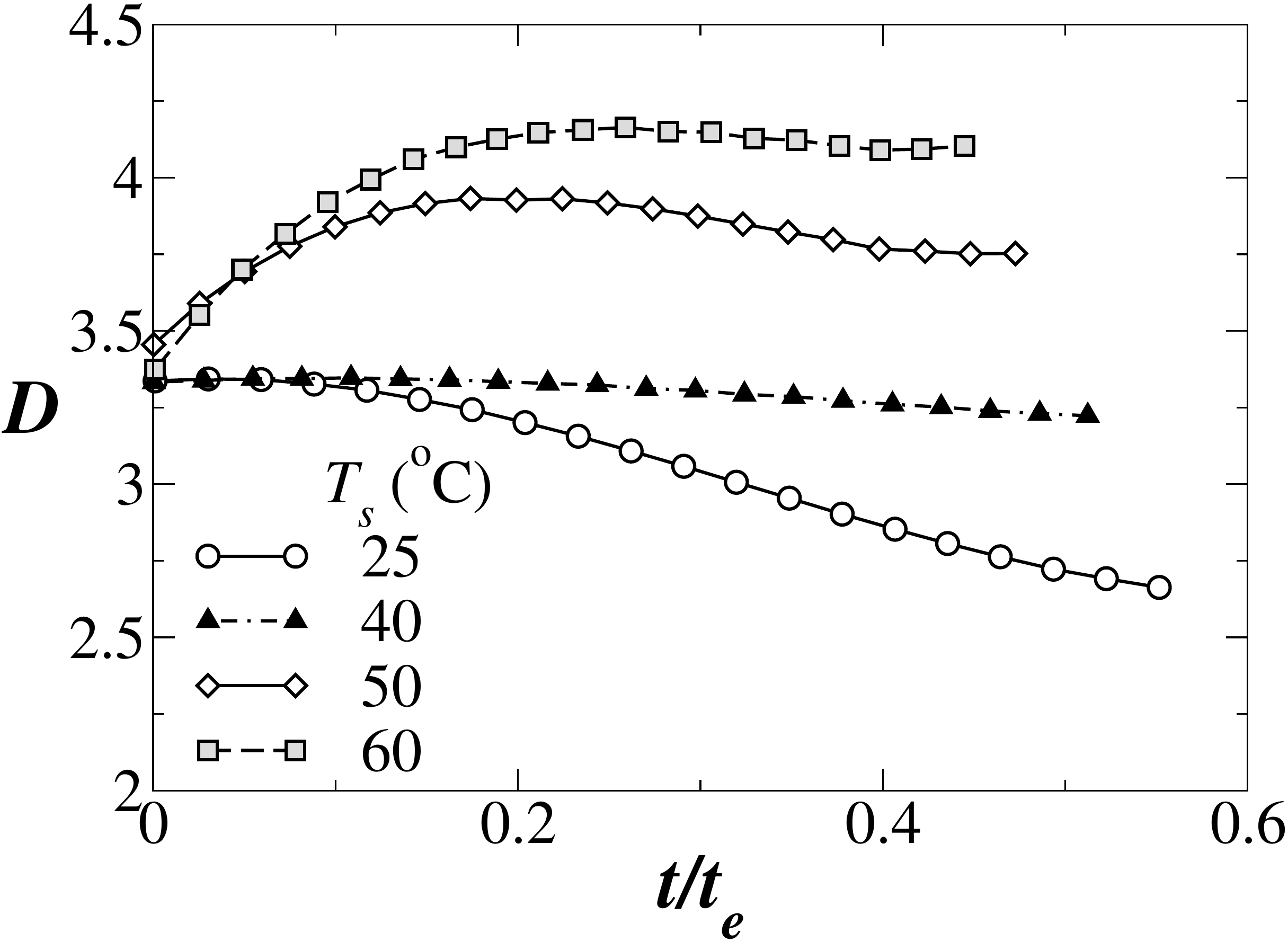} \\
 \hspace{0.6cm}  (c) \hspace{6.0cm} (d) \\
\includegraphics[width=0.46\textwidth]{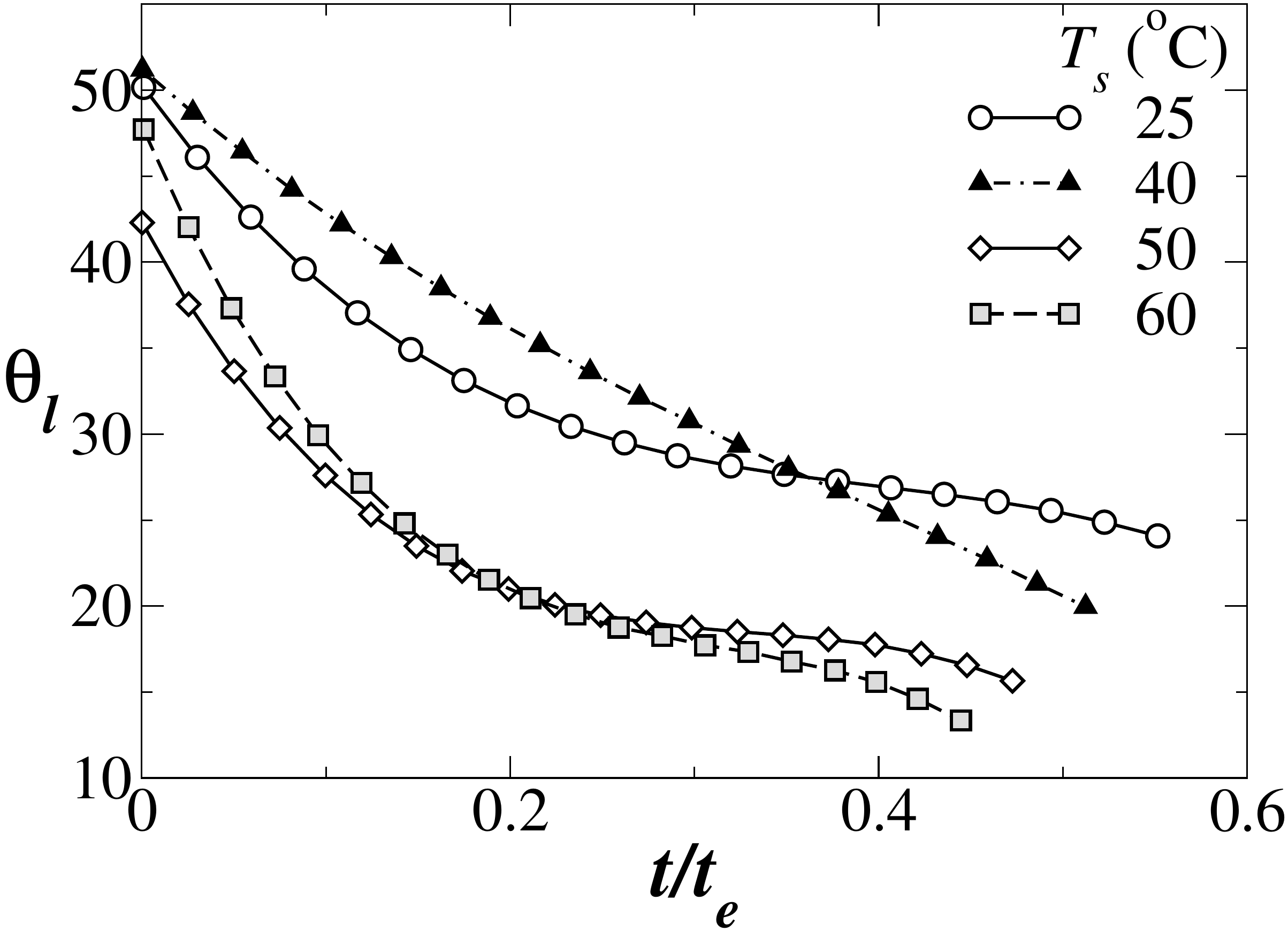} \hspace{2mm} \includegraphics[width=0.46\textwidth]{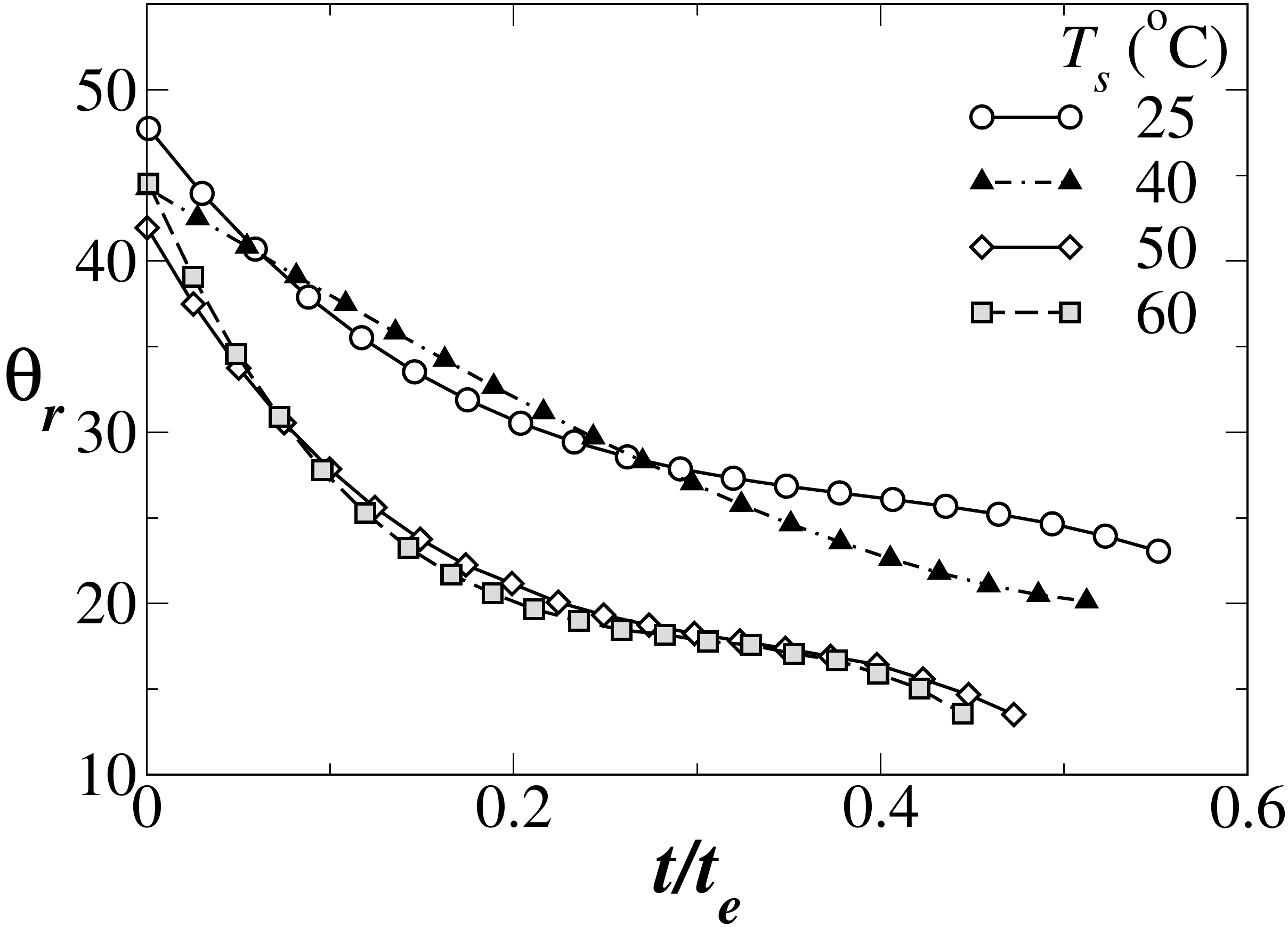} \\
 \hspace{0.6cm} (e) \\
\includegraphics[width=0.46\textwidth]{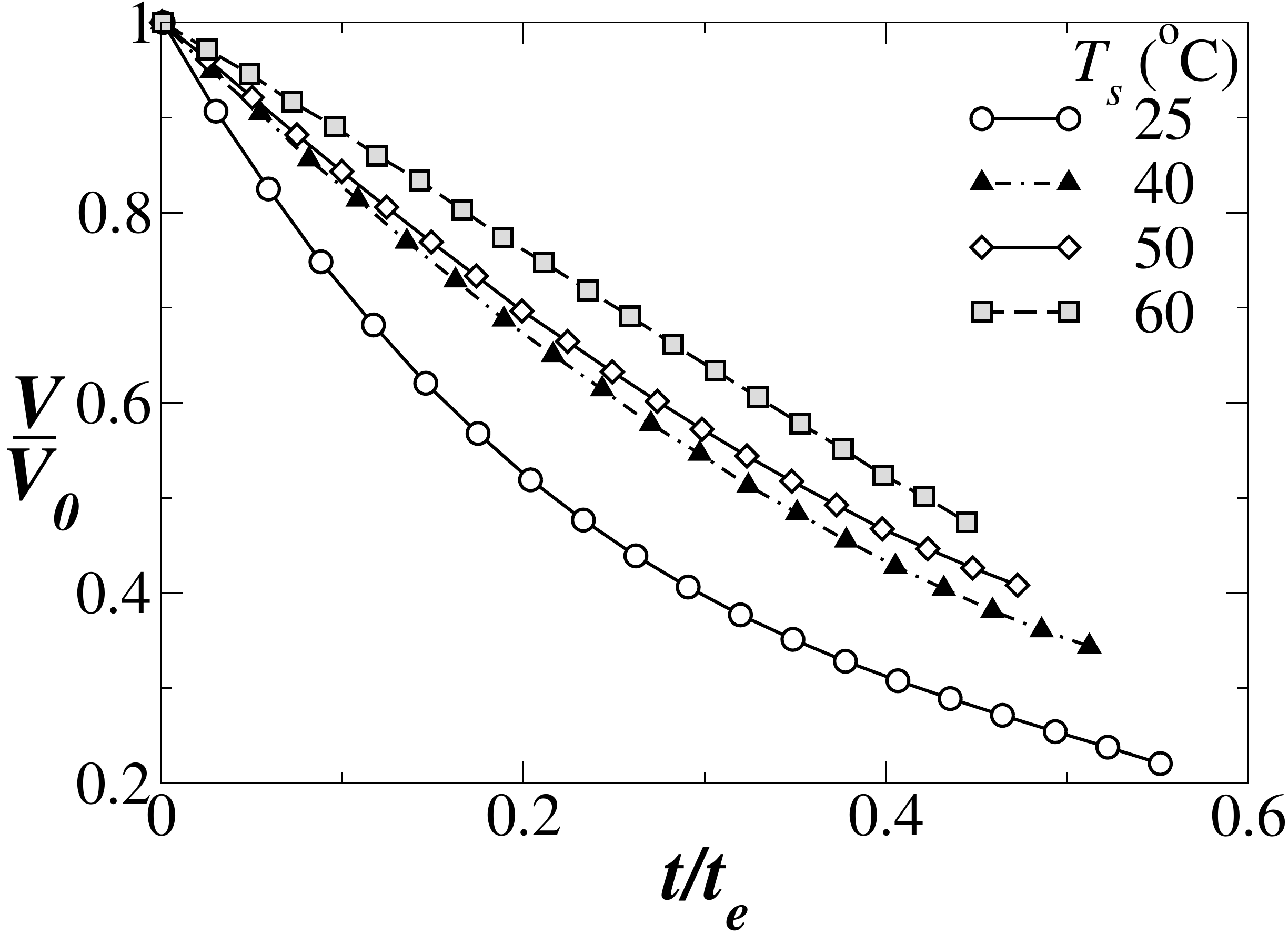}  \\
\caption{Variations of (a) the height ($h$) in mm, (b) the wetting diameter of the droplet ($D$) in mm, (c) the left contact angle ($\theta_l$) in degree, (d) the right contact angle ($\theta_r$) in degree and (e) the normalised volume with initial volume of the droplet $\left ({V / V_0} \right)$ versus time normalised with the life time of the droplet $\left(t/t_e\right)$ at different substrate temperatures for (E 50\%  + W 50\%) solution. }
\label{fig:fig7}
\end{figure}

In Figures \ref{fig:fig7}(a) - (e), we present the variations of the droplet height ($h$), the droplet wetting diameter ($D$), the left and right contact angles ($\theta_l$ and $\theta_r$) and the normalised volume of the droplet ($V/V_0$) obtained at different values of $T_s$ versus $t/t_e$, respectively for a droplet of (E 50\% + W 50\%). It can be seen in Figure \ref{fig:fig7}(a) that the height of the droplet decreases rapidly at the early time due to the initial spreading of the droplet on the substrate. After the initial spreading stage, the droplet height decreases at a slower rate due to evaporation. Inspection of Figure \ref{fig:fig7}(b) reveals that at $T_s=25^\circ$C, the droplet does not undergo the initial spreading stage and the wetting diameter decreases during the entire evaporation process. At $T_s=40^\circ$C, the wetting diameter remains almost constant, which is responsible for the non-monotonic behaviour observed at $T_s=40^\circ$C in Figures \ref{fig:fig7}(a), (c) and (d). In contrast, at $T_s=50^\circ$C and $60^\circ$C, the (E 50\% + W 50\%) droplet spreads faster at early times ($D$ increases) and then $D$ becomes constant (pinned phase) as shown in Figure \ref{fig:fig7}(b). Unlike pure water at $T_s=60^\circ$C, in Figures \ref{fig:fig7}(c) and (d), it is observed that a (E 50\% + W 50\%) droplet exhibits hydrophilic behaviour at all the substrate temperatures. In Table \ref{T3} and Figure \ref{fig:te2}, it can be seen that with the increase in substrate temperature from $T_s=25^\circ$C to $60^\circ$C, $t_e$ decreases from 1035 s to 38 s for the ( E 50\% + W 50\%) mixture droplets. In Figure \ref{fig:fig7}(e) the variations of $V/V_0$ versus $t/t_e$ are shown at different values of $T_s$. Close inspection of Figure \ref{fig:fig7}(e) reveals that, at the high substrate temperature ($T_s=60^\circ$C), the normalised volume of a (E 50\% + W 50\%) droplet decreases linearly with normalised time. However, the curve becomes nonlinear as we decrease the substrate temperature. Thus, it can be concluded that a droplet of binary mixture (E 50\% + W 50\%) at $T_s=60^\circ$C behaves like a droplet of another pure fluid with a higher volatility.

\section{Theoretical modelling of droplet evaporation rates}
\label{sec:theory_drop}

In order to gain a greater insight into the physics of the evaporation dynamics, in this section, we have developed theoretical evaporation models for sessile droplets of pure and binary fluids at room and elevated temperatures. {Section \ref{sec:theory_drop_pure} develops an analytical model of droplet evaporation for pure fluids which is subsequently extended to the case of binary mixture droplets in Section \ref{sec:theory_drop_binary}.} {In Section \ref{theory_expt_comp}, we compare our experimental results with the theoretical predictions obtained from the following analysis.}

\subsection{Evaporation of pure fluid droplets}
\label{sec:theory_drop_pure}
The diffusion and free convection processes from the droplet interface are the primary mechanisms governing the droplet evaporation rates in an initially quiescent and unsaturated atmosphere. {For substrates at the room or nearly the room temperature, the diffusion based vapour transport mechanism is expected to dominate \citep{semenov2014simultaneous}.} However, for heated substrates, the free convection driven fluxes are observed to play a significant role \citep{carle2013experimental}. Here, first, we discuss the diffusion limited evaporation model and then extend it to incorporate the convection flux.
\subsubsection{The diffusion limited model}
\label{sec:theory_diff}
\begin{figure}[h]
\centering
\includegraphics[width=0.6\textwidth]{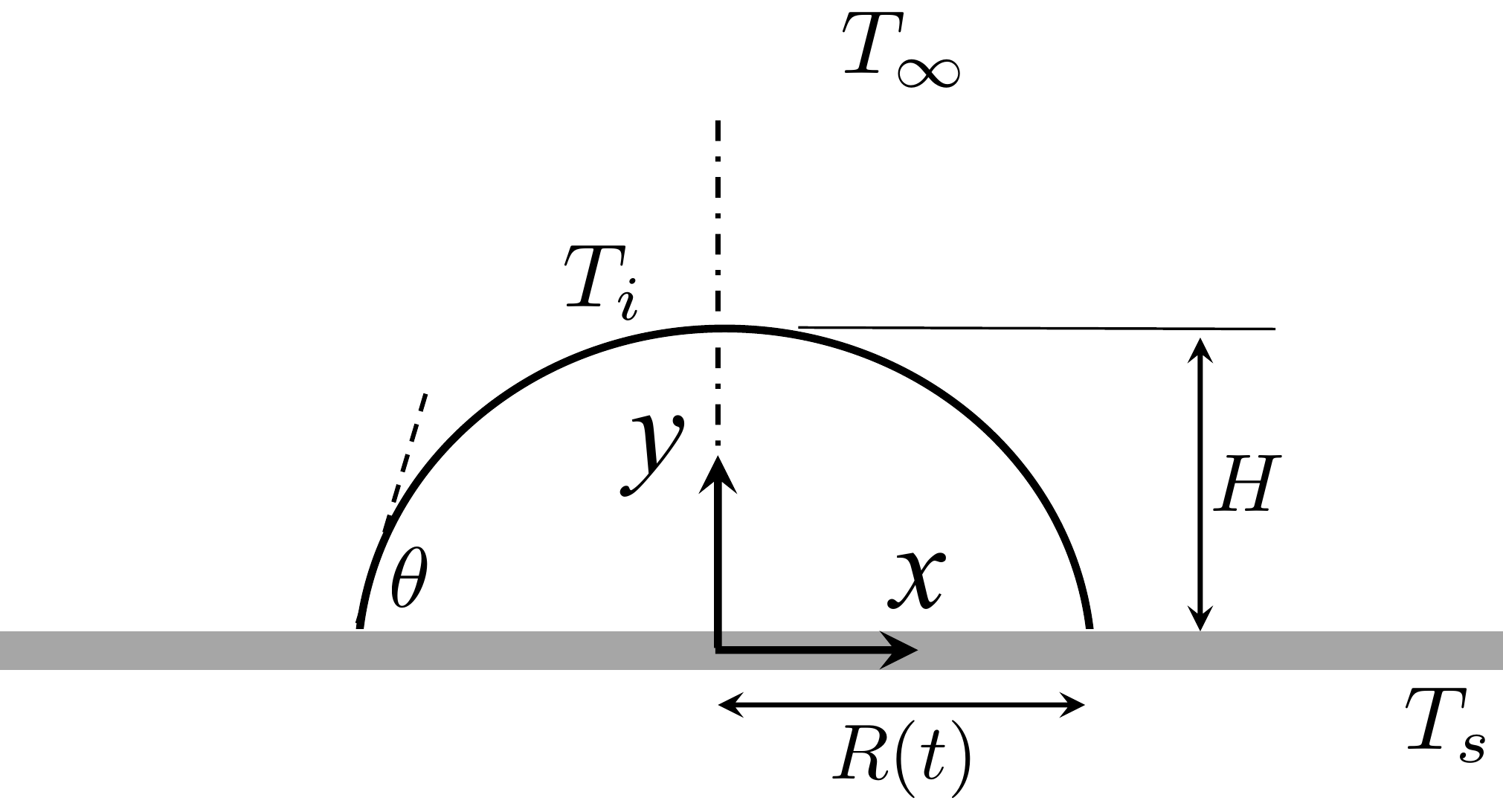} 
\caption{Schematic diagram of a sessile droplet on a substrate maintained at temperature, $T_s$. $H$ and $R(t)$ are the height and the wetting radius of the droplet; $\theta$ is the contact angle; $T_i$ and $T_\infty$ are the temperature at the liquid-vapour interface and the temperature of the ambient far away from the droplet, respectively.}
\label{fig:drop}
\end{figure}

The diffusion limited droplet evaporation model is relatively well-developed and widely used by {several} research groups to model evaporation flux of pure droplets at room temperatures \citep{brutin2015droplet}. In our case, all the droplets can be assumed to have a spherical cap profile during the evaporation process, which in turn implies that $\theta_l = \theta_r = \theta$. Thus we can calculate the experimental droplet volume as 
\begin{equation}
V (t) = {\pi R^3 \over 3} {(1-\cos \theta)^2 (2 + \cos \theta) \over \sin^3 \theta},
\end{equation}
where $R$ is the wetting radius of the droplet as shown in Figure \ref{fig:drop}{, that provides a schematic diagram of a sessile droplet and other parameters used in the theoretical modelling.}

As suggested by \cite{sobac2012}, when the substrate temperature $(T_s)$ and surrounding temperature $(T_\infty)$ are nearly the same, the evaporation is dominated by the quasi-steady state diffusion process as the diffusion time scale, $t_{D} (\sim R^2/{\cal D}) \ll t_e$, the total evaporation time. Here, $\cal D$ is the diffusion coefficient. In our experiments, the maximum values of $t_D/t_E$ for droplets of pure water and pure ethanol at the substrate temperature of $T_s=25^\circ$C are $\mathcal{O}(10^{-5} {\rm s})$ and $\mathcal{O}(10^{-3} {\rm s})$, respectively.  

Secondly, since the droplet volume is small and the droplet is placed over a metal plate, one may assume that the effect of internal temperature gradients on the evaporation flux is small. Hence our analysis assumes the droplet to be isothermal and at the same temperature as that of the substrate throughout the evaporation process. {Strictly speaking, the isothermal assumption is not valid for a droplet evaporating even at room temperature conditions \citep{david2007experimental,kovalchuk2014evaporation}. However, it has been shown previously \citep{sobac2012,carle2016} that the temperature differences are small for droplets deposited on substrates with high thermal conductivities, such as the aluminium plate used in our experiments.}

Thirdly, in our experiments, the contact angle, $\theta$ is less than $90^\circ$ at $T_s=25^\circ$C. Further, it is assumed that the vapour concentration at the liquid-vapour interface is at the saturated condition, $c_{sat}(T_s)$. The vapour concentration in the region far away from the droplet $c_\infty(T_\infty)$ is ${\cal H} c_{sat}(T_s)$, where ${\cal H}$ is the relative humidity of ambient air. In the case of pure ethanol, ${\cal H}=0$. Under the above-mentioned assumptions, the rate of evaporation for a pure droplet due to diffusion is given by \citep{sobac2012,carle2016}
\begin{equation}
\left(\frac{dm}{dt}\right)_d =  \pi R {\cal D} {\cal M} \left[c_{sat} (T_s) - c_\infty (T_\infty)\right] f(\theta), \label{case1a}
\end{equation}
where $\left(\frac{dm}{dt}\right)_d$ is the mass evaporation rate due to diffusion, ${\cal M}$ is the molecular weight of the fluid, ${\cal D}$ is the vapour diffusion coefficient at the mean temperature $(T_s + T_{\infty})/2$ and the expression for $f(\theta)$ for $\theta \le 90^\circ$ is given by $f(\theta) = 1.3 + 0.27 \theta^2$ \citep{hu2002}.  

\subsubsection{Model for free convection } \label{sec:theory_drop_conv}
While the evaporation of a droplet on an unheated substrate is primarily governed by the diffusive fluxes, the evaporation dynamics of a droplet on a heated substrate depends on both the Stefan flow and the natural convection, particularly for the more volatile ethanol liquid that boils at 78$^\circ$C \citep{lemmon2007nist}. The impact of the natural convection on droplet evaporation was studied by \cite{sobac2012}. They found that for light alcohols and hydrocarbons, the pure diffusion model may underestimate the evaporation flux by over 50\%. In our analysis, we find that the steady state diffusion model discussed in Section \ref{sec:theory_diff} underpredicts the evaporation rate of pure ethanol droplet at $T_s=60^\circ$C (see Figure \ref{fig:comp}), even though the model remains satisfactory for the evaporation of the water droplet. Hence, for the pure ethanol droplet, one needs to incorporate at least a simplified model to account for the effect of natural convection.

Usually, numerical methods are used for incorporating the effect of evaporation fluxes other than the steady state diffusion model (see, for instance, \cite{saenz2017dynamics}). Four effects are responsible for the final evaporation flux of a sessile droplet. These are: (i) the mass flux due to the diffusion of vapour in the ambient air, (ii) the mass flux due to the Stefan flow, (iii) the mass flux due to the free convection of vapour from the saturated interface to the unsaturated ambient, and (iv) the passive mass transport flux due to the natural convection of air from the hot substrate to the cool ambient. Moreover, these effects are coupled to each other creating a complex mass flux field around the droplet interface. Thus, it is challenging to develop a satisfactory analytical correlation to calculate the evaporation flux for such a complex coupled phenomenon. Nevertheless, some correlations based on experimental data can be found in the literature. For example, \cite{carle2016} developed a thermal Rayleigh number based correlation by fitting the data of the mean evaporation rate over the entire droplet lifetime at different substrate temperatures for several alcohols and hydrocarbons. While the correlation fit is extremely useful, the time evolution behaviour was not modelled, and further, a temperature gradient based Rayleigh number does not take into account the convection due to the concentration gradients. Recently, \cite{kelly2018correlation} have developed a correlation that takes into account both the diffusive and the concentration driven natural convective mass fluxes from an evaporating sessile droplet for several types of pure hydrocarbon species. Since these mass flux correlations are expressed as a function of the droplet wetting radius, they may be used to model the decrease of the droplet mass (and hence volume) over its lifetime. We have incorporated this model to calculate the droplet volume $V(t)$ during the evaporation process and compared with our experimental results. Unfortunately, all the correlations developed by \cite{kelly2018correlation} are for unheated substrates, and thus does not take into account the temperature gradient driven natural convection of air from the heated substrate to the cooler ambient that is likely to enhance the evaporation flux by increasing the velocity in the bulk region. In the present study, a modified version of the \cite{kelly2018correlation} correlation has been implemented in order to account for the evaporation of an ethanol droplet from a heated substrate. {Our modified evaporation model is discussed below.}

The evaporation mass transport rate due to the free convection can be expressed as
\begin{equation}\label{eq_conv}
\left(\frac{dm}{dt}\right)_c = h_m A_s (\rho_{v,s} - \rho_{v,\infty}),
\end{equation}
where $h_m$ is the convective mass transfer coefficient, $\rho_{v,s}$ is the density of the air-vapour mixture just above the droplet free surface and $\rho_{v,\infty}$ is the density of the ambient medium. The liquid-vapour interface area, $A_s$ is given by (assuming a spherical cap profile)
\begin{equation}
A_s = {2 \pi R^2 \over 1 + \cos \theta}.
\end{equation}

Neglecting the Stefan flow, the total mass transport rate can be calculated as
\begin{equation}\label{eq_massrate}
\left(\frac{dm}{dt}\right) = \left(\frac{dm}{dt}\right)_d + \left(\frac{dm}{dt}\right)_c + \left(\frac{dm}{dt}\right)_t,
\end{equation}
where the first, second and third terms in the right hand side {of Eq. (\ref{eq_massrate})} are the diffusion mass transfer rate, the free convective mass transfer rate and the vapour mass transport rate due to air convection, respectively. 

The convective mass transfer coefficient is usually expressed in terms of the convective Sherwood number, $Sh_c \equiv  {h_{m}R/ {\cal D}}$, where ${\cal D}$ is the vapour diffusion coefficient. To incorporate the convective and the diffusive mass transfers in a single expression, we can define a diffusion Sherwood number, $Sh_d \equiv {h_{d}R/{\cal D}}$. The diffusive mass transfer coefficient, $h_{d}$ can then be calculated from the following relation
\begin{eqnarray}
\left(\frac{dm}{dt}\right)_d &=& \pi R {\cal D} \mathcal{M}_E(c_{sat}(T_s) - c_{\infty}(T_{\infty}))f(\theta), \nonumber \\ &=& h_{d} A_s \mathcal{M}_{E}(c_{sat}(T_s) - c_{\infty}(T_{\infty})),
\end{eqnarray}
where $\mathcal{M}_E$ is the molecular weight of ethanol vapour, $c_{sat}(T_s)$ and $c_{\infty}(T_{\infty})$ are the saturated vapour concentration at the substrate temperature (i.e the temperature at the droplet interface under the isothermal droplet assumption) and the vapour concentration in the ambient. Noting that the concentration of ethanol is zero in the ambient, the expression for the diffusion Sherwood number reduces to 
\begin{equation}\label{eq_Sh_diff}
Sh_d = \frac{f(\theta)(1+\cos \theta)}{2}.
\end{equation}
The correlation for the effective Sherwood number is given by  
\begin{equation}\label{eq_Sh_corr}
Sh_{cor} = Sh_{d}^{*} + Sh_{c}^{*},
\end{equation}
where $Sh_{d}^{*}$ and $Sh_{c}^{*}$ are the modified diffusion and the modified convective Sherwood numbers, respectively after taking into account of the coupling between the diffusion flux and the convective flux when both the evaporation modes are active.

\begin{table}   
\begin{center}
 \begin{tabular}{c|c}
 Constants used  & Values \\
in Eqs. (\ref{315}) and (\ref{316}) &  \\
    &   \\
$a$ &   $1.23 \times 10^{-3}$\\
$b$ &    $0.648$ \\
$c$ &   $-0.14$ \\
$d$ &   $8.44 \times 10^{-2}$ \\
$e$ &  $0.737$ \\
$f$ & $0.478$\\
$i$ &    0.375 \\
$j$ &    0.212  \\
\end{tabular}
\end{center}
\caption{The values of the fitting parameters in Eqs. (\ref{315}) and (\ref{316}).} \label{Tp}
\end{table}

The correlations developed for the modified Sherwood numbers are adapted from \cite{kelly2018correlation}, and they are given by
\begin{equation}
Sh_{d}^* = Sh_d \left[1+a\left(\frac{gR_{0}^{3}}{\nu_{o}^{2}}\right)^{-b}\left(\frac{\rho_{m}-\rho_{a}}{\rho_a}Sc\right)^{c-b}Ra^b\right], \label{315}
\end{equation}
\begin{equation}
Sh_{c}^{*}=d\left(\frac{gR_{0}^{3}}{\nu_{o}^{2}}\right)^i\left(\frac{\nu_o}{\nu}\right)^{2(i-j)}\left(\frac{\rho_{m}-\rho_{a}}{\rho_a}\right)^{f-j}Sc^{e-j}Ra^j,\label{316}
\end{equation}
where $R_0$ is a nominal drop radius ($=1$ mm); $\rho_m$ and $\rho_a$ denote the density of the air-vapour mixture near the interface and of ambient air, respectively; $\nu_o$ is the ambient air viscosity at the standard condition, i.e $25^\circ$C and one atmospheric pressure; $\nu$ is the viscosity of the ethanol vapour at $(T_s + T_\infty)/2$; $g$ is the acceleration due to gravity; $Sc (\equiv {\nu / {\cal D}})$ is the Schmidt number of the vapour at $(T_s + T_\infty)/2$. The Rayleigh number, $Ra$ associated with the convective mass transfer is defined as
\begin{equation}
Ra=GrSc = \left(\frac{\rho_{m}-\rho_{a}}{\rho_a}\right)\left(\frac{g{R}^3}{\nu^2}\right)\times\left(\frac{\nu}{\cal D}\right),\label{eq_Ra}
\end{equation}
where $Gr$ is the Grashof number. The fitting parameter values are taken from \cite{kelly2018correlation} and are given in Table \ref{Tp}. The values of the relevant physical properties of ethanol vapour, water vapour, liquid ethanol, liquid water and air are provided in Table \ref{table:props}. Once the value of $Sh_{cor}$ is evaluated from Eq. (\ref{eq_Sh_corr}), the combined evaporation mass transfer rate due to  diffusion and convection from the droplet interface can be evaluated from the following relation
\begin{equation}
\left(\frac{dm}{dt}\right)_d + \left(\frac{dm}{dt}\right)_c = h_{d+c}A_s\mathcal{M}(c_{sat}(T_s) - c_{\infty}(T_{\infty})),\label{eq_conv+diff}
\end{equation}
where the combined diffusion and convection mass transfer coefficient, $h_{d+c}$ is given by
\begin{equation}
h_{d+c} = \frac{Sh_{cor}\mathcal{D}}{R}.\label{eq_h_d+f}
\end{equation} 
The above expression (Eq. (\ref{eq_h_d+f})) can be used for the evaporation of both pure ethanol and pure water droplets at room temperature. The volumetric change can be evaluated by dividing with the density of the corresponding liquids.

\begin{table}
\centering
\includegraphics[width=0.95\textwidth]{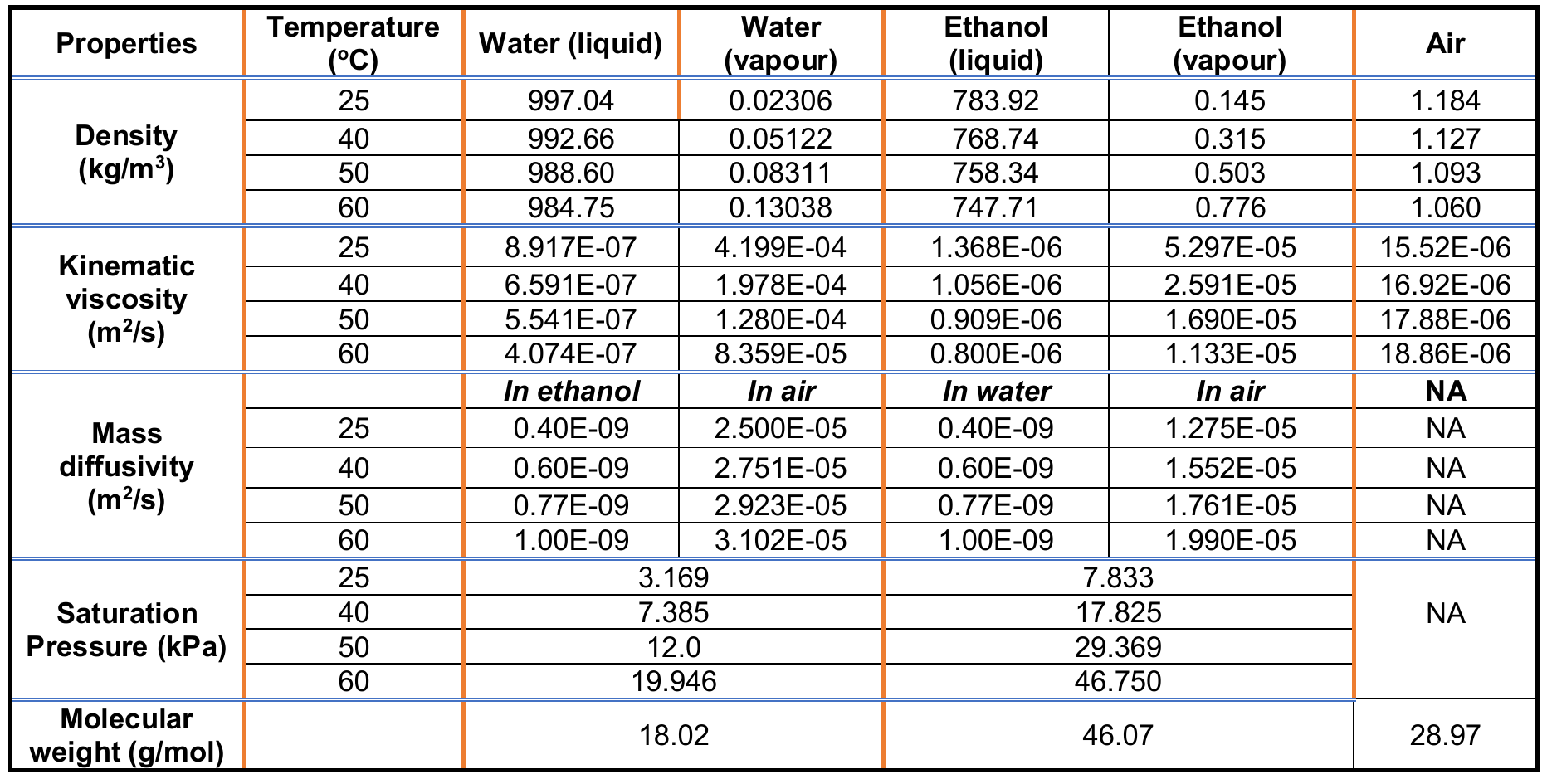}  
\caption{The properties of fluids considered in the theoretical modelling. The values are taken from various sources \citep{pal1968intermolecular,teske2005viscosity,lapuerta2014equation,hall1976survival}.}
\label{table:props}
\end{table}

At elevated substrate temperatures, the presence of free convection flux of air adds an additional complication that needs to be accounted for, particularly for the evaporation of an ethanol droplet as ethanol vapour is heavier than air, and does not rise up on its own like water vapour. Therefore, we consider the effect of passive transport of ethanol vapour due to air convection in the next section.

\subsubsection{Passive transport due to free convection of air} \label{sec:theory_drop_transp}
The relation considered so far incorporates the effects of both the diffusive mass flux and the convective mass flux due to the gradients in vapour concentration. But in the case of a sessile droplet on a heated substrate, additional terms arise due to the temperature gradient driven free convection of air. While water vapour can spontaneously rise up due to its natural buoyancy in air, the heavier ethanol vapour is passively transported along with the free convective air flow in the upward direction. This transport mass flux, denoted by $\left(dm/dt\right)_{t}$, can be expressed in terms of the mass flow rate of air as 
\begin{equation}
\left(\frac{dm}{dt}\right)_{t}=Y_{v}^s \left(\frac{dm}{dt}\right)_{a},
\end{equation}
where $Y_{v}^s$ is the mass fraction of (ethanol) vapour above the free surface of the droplet. The mass convection of air over the area of the heated substrate covered by the droplet can be expressed as
\begin{equation}
\left(\frac{dm}{dt}\right)_{a} = h_{m}^{a} \pi R^2 \frac{\mathcal{M}_a}{{\cal R}_u}\left(\frac{p_{\infty}^{a}}{T_{\infty}} - \frac{p_{s}^{a}}{T_{s}}\right).
\end{equation}
Here, air has been approximated as an ideal gas; ${\cal R}_u$ is the universal gas constant; $h_{m}^{a}$ denotes the mass transfer coefficient for air; ${\cal M}_a$ is the molecular weight of air; $p_{\infty}^a$ and $p_{s}^{a}$ are the partial pressures at the ambient and the plate surface, respectively. The Sherwood number for air {is} given by $Sh_{a} = (h_{m}^{a} R)/{\cal D}_{a}$, wherein ${\cal D}_{a}$ is the diffusion coefficient of air. For natural convection over a horizontal flat surface, $Sh_{a}$ is related with the air Rayleigh number, $Ra_{a}$ as follows \citep{lloyd1974natural},
\begin{equation}
Sh_{a}=0.54Ra_{a}^{1/4},
\end{equation}
where 
\begin{equation}
Ra_a=Gr_a Sc_a = \left(\frac{\rho_{a} (T_s) -\rho_a (T_\infty)}{\rho_a (T_\infty)}\right)\left(\frac{g{L}^3}{\nu_a^2}\right)\times\left(\frac{\nu_a}{\cal D}_a\right).
\end{equation} 
Here all the properties are with respect to air; the characteristic length, $L = A_p/{\cal P}_p$, wherein $A_p$ and ${\cal P}_p$ are the area and perimeter of the heated plate, respectively.

The final mass evaporation rate for ethanol from the substrates at elevated temperatures is thus the sum of diffusion, convection and passive transport terms given by Eq. (\ref{eq_massrate}) which can now be expressed as
\begin{eqnarray}\label{eq_ethanolhightemp}
\left(\frac{dm}{dt}\right) &=& \left(\frac{dm}{dt}\right)_d + \left(\frac{dm}{dt}\right)_c + \left(\frac{dm}{dt}\right)_t \nonumber \\ &=& h_{d+c}A_s\mathcal{M}(c_{sat}(T_s) - c_{\infty}(T_{\infty})) + Y_{v}^s \left(\frac{dm}{dt}\right)_{a}
\end{eqnarray}
 and is evaluated for the pure ethanol droplet at a given substrate temperature, $T_s$. The density of ethanol is calculated by assuming that the temperature of the droplet is equal to the substrate temperature. The calculated density is used to compute $V/V_0$ at various substrate temperatures. For water vapour, the passive transport term is not included as water vapour spontaneously rises up in air due to buoyancy. In the next section, we extend this model to study the evaporation of droplets of ethanol-water binary mixture of different compositions.

\subsection{Evaporation of binary fluid droplets} \label{sec:theory_drop_binary}

\begin{figure}[h]
\centering
\hspace{0.6cm}  (a) \hspace{6.0cm} (b) \\
\includegraphics[width=0.46\textwidth]{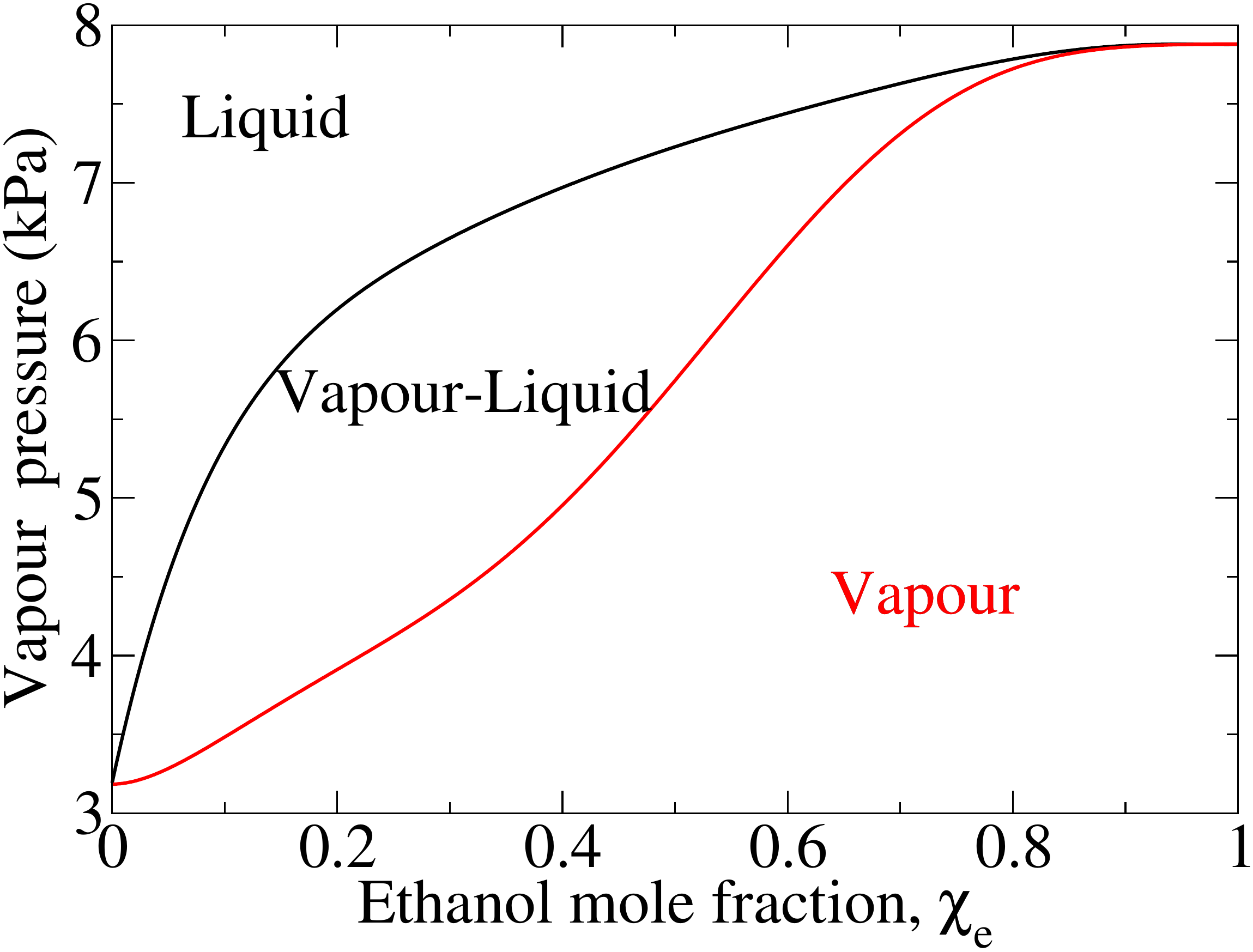} \hspace{2mm} \includegraphics[width=0.46\textwidth]{Figure19a.pdf} 
\caption{The vapour-liquid pressure curves for the binary mixture at (a) $25^\circ$C and (b) $60^\circ$C \citep{williams1998industrial}.}
\label{fig:vapourpressureT25}
\end{figure}

The evaporation of binary water-ethanol droplets is dependent on the same three processes, {namely} the molecular diffusion, the convective mass transfer and the convective air flow induced passive transport, which are discussed in the theoretical development for pure fluids (Section \ref{sec:theory_drop_pure}). However, for binary mixtures, the drop volatility and the component mole-fractions in the evaporating vapour now varies with the variation of the concentration of liquid water and liquid ethanol inside the evaporating droplet. Thus, the evaporation dynamics is governed by the vapour-liquid equilibrium of the binary mixtures. Details of the vapour-liquid equilibrium diagrams for the ideal and non-ideal solutions can be referenced from a chemical thermodynamics book (e.g. \cite{sandler2017chemical} ). The water-ethanol binary mixture in the liquid state deviates significantly from the ideal solution model and hence does not follow Raoult's law. However the vapour-liquid equilibrium (VLE) diagram for non-ideal solutions can be evaluated using the widely used functional group based semi-empirical UNIFAC and modified UNIFAC models \citep{gmehling1993modified}. {Sample plots of the VLE diagrams of water-ethanol binary mixture at $T_s = 25^\circ$C and $T_s = 60^\circ$C are provided in Figures \ref{fig:vapourpressureT25} (a) and (b), respectively. We used the isothermal droplet assumption such that the temperature of the entire droplet is  the same as the substrate temperature.}

In Figures \ref{fig:vapourpressureT25}(a) and (b), the ethanol mole-fraction $(\chi_e)$ and the vapour pressure are plotted along the $x$ and $y$-axes, respectively. The saturated liquid line (bubble line) and the saturated vapour line (dew line) are shown, which separate the pure vapour, the pure liquid and the two-phase region in between. For a given initial mole-fraction based on the droplet composition and at a given substrate temperature, the saturated liquid line provides the vapour pressure of the evaporating binary mixture. Then the tie line intercept with the saturated vapour line provides the molar composition of the newly evaporated vapour \citep{sandler2017chemical}. These vapour pressure and vapour phase mixture composition data from the VLE plot are used to calculate the instantaneous mass evaporation rate of the individual components (water and ethanol) via the relations developed in Section \ref{sec:theory_drop_pure}. The new molar composition of the liquid in the droplet for the next time step is calculated subsequently and is used in conjunction with the VLE diagram to evaluate the new vapour pressure and the bubble point composition. This iterative process is continued till {the end of evaporation}. The liquid solution density is evaluated at every time step, which is used to calculate the instantaneous droplet volume as shown below.
The total mass of the droplet at any instant, $m_{droplet}(t)$ is given by 
\begin{equation}
m_{droplet} (t)  = m_w(t) + m_{e} (t), \label{case1c}
\end{equation}
where $m_w(t)$ and $m_e(t)$ are the masses of water and ethanol present in the droplet of the binary mixture at any instant $t$. 
The mass fractions of water $Y_w (t)$ and ethanol $Y_e (t)$ in the droplet at any time, $t$ are given by
\begin{equation}
Y_w (t) = {m_w (t)  \over m_w (t)  + m_e (t)} , ~~ {\rm and} ~~ Y_e (t) = 1 - Y_w (t),
\end{equation} 
respectively.  The mole fractions of water, $\chi_w (t)$ and of ethanol, $\chi_e (t)$ can also be evaluated from the following relations
\begin{equation}
\chi_w (t) = {m_w (t)/\mathcal{M}_w  \over m_w (t)/ \mathcal{M}_w + m_e(t)/\mathcal{M}_e} , ~~ {\rm and} ~~ \chi_e (t) = 1 - \chi_w (t).
\end{equation}
The ethanol-water mixture is a non-ideal solution and requires an estimation of the excess molar volume of mixing $V_{\epsilon}$ \citep{marsh1980excess}. The density of the non-ideal mixture, $\rho_{m}$ can be evaluated from the expression,
\begin{equation}
\rho_m(t) = {\chi_w(t) \mathcal{M}_w +\chi_e(t) \mathcal{M}_e \over V_{\epsilon} + \frac{\chi_w(t)\mathcal{M}_w}{\rho_w} +\frac{\chi_e(t)\mathcal{M}_e}{\rho_{e}} }. \label{case1d}
\end{equation}
where $\rho_w$ and $\rho_e$ are the densities of water and ethanol, respectively. The values of the excess volume vary with mixture composition and mixture temperature, and are either tabulated \citep{marsh1980excess} or expressed in terms of Redlich-Kister (R-K) correlations \citep{jimenez2004effect}. For our calculations of $V_{\epsilon}$, we have used the R-K polynomial expansion coefficients provided in a recent study by \cite{danahy2018computing}.  Using Eqs. (\ref{case1c}) and (\ref{case1d}), we obtain the volume of the droplet of the ethanol-water mixture at any instant as
\begin{equation}
V(t) = {m_{droplet} (t) \over \rho_{m} (t)}.\label{case1e}
\end{equation}
The theoretically evaluated droplet volumes are then compared with the experimentally obtained values of the same. The comparisons between theoretical and experimental $(V/V_o)$ against $(t/t_e)$ for the various cases are presented in the next section.


\section{Comparison of the theoretical and experimental results }\label{theory_expt_comp}
In this section, the  theoretically obtained $(V/V_0)$ versus $t/t_e$ for droplets of pure ethanol and water  as well as ethanol-water binary solutions at a near-ambient and at elevated substrate temperatures are compared against the experimental results. 
\subsection{Droplets evaporating at a near ambient temperature}
\label{sec:theory_amb}
\begin{figure}
\centering
 \hspace{0.6cm}  (a) \hspace{6.0cm} (b) \\
\includegraphics[width=0.46\textwidth]{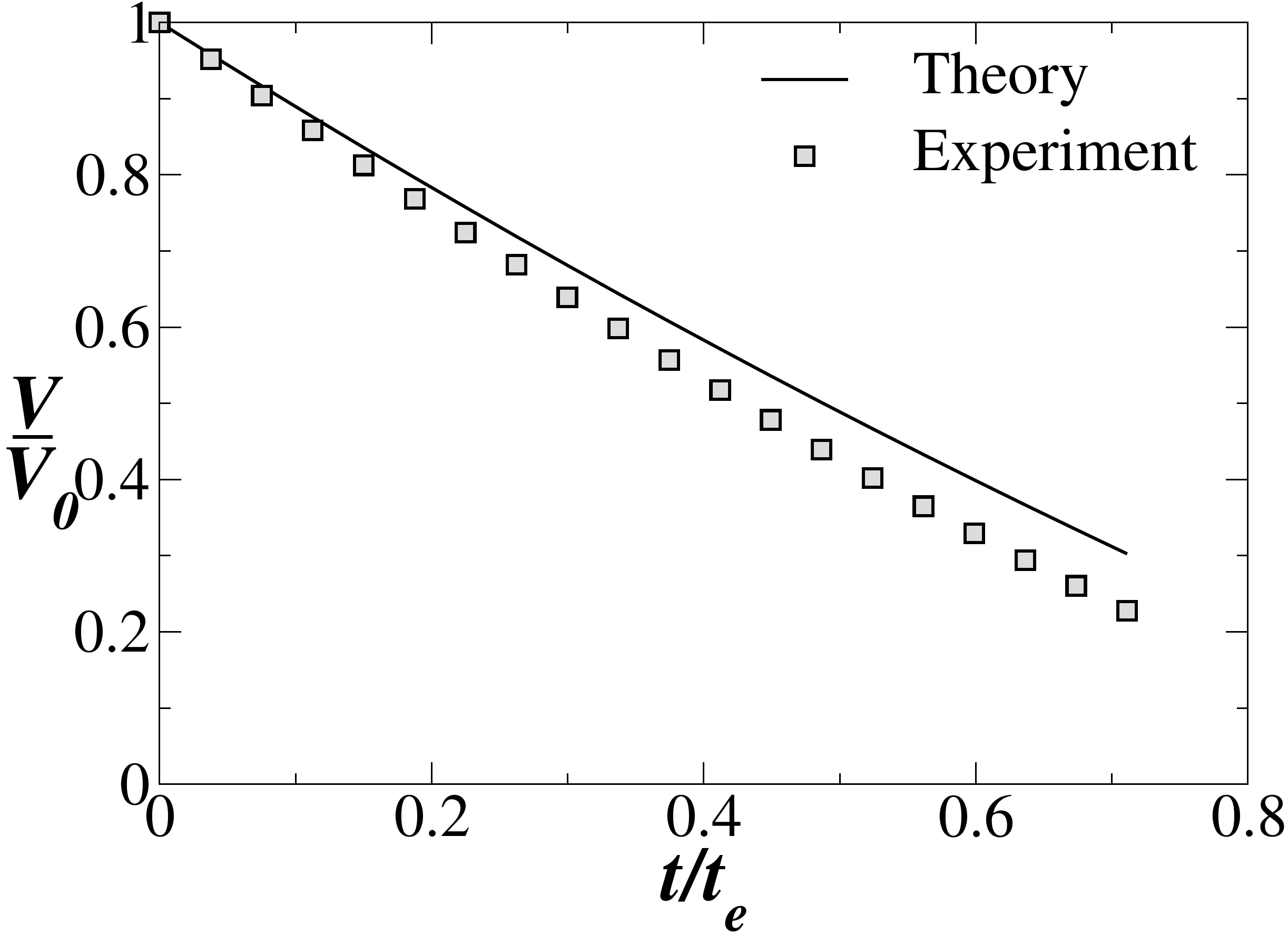}  \hspace{2mm} \includegraphics[width=0.46\textwidth]{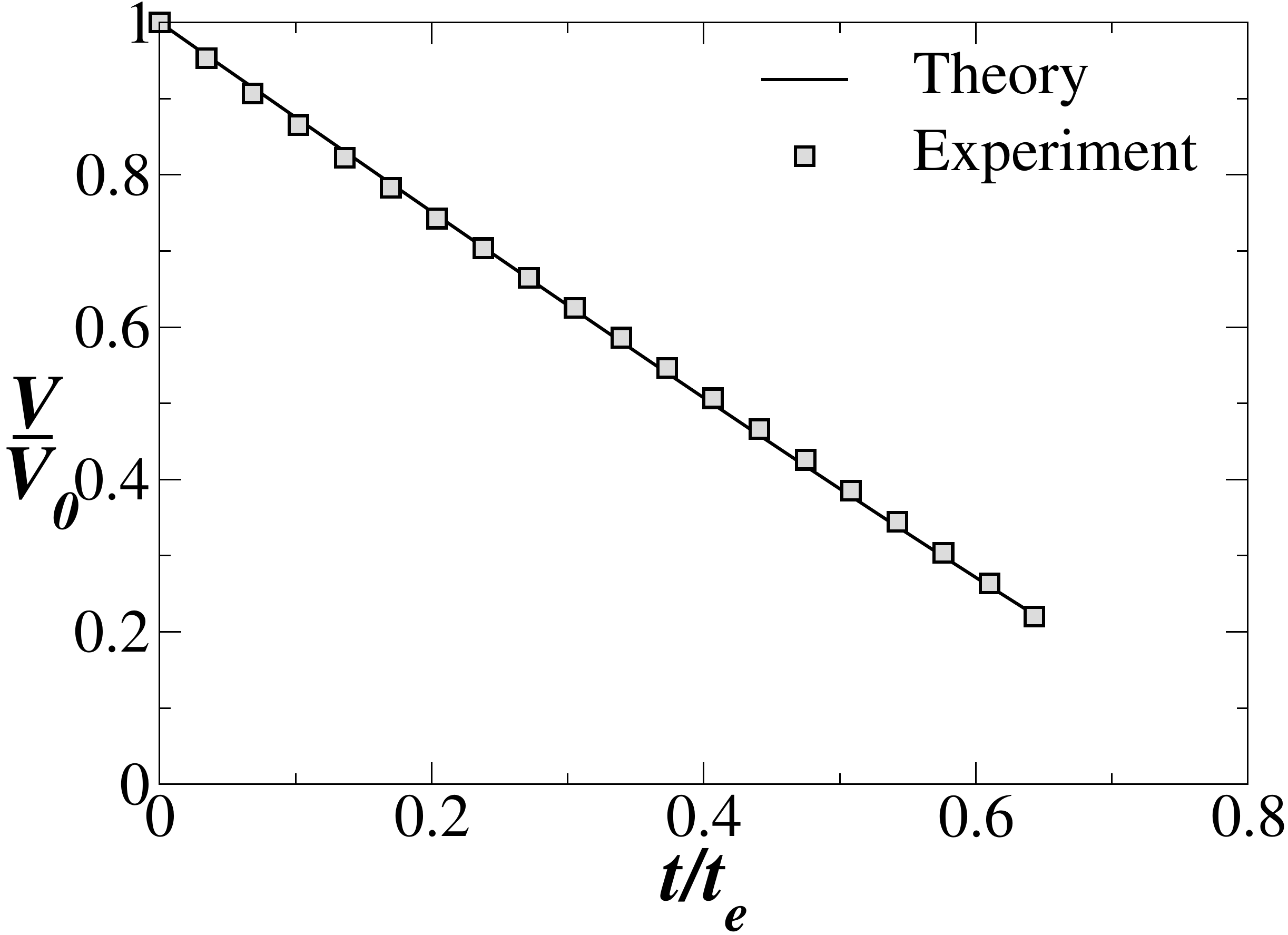} \\
 \hspace{0.6cm}  (c) \hspace{6.0cm} (d) \\
\includegraphics[width=0.46\textwidth]{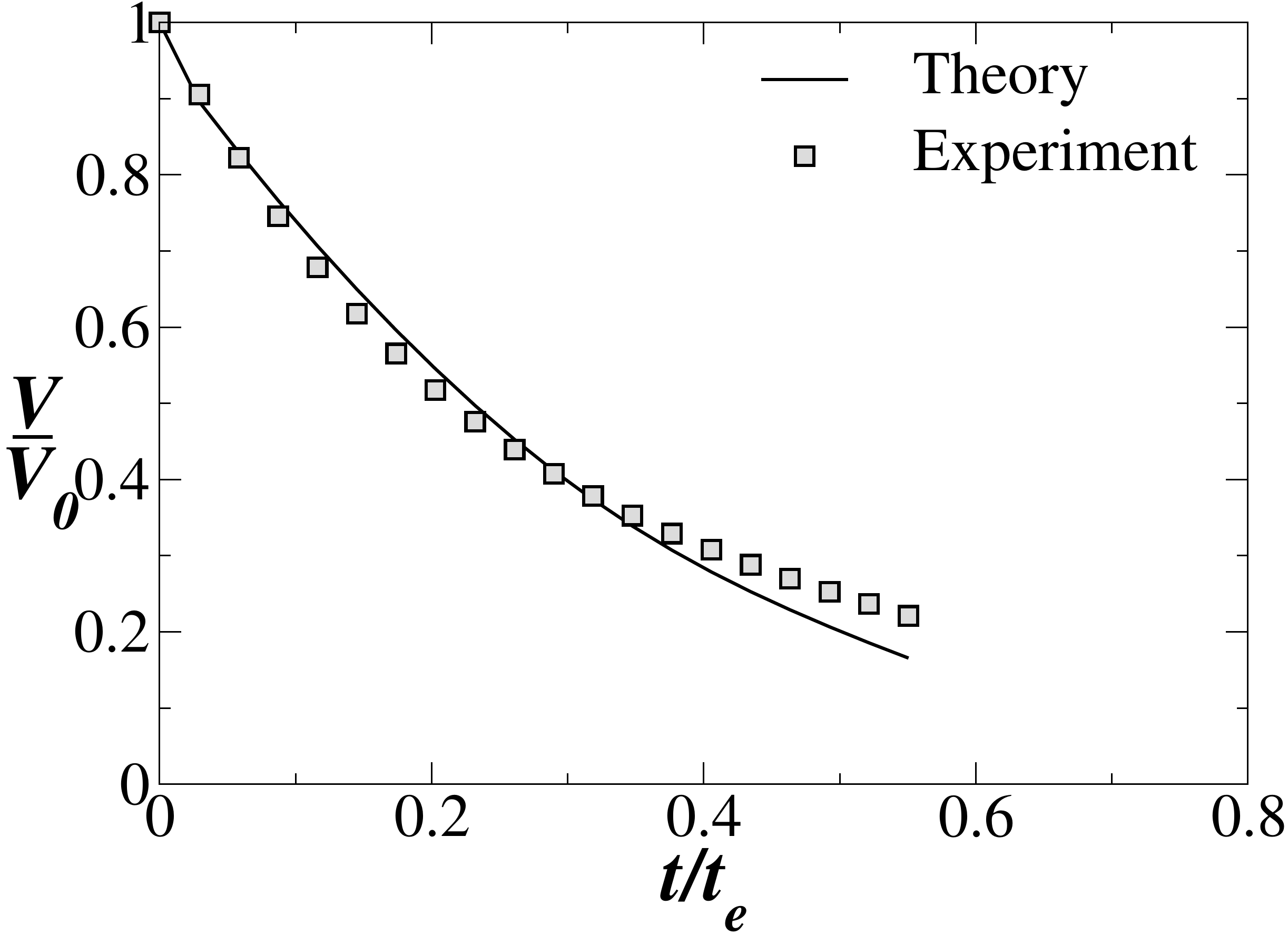}  \hspace{2mm} \includegraphics[width=0.46\textwidth]{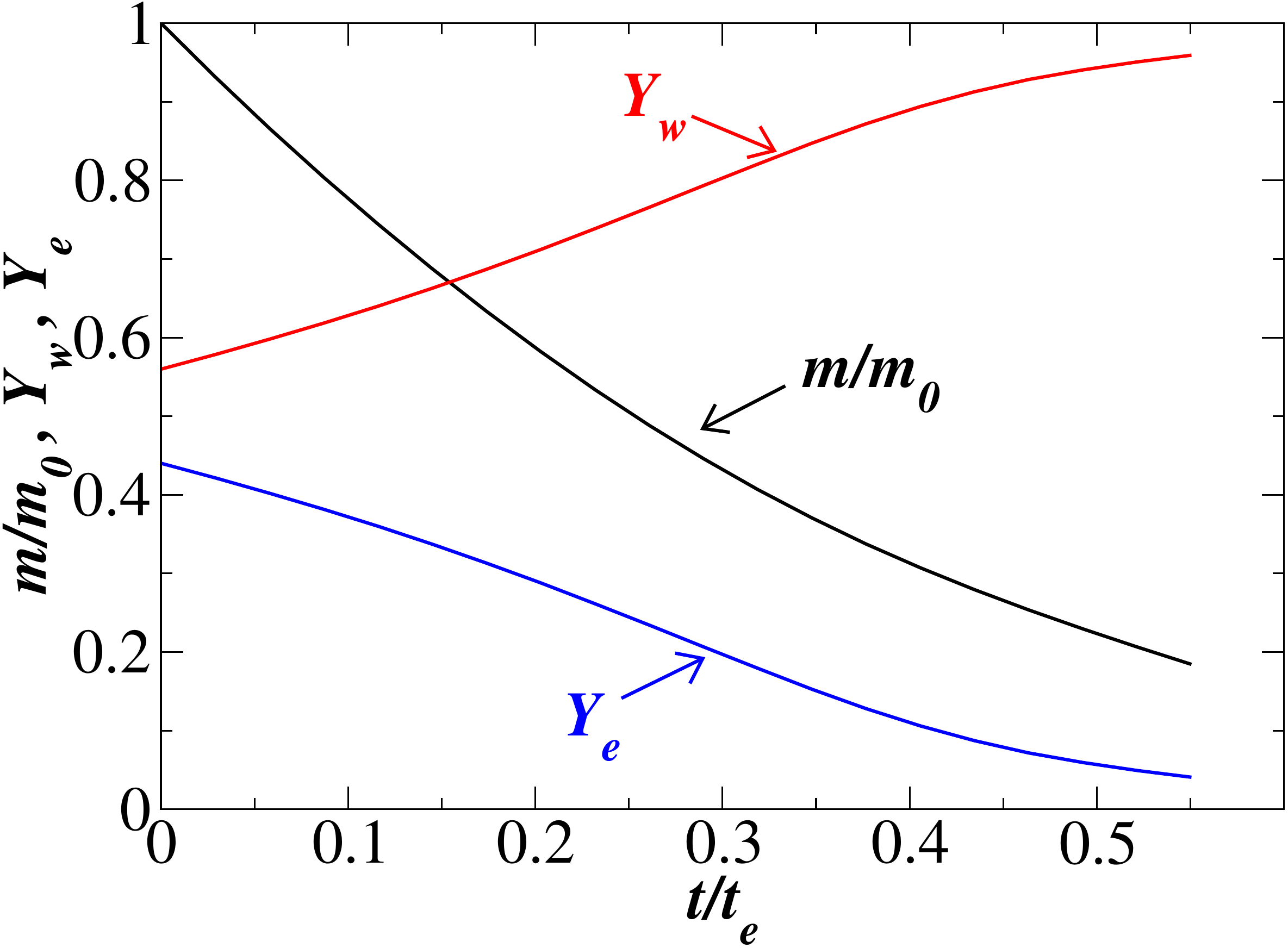} 
\caption{Comparison of the experimental and the theoretically obtained $\left ({V / V_0} \right)$ versus $t/t_e$. (a) Pure water (E 0\% + W 100\%), (b) pure ethanol (E 100\% + W 0\%) and (c) droplet of (E 50\% + W 50\%) binary mixture. (d) The variations of the normalised mass of the droplet with the initial mass of the droplet $(m/m_0)$, water and ethanol mass fractions, $Y_w$ and $Y_e$ versus $t/t_e$. Here substrate temperature, $T_s$ is $25^\circ$C}
\label{fig:theory1}
\end{figure}

The evolution of $\left ({V / V_0} \right)$ versus $t/t_e$ for droplets of pure water, pure ethanol and the (E 50\% + W 50\%) binary mixture evaporating at a near ambient substrate temperature of $25^\circ$C are evaluated from the theoretical models (Section \ref{sec:theory_drop}) and are compared with the data obtained from our experiments (Section \ref{sec:exptdis_rt}) in Figures \ref{fig:theory1}(a), (b) and (c), respectively. It is  seen in Figures \ref{fig:theory1}(a) and (b) that the diffusion-convection model (Eq. (\ref{eq_conv+diff})) predicts the evaporation behaviours of droplets of pure water and pure ethanol well at $T_s=25^\circ$C.  It is also seen in Figure \ref{fig:theory1}(c) that the implementation of the binary VLE based iterative algorithm, discussed in Section \ref{sec:theory_drop_binary}, provides an excellent theoretical match with the observed droplet volume curve for the (E 50\% + W 50\%) binary mixture. It is to be noted here that the experimental {result for} (E 50\% + W 50\%) is on the volume/volume basis and is converted to the mole/mole basis for the implementation in the VLE based calculations. The theoretically evaluated  mass-fraction histories of the individual  constituents along with the decrease in the normalised mass $(m/m_0)$ for the (E 50\% + W 50\%) droplet with $t/t_e$ are shown in Figure \ref{fig:theory1}(d). The figure demonstrates the the continuous decrease in the concentration of ethanol in the liquid solution with time, such that the droplet is almost entirely pure water for $t/t_e > 0.6 $. From the VLE diagram shown in Figure \ref{fig:vapourpressureT25}(a), one sees that a decreasing fraction of ethanol in the solution with dilute ethanol concentration ($\chi_e <0.1$) leads to a steep fall in the vapour pressure of the solution which explains the experimentally observed flattening curvature of the $(V/V_0)$ versus $t/t_e$ curve in the later stages of the droplet lifetime. Thus, apart from providing reliable predictions, the theoretical model is also capable of providing physical insights into the system dynamics that could not be gleaned from the experimental data alone.

\subsection{Comparison for pure droplets evaporating at an elevated substrate temperature}
\label{sec:theory_pure}
\begin{figure}[h]
\centering
 \hspace{0.6cm}  (a) \hspace{6.0cm} (b) \\
\includegraphics[width=0.46\textwidth]{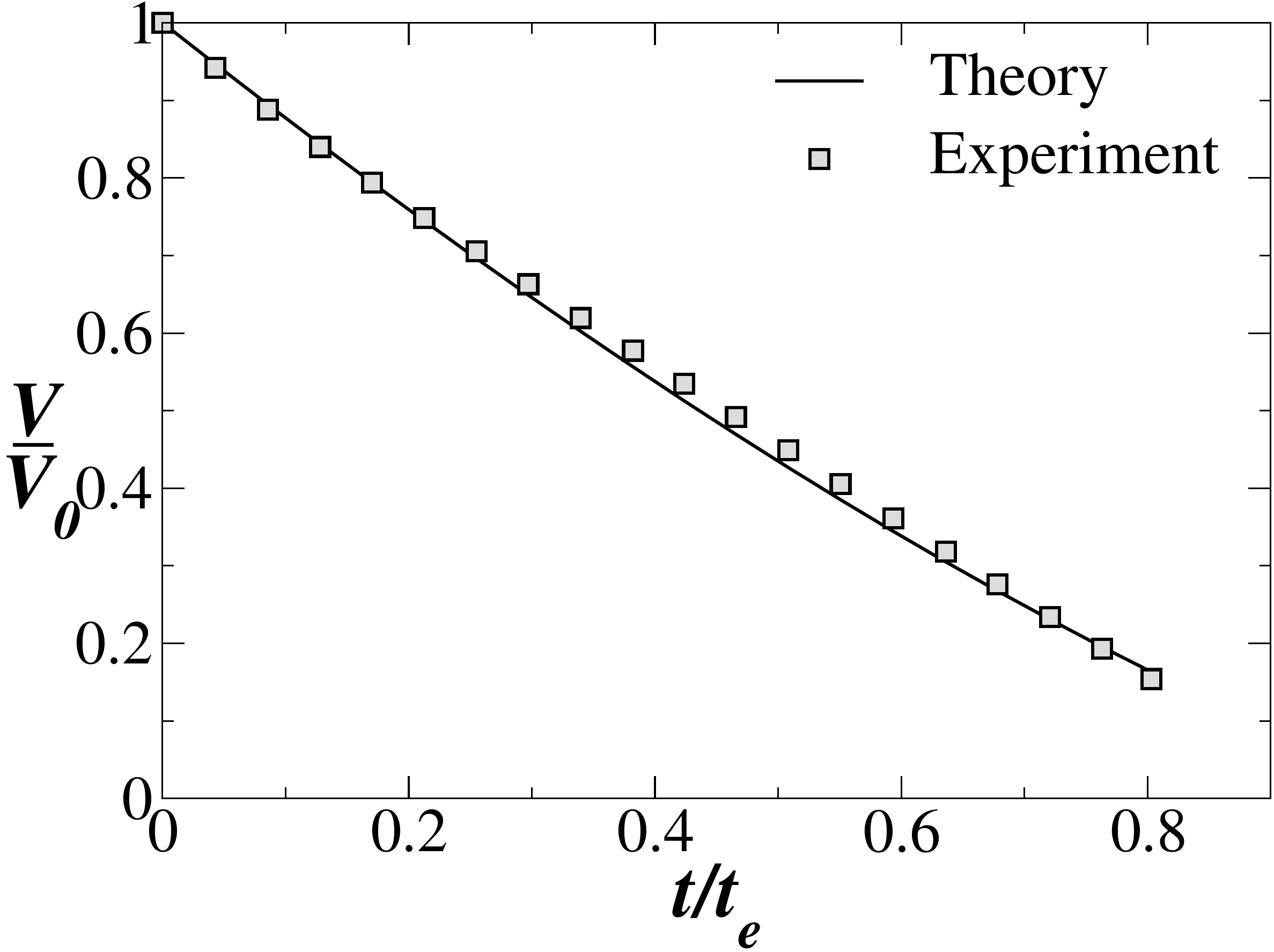}  \hspace{2mm} \includegraphics[width=0.46\textwidth]{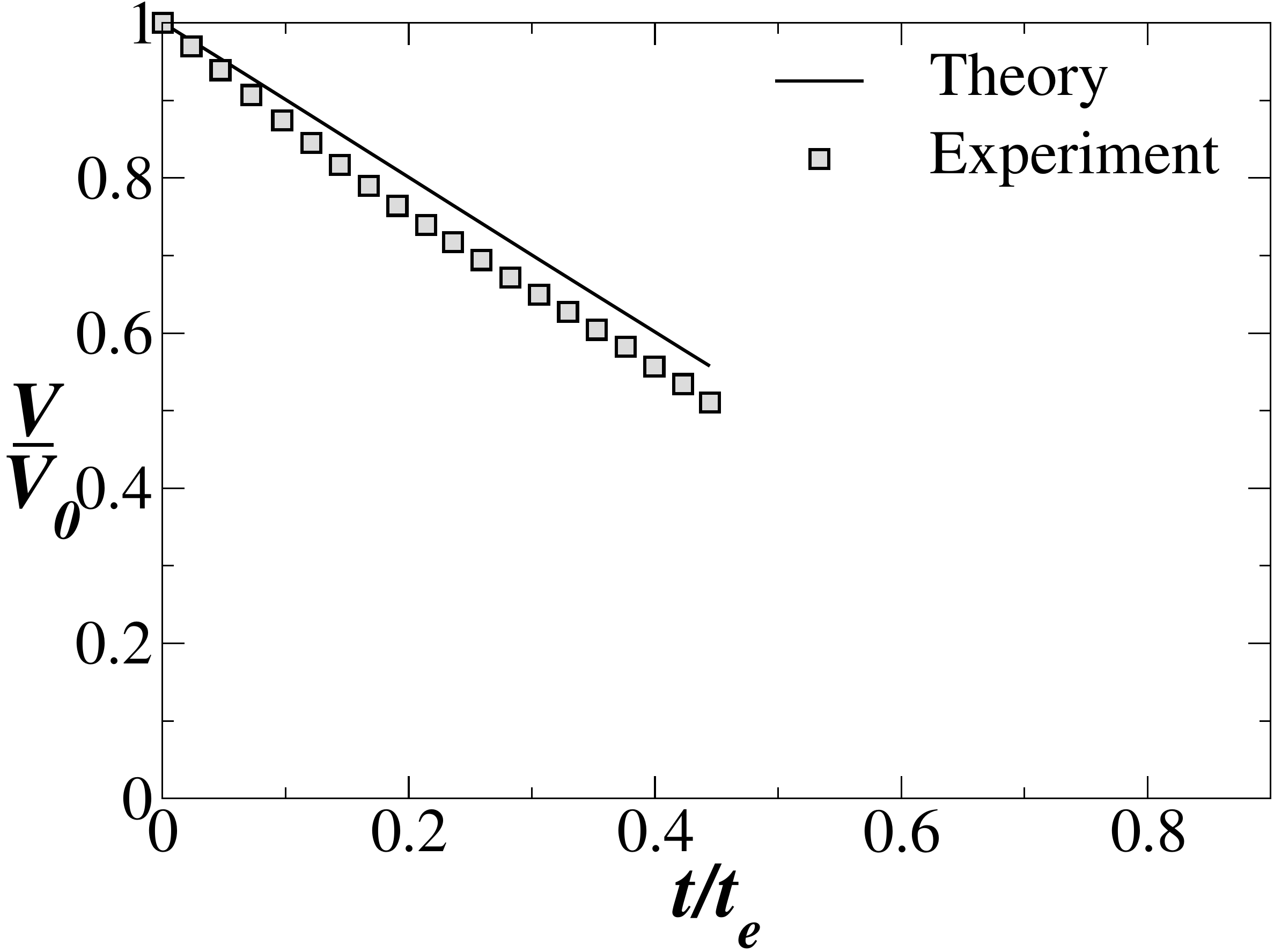} 
\caption{Comparison of the experimental and theoretically obtained $\left ({V / V_0} \right)$ versus $t/t_e$ at $T_s=60^\circ$C for pure fluids. (a) Pure water (E 0\% + W 100\%), and (b) pure ethanol (E 100\% + W 0\%).}
\label{fig:theory2}
\end{figure}

\begin{figure}[h]
\centering
\includegraphics[width=0.46\textwidth]{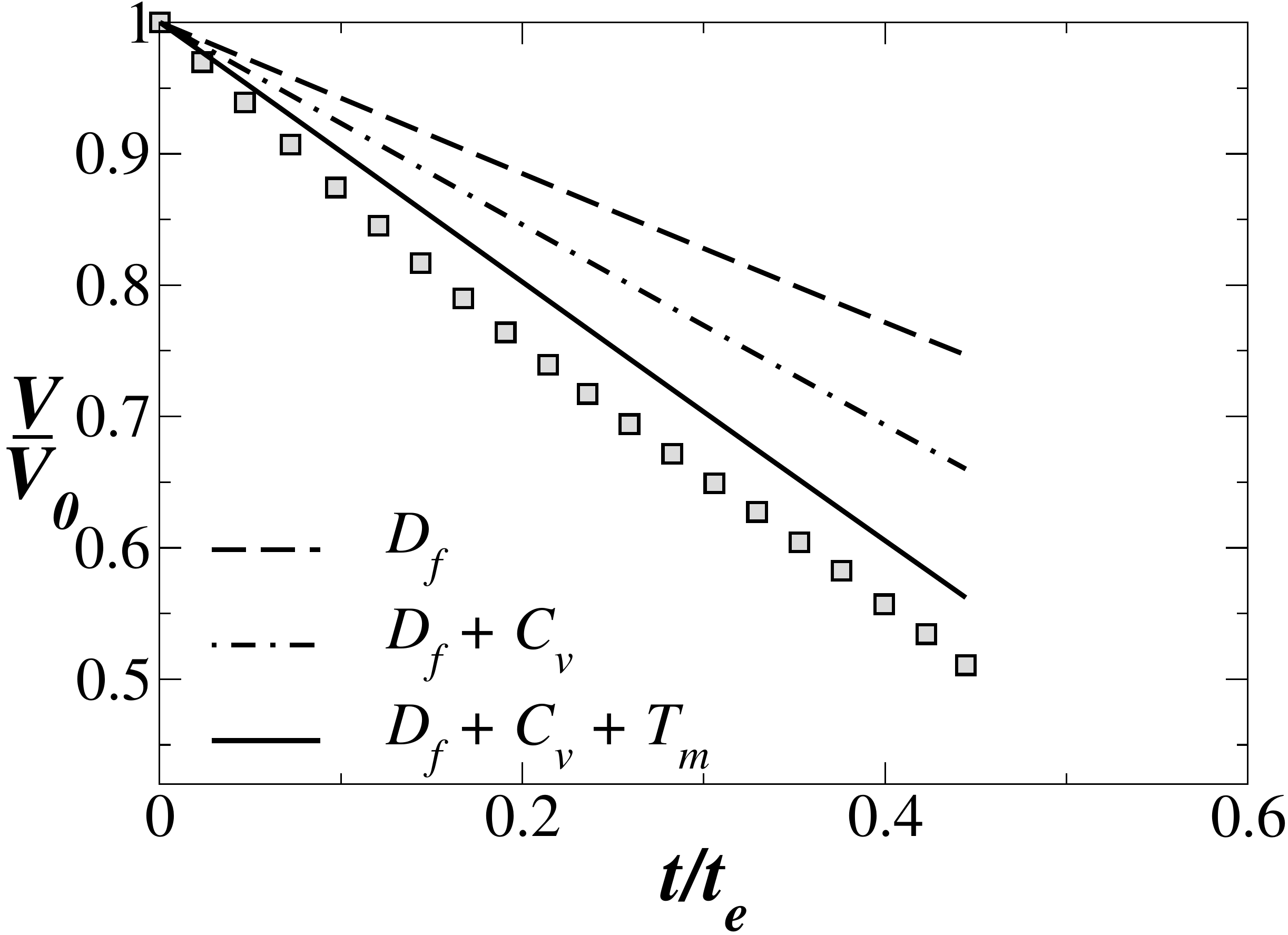} 
\caption{Comparison of the experimental and theoretically obtained $\left ({V / V_0} \right)$ versus $t/t_e$ at $T_s=60^\circ$C for pure ethanol (E 100\% + W 0\%) calculated using diffusion ($D_f$), diffusion + convection ($D_f+C_v$) and diffusion + convection + transport ($D_f+C_v+T_m$) models.}
\label{fig:comp}
\end{figure}

Next we compare the theoretical predictions for droplet volume with time for pure droplets at an elevated substrate temperature ($T_s=60^\circ$C). Figures \ref{fig:theory2}(a) and (b) show the comparison of the theoretical and experimental $(V/V_0)$ against $t/t_e$ at $T_s = 60^\circ$C for the droplets of pure water and pure ethanol, respectively. It can be seen in Figure \ref{fig:theory2}(a) that the diffusion-convection model agrees well with the experimental result for pure water, while Figure \ref{fig:theory2}(b) shows that the combined diffusion-convection-transport model outlined in Section \ref{sec:theory_drop_pure} (Eq. (\ref{eq_ethanolhightemp})) predicts the ethanol evaporation behaviour satisfactorily. In Figure \ref{fig:comp}, we plot the relative contributions of the three evaporation mechanisms for a pure ethanol droplet at $T_s=60^\circ$C. It is seen that pure diffusion alone {($D_f$)} grossly underpredicts the rate of evaporation mass loss. The $(V/V_o)$ slope remains underpredicted even when convection is coupled with diffusion {($D_f+C_v$)} using the correlations developed by \cite{kelly2018correlation}. Only after adding the natural convection of air induced passive mass transport {($D_f+C_v+T_m$)}, do we see a good match with the experimental results {(Figure \ref{fig:comp})}. Thus our experiments provide good validation of the importance of the three above mentioned mechanisms in jointly determining the evaporation history of ethanol from a sessile {ethanol} droplet deposited on a heated substrate. For water, the coupling {between the} diffusion and the convection alone is found to be sufficient as water vapour, being lighter than air, rises up by its inherent buoyancy.  

\subsection{Comparison for binary droplets evaporating at elevated substrate temperatures}
\label{sec:theory_binary}

\begin{figure}[h]
\centering
 \hspace{0.6cm}  (a) \hspace{6.0cm} (b) \\
\includegraphics[width=0.46\textwidth]{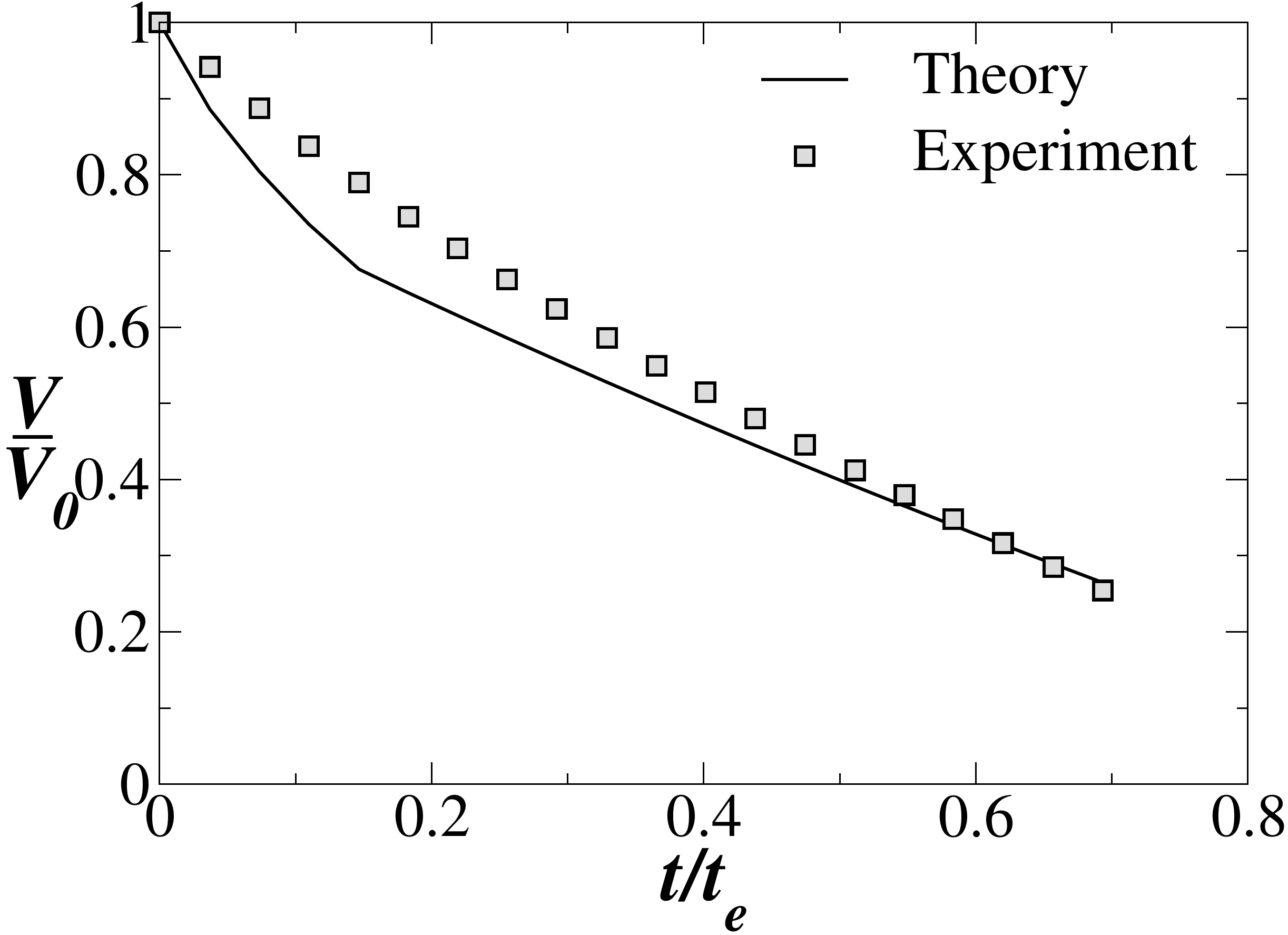}  \hspace{2mm} \includegraphics[width=0.46\textwidth]{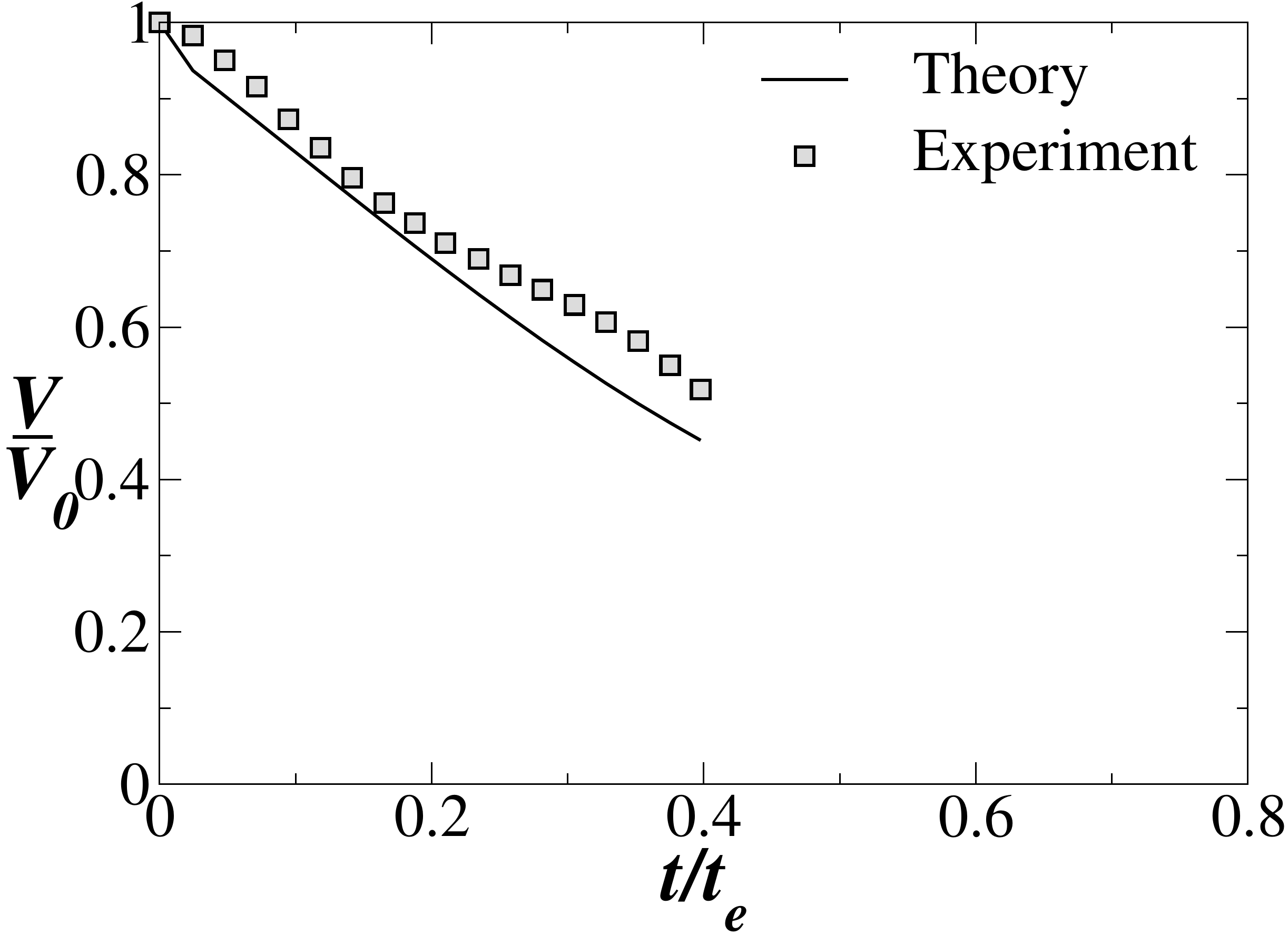}  \\
 \hspace{0.6cm}  (c) \hspace{6.0cm} (d) \\
\includegraphics[width=0.46\textwidth]{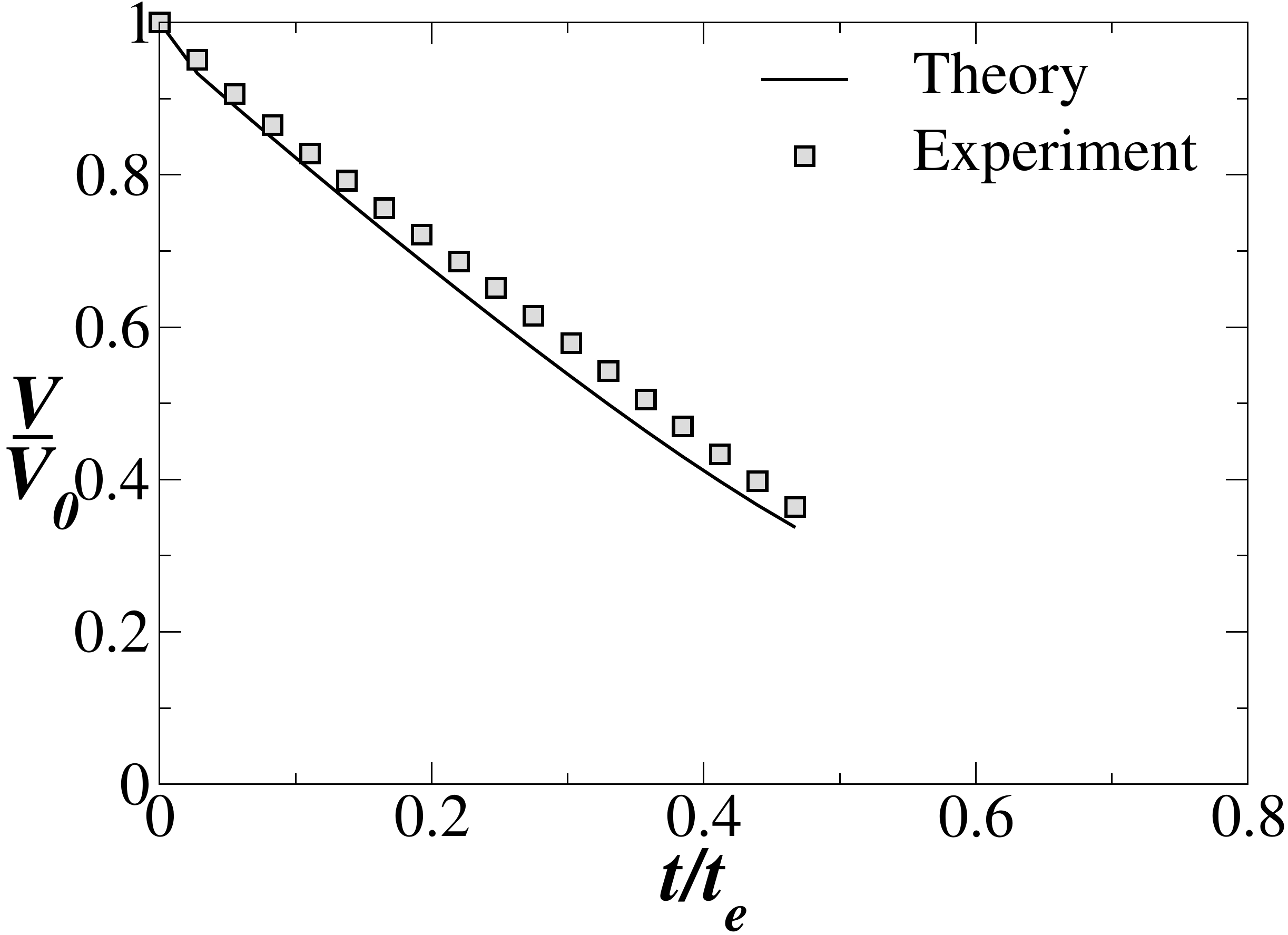} \hspace{2mm} \includegraphics[width=0.46\textwidth]{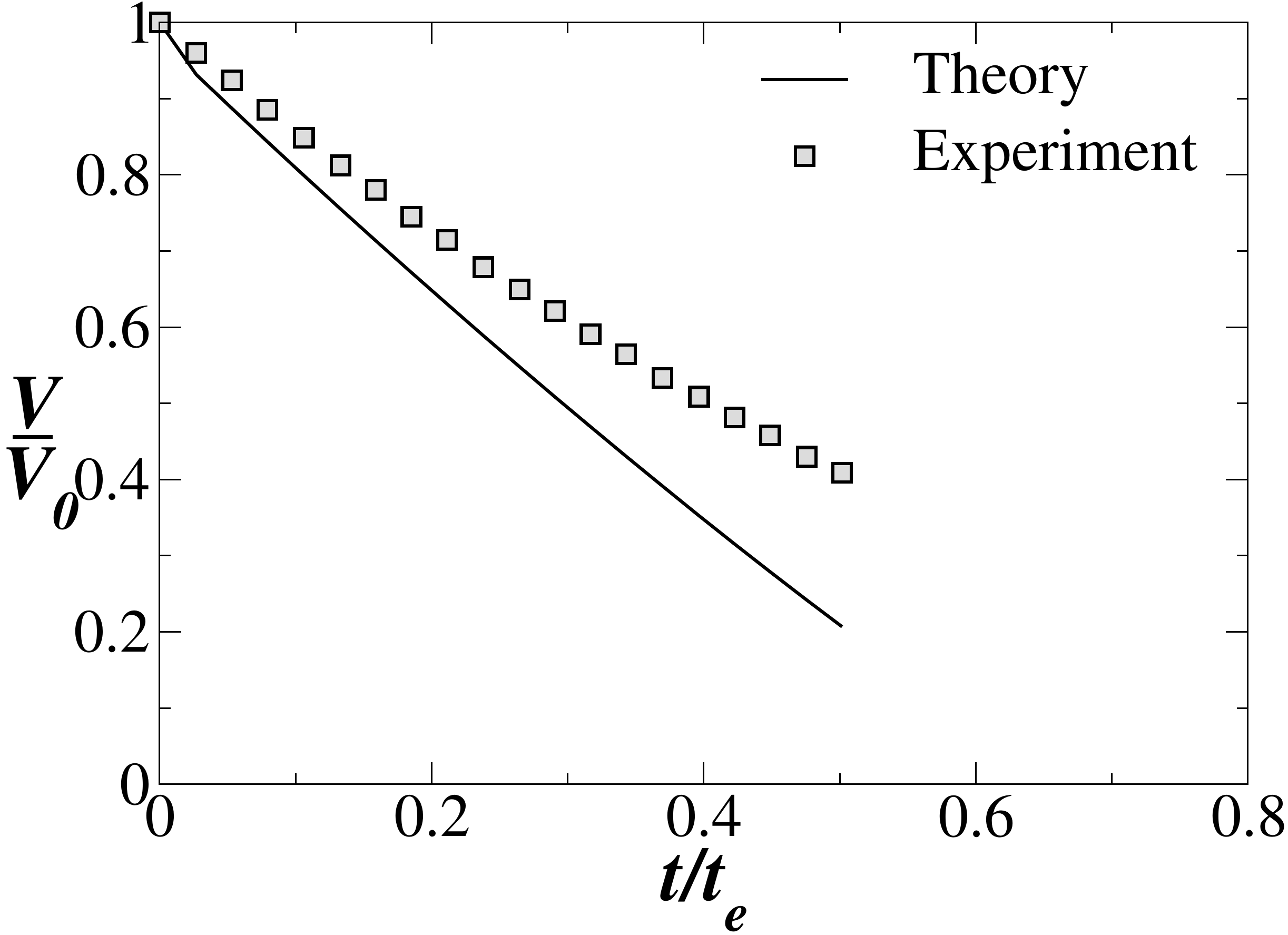} 
\caption{Comparison of the experimental and theoretically obtained $\left ({V / V_0} \right)$ versus $t/t_e$ at $T_s=60^\circ$C for droplet of binary mixtures (a) (E 20\% + W 80\%), (b) (E 50\% + W 50\%), (c) (E 60\% + W 40\%) and (d) (E 80\% + W 20\%).}
\label{fig:theory2b}
\end{figure}

Figures \ref{fig:theory2b}(a)-(d) compares the theoretical and experimentally obtained $(V/V_0)$ against $t/t_e$ for droplets of ethanol-water binary mixture of (E 20\% + W 80\%), (E 50\% + W 50\%), (E 60\% + W 40\%) and (E 80\% + W 20\%) compositions, respectively. It can be seen in Figure \ref{fig:theory2b}(a), (b) and (c) that for the (E 20\% + W 80\%), (E 50\% + W 50\%) and (E 60\% + W 40\%) droplets, a reasonably good agreement between the theoretical predictions and the experimental results is achieved. However, there is significant late stage divergence between the experimental and the theoretical $V/V_0$ values for the (E 80\% + W 20\%) droplet evaporation case. We suspect that this divergence is associated with the early onset of vigorous interfacial instability waves and eventual droplet break-up that was observed for this composition as has been noted earlier in Section \ref{sec:exptdis_ht}. Specifically, Figure \ref{new} shows that even before the eventual droplet break-up, the profile of the (E 80\% + W 20\%) droplet is visibly deviating from the spherical cap shape, while the (E 60\% + W 40\%) droplet retains the spherical cap profile till the very end of evaporation. Thus it is not surprising that the theoretical evaluations begin to deviate away from the experimental results as the droplet departs from the assumed spherical cap profile due to the increasingly intense waves and undulations on its free surface. Thus, while the simple analytical model remains adequate for the predictions of the steady-state droplet evaporation dynamics of the binary solutions on heated substrates, instability induced oscillations do have a non-negligible impact on the evaporation behaviour for some (though not all) binary compositions. More experimental and theoretical studies are required to predict the onset of such instabilities and assess their impact on the binary droplet evaporation.

\begin{figure}[h]
\centering
 \hspace{0.6cm}  (a) \hspace{6.0cm} (b) \\
\includegraphics[width=0.46\textwidth]{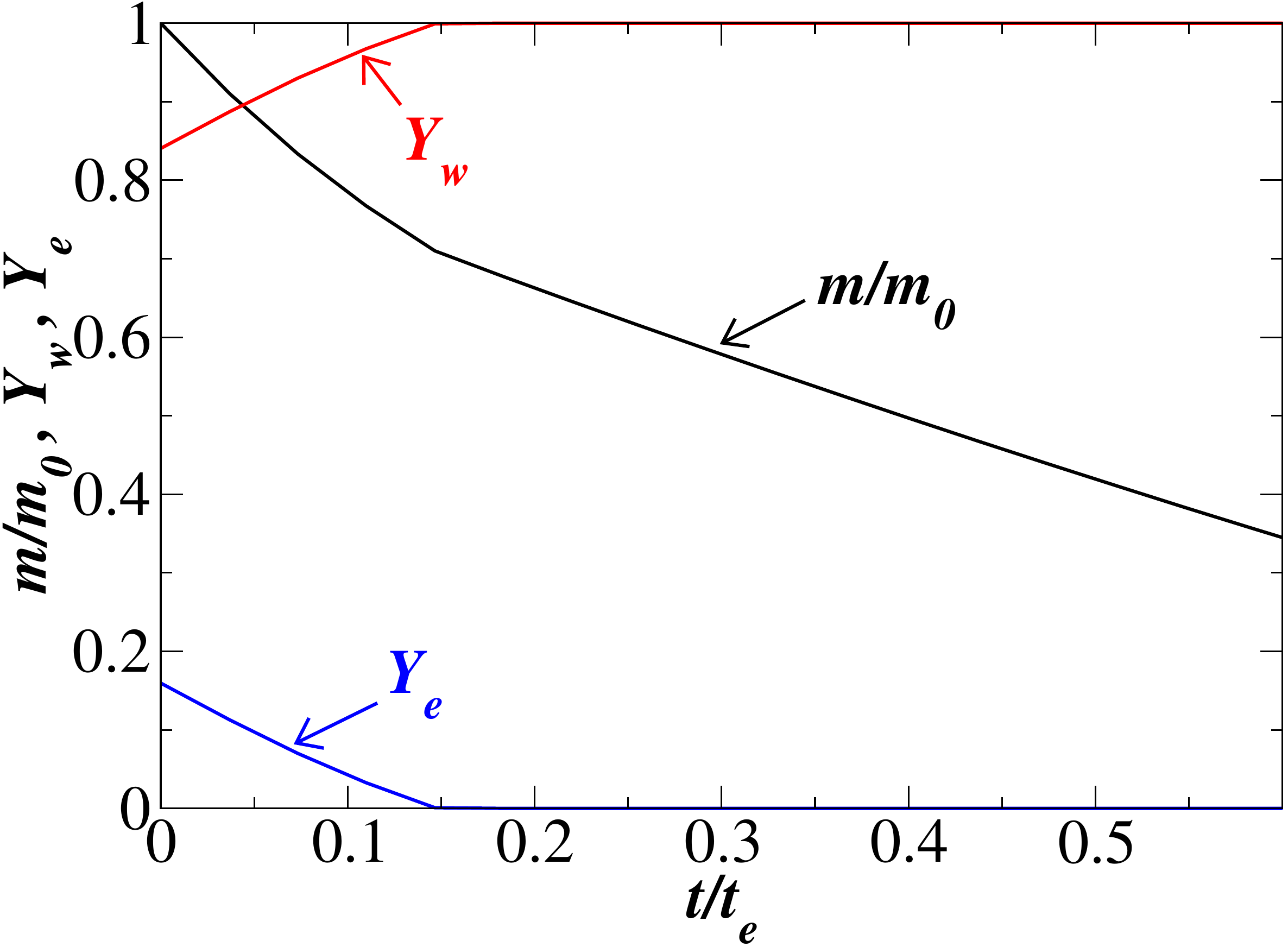}   \hspace{2mm} \includegraphics[width=0.46\textwidth]{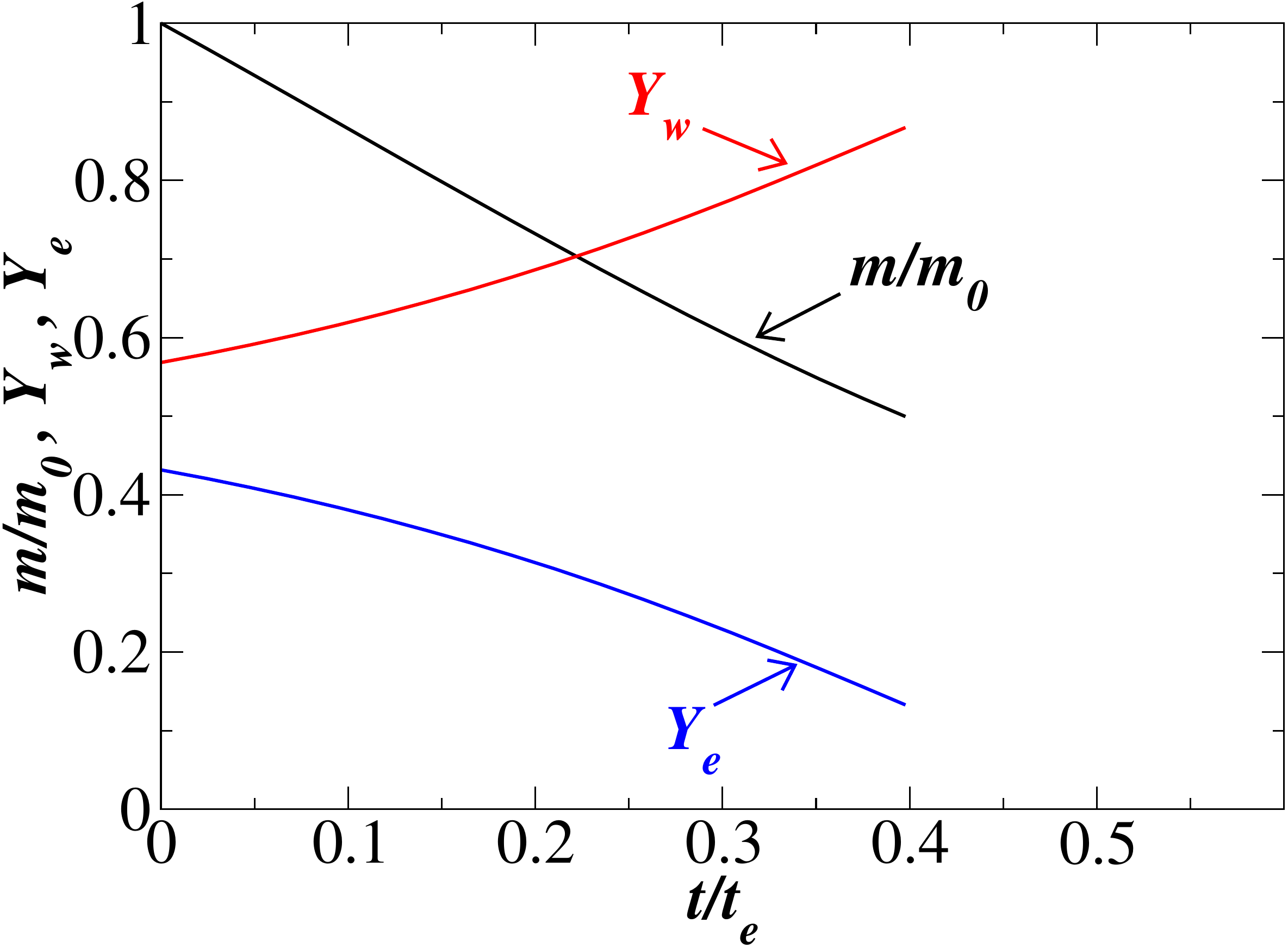}  \\
 \hspace{0.6cm}  (c) \hspace{6.0cm} (d) \\
\includegraphics[width=0.46\textwidth]{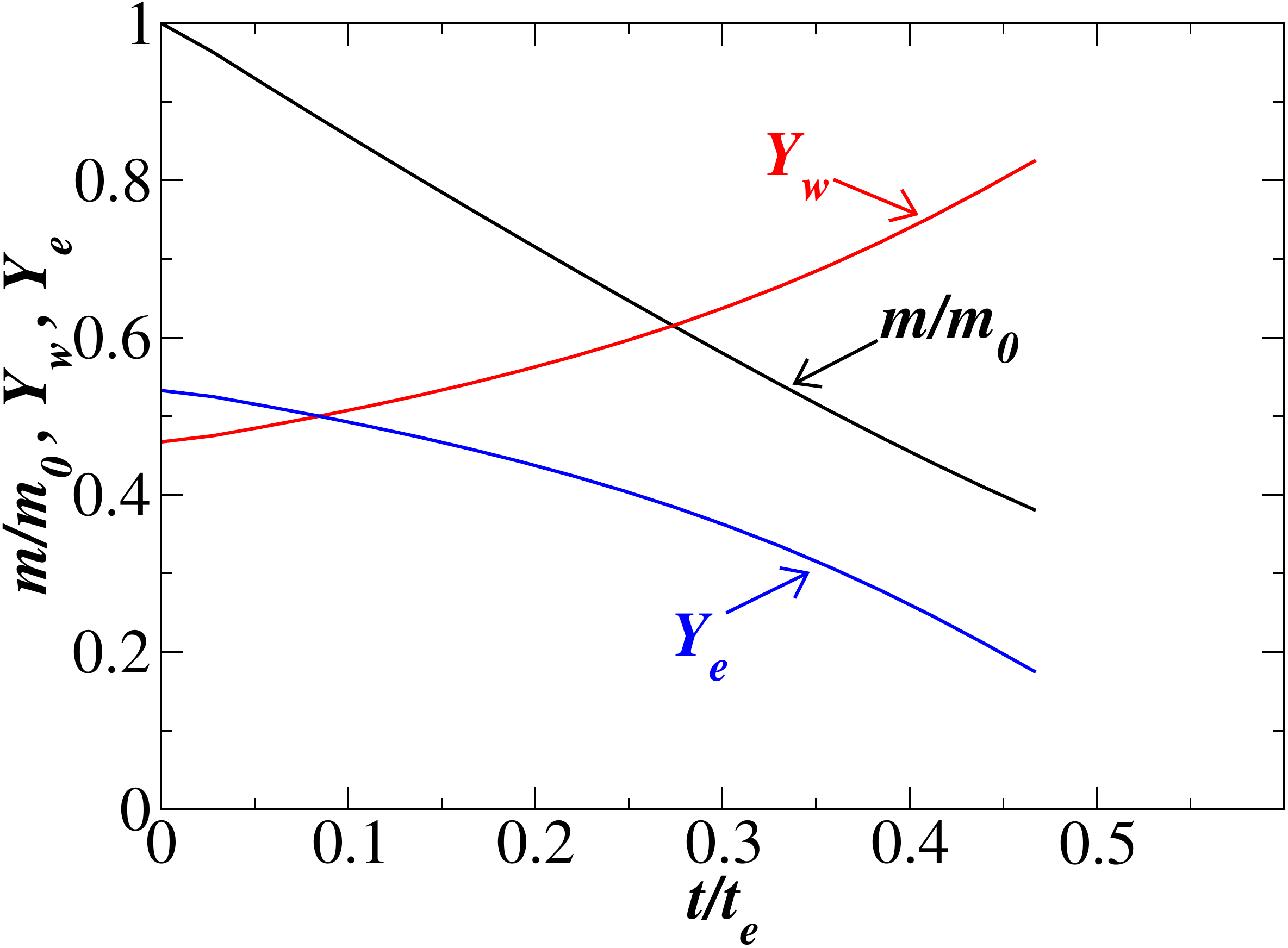} \hspace{2mm} \includegraphics[width=0.46\textwidth]{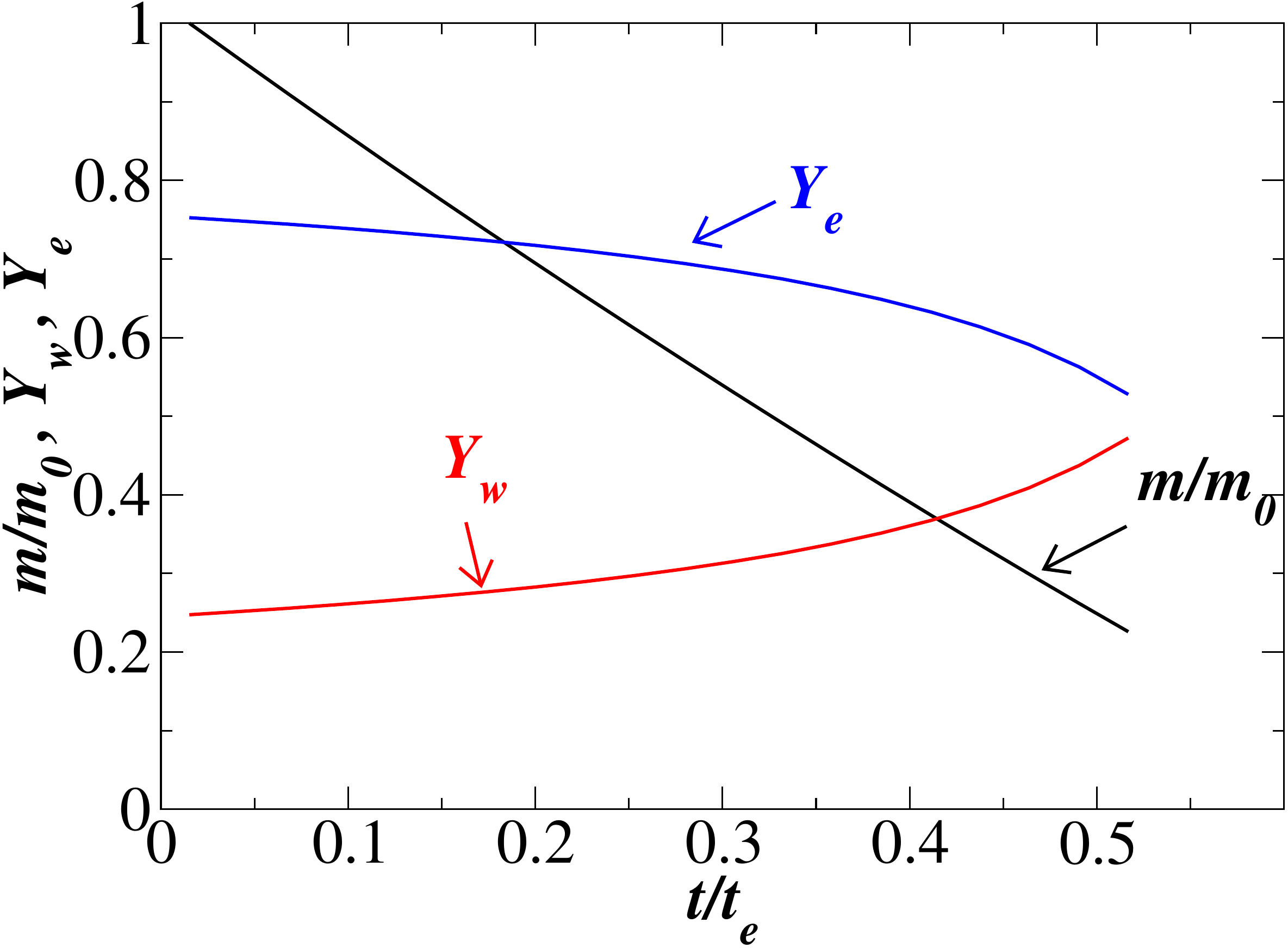} 
\caption{The variations of normalised mass of the droplet with the initial mass of the droplet $(m/m_0)$, $Y_w$ and $Y_e$ versus $t/t_e$ at $T_s=60^\circ$C for droplets of (a) (E 20\% + W 80\%), (b) (E 50\% + W 50\%), (c) (E 60\% + W 40\%) and (d) (E 80\% + W 20\%).}
\label{fig:massdiffC}
\end{figure}

Figures \ref{fig:massdiffC}(a)-(d) compare the evolution of the mass fractions of the individual components as well as the total normalised droplet mass $(m/m_0)$ with respect to the normalised evaporation time, $t/t_e$ for the (E 20\% + W 80\%), (E 50\% + W 50\%), (E 60\% + W 40\%) and (E 80\% + W 20\%) compositions, respectively. For the (E 20\% + W 80\%) case, the droplet is predicted to completely loose all its ethanol by $t/t_e = 0.15$, which is associated with the inflection point seen in the theoretical $(V/V_0)$ versus $t/t_e$ curve at this value of $t/t_e$ (Figure \ref{fig:theory2b}(a)). The real behaviour however shows a much gradual change in the slope curvature (experimental curve of Figure \ref{fig:theory2b} (a)), indicating that some ethanol remains in the droplet at a very dilute concentration for a time period considerably longer than that predicted by the mass loss calculation alone. The presence of the low concentrations of ``residual'' alcohol during the late stage evaporation of binary alcohol-water mixtures has been indicated in earlier studies as well \citep{sefiane2008wetting}. This observation is related to the fact that at very low alcohol concentrations the thermo-solutal convection flows {are switched off} inside the droplet \citep{christy2011flow}  and  the evaporation of ethanol becomes limited by the slow diffusion of ethanol molecules to the interface from deep inside the droplet as has been discussed in greater detail by \cite{liu2008evaporation}. {The ethanol component lasts longer for higher initial concentrations, though Figures \ref{fig:massdiffC}(a)-(d) show that mass fraction of water increases inside the droplet as evaporation proceeds.} Interestingly, for the (E 80\% + W 20\%) case, Figure \ref{fig:massdiffC}(d) shows that ethanol remains {as} the dominant component in the mixture till near the end stages of evaporation, with $Y_e \approx 0.5$ when $m/m_0 \approx 0.2$. Whether the large concentration of ethanol and the corresponding low surface tension of the droplet interface near the end stages of evaporation {are} conducive to the observed larger end-stage instability and breakup propensity of the (E 80\% + W 20\%)  evaporating droplet warrants further investigation. {Nevertheless}, once again, the simple analytical theory is seen to provide invaluable insights into the physics behind some of the more complex dynamics observed in our experiments for binary sessile droplets on heated substrates . 

\begin{figure}[h]
\centering
 \hspace{0.6cm}  (a) \hspace{6.0cm} (b) \\
\includegraphics[width=0.46\textwidth]{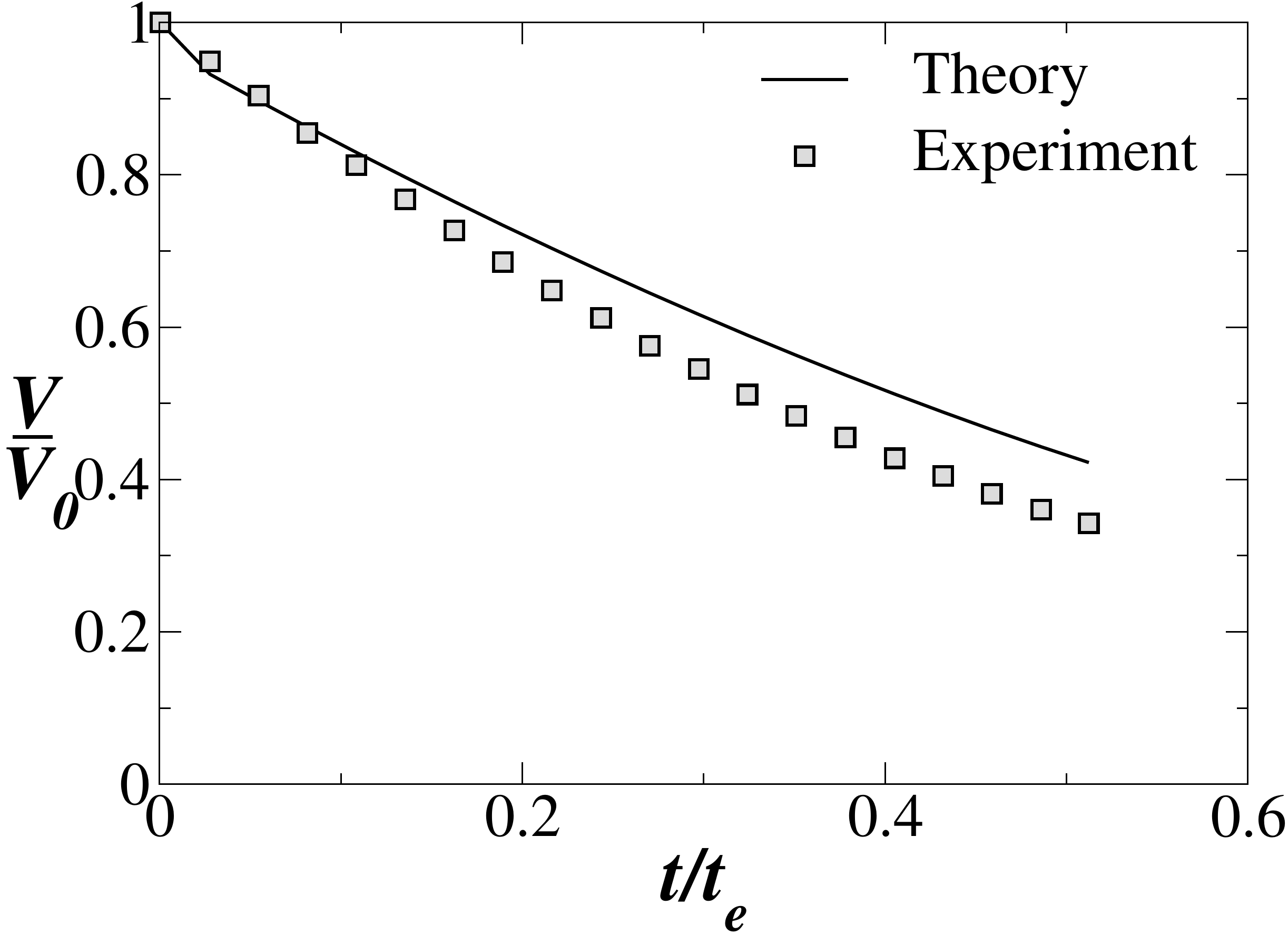}   \hspace{2mm} \includegraphics[width=0.46\textwidth]{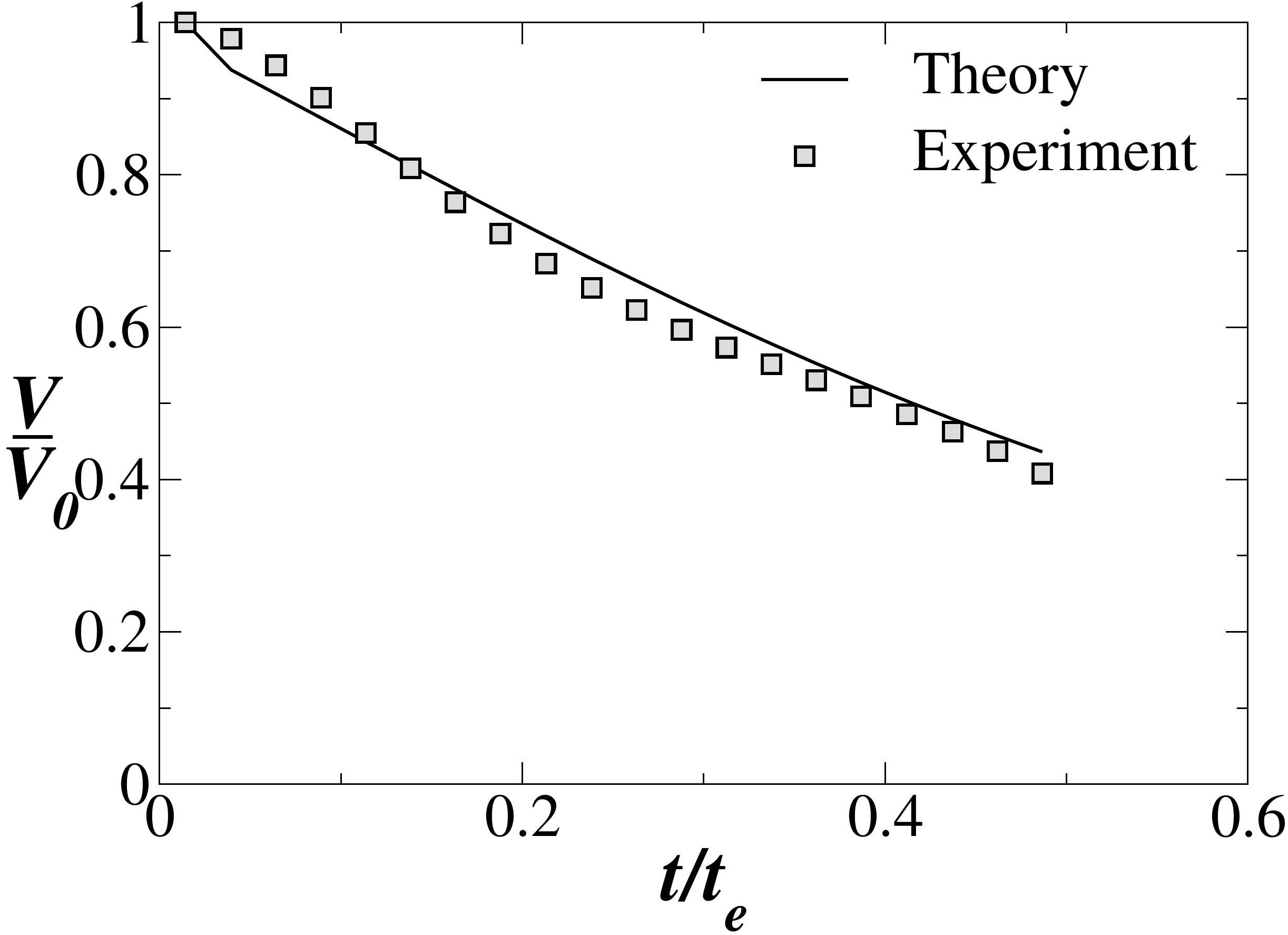} 
\caption{Comparison of the experimental and theoretically obtained $\left ({V / V_0} \right)$ versus $t/t_e$ for a droplet of (E 50\% + W 50\%). (a) $T_s = 40^\circ$C and (b) $T_s = 50^\circ$C.}
\label{fig:theory3}
\end{figure}

\begin{figure}[h]
\centering
 \hspace{0.6cm}  (a) \hspace{6.0cm} (b) \\
\includegraphics[width=0.46\textwidth]{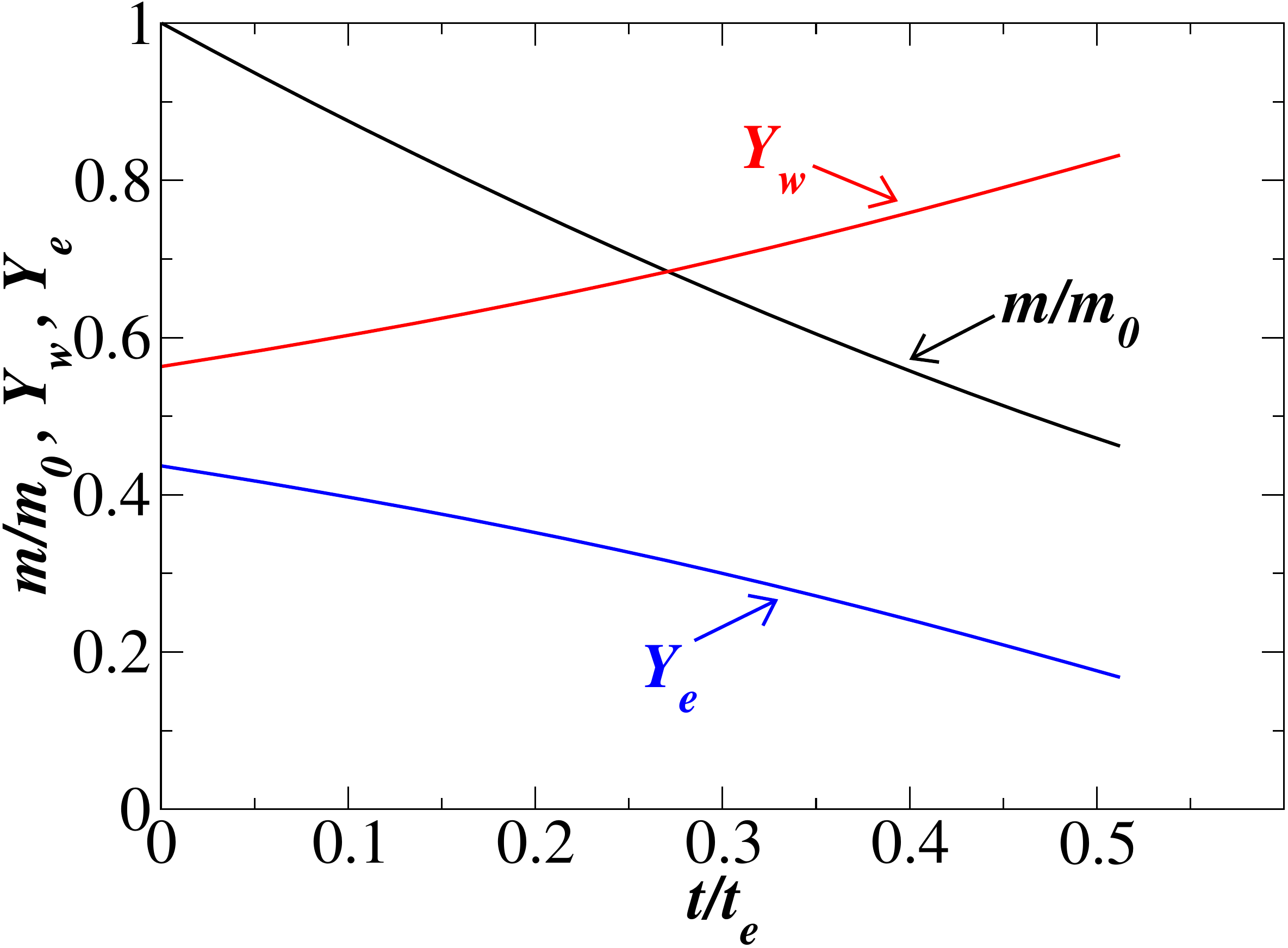}  \hspace{2mm} \includegraphics[width=0.46\textwidth]{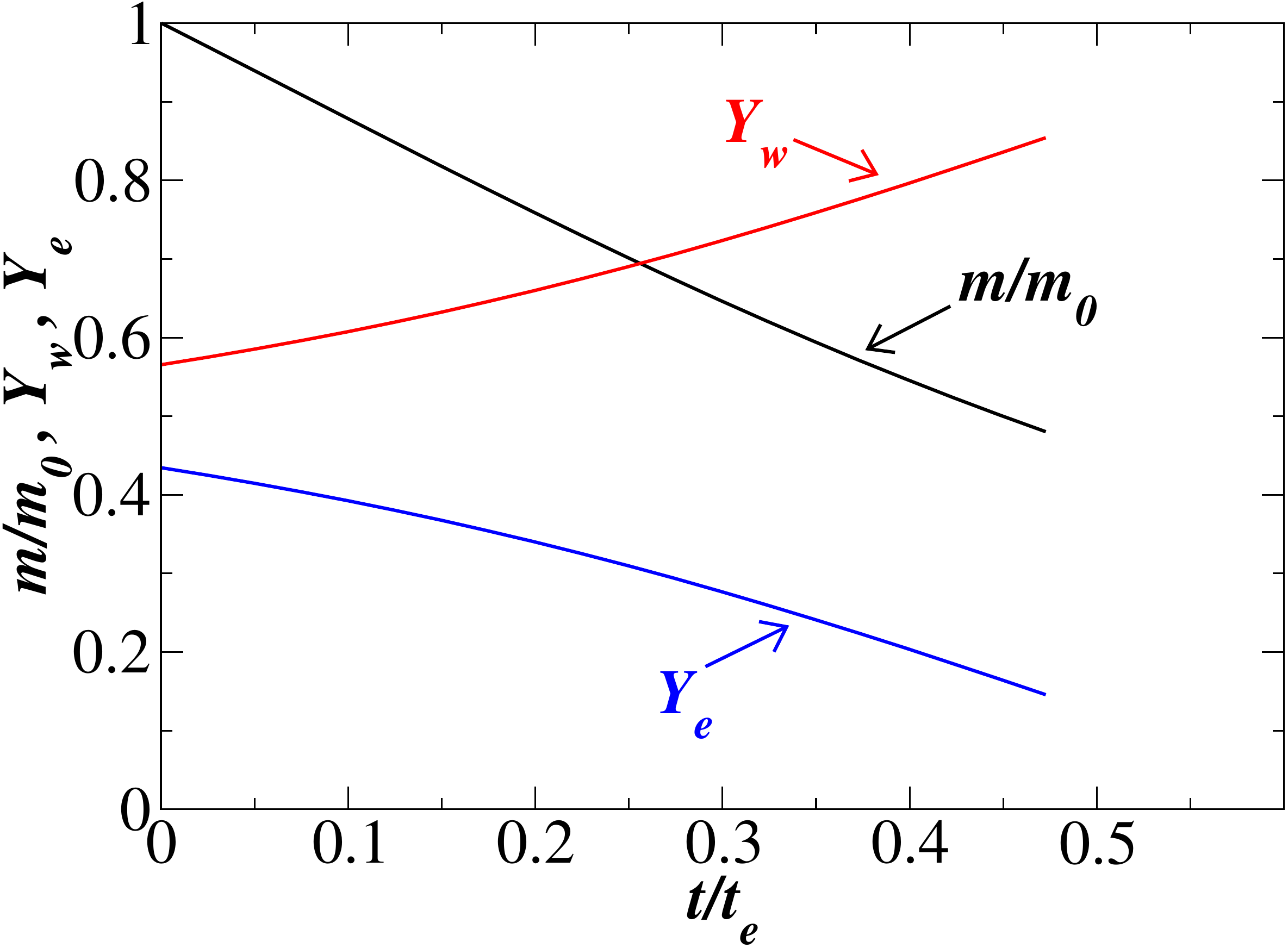}  
\caption{The variations of the normalised mass of the droplet with the initial mass of the droplet $(m/m_0)$, water and ethanol mass fractions, $Y_w$ and $Y_e$ versus $t/t_e$ at different substrate temperatures for a droplet of (E 50\% + W 50\%) binary mixture. (a) $T_s=40^\circ$C and (b) $T_s=50^\circ$C }
\label{fig:massdiffT}
\end{figure}

Finally, we have compared the theoretical and the experimental $(V/V_0)$ versus $t/t_e$ curves for a (E 50\% + W 50\%) droplet at $T_s=40^\circ$C and $50^\circ$C in Figures \ref{fig:theory3}(a) and (b), respectively. Once again, good agreements between the theoretical and experimental results are evident at the different substrate temperatures. The mass fraction plots for the two cases are shown in Figure \ref{fig:massdiffT}. 

It is to be noted that the normalisation of the time axis for ease of visualisation obscures the fact that the absolute evaporation times of the (E 50\% + W 50\%) droplet between the substrate at $25^\circ$C and the substrate at $60^\circ$C are widely divergent, spanning over two orders of magnitude, as can be {seen} in Figure \ref{fig:te2}. The ability of the model to accurately predict the evaporation rates over such a wide range of evaporation times and different substrate temperatures points to its inherent robustness. As the theory also incorporates the correlation from \cite{kelly2018correlation} that had been well validated for a wide range of hydrocarbons, we believe that the current methodology can be easily extended to other types of binary droplets as well, though that is left as a topic of future research.

\section{Summary}
\label{sec:conc}
The evaporation dynamics of sessile droplets of different compositions of ethanol-water binary mixture on a cellulose-acetate tape and at different substrate temperatures have been investigated experimentally using a customised goniometer fabricated by Holmarc Opto-Mechatronics Pvt. Ltd. Eight compositions of the ethanol-water binary mixture: pure water (E 0\% + W 100\%), (E 20\% + W 80\%), (E 50\% + W 50\%), (E 60\% + W 40\%), (E 80\% + W 20\%) and pure ethanol (E 100\% + W 0\%), and four values of substrate temperatures: $T_s=25^\circ$C, $40^\circ$C, $50^\circ$C and $60^\circ$C, have been considered in our experiments. We did not increase the substrate temperature further as the boiling temperature of pure ethanol is about $78^\circ$C \citep{lemmon2007nist}. The volume of the droplets is fixed at 5$\mu$l. In order to ensure the repeatability of the experiments, each case is repeated about 6 times. The experimental results are compared against the predictions obtained from theoretical models. To the best of our knowledge, the evaporation dynamics of a droplet consisting of a binary mixture have not been studied at elevated temperatures. \cite{saenz2017dynamics} have mentioned about the evaporation dynamics at elevated temperatures but mainly focused on evaporation of droplets of complex shapes at room temperature. Even at room temperature and for a fixed composition of ethanol-water binary mixture, previous experimental studies have predicted different spreading and wetting behaviours of a sessile droplet. These differences in the droplet evaporation dynamics observed by the previous investigations were due to the variations in the property of the substrates used and due to different local ambient conditions \citep{cheng2006evaporation}. As the surface roughness and the local ambient conditions are known to play a significant role in the evaporation dynamics, in order to obtain a physically grounded understanding of this complex phenomenon, it is necessary to perform systematic experiments at different substrate temperatures and for different compositions of the binary mixture in a fixed condition. Such an attempt is made in the present study for the first time. 

In the present work, we have chosen the substrate on the basis of roughness studies by an atomic force microscope (AFM) and micro-scale images taken by a scanning electron microscope (SEM) at room and elevated temperatures. We found that the cellulose-acetate tape is stable even at elevated temperatures, and thus has been used in the present study. The experiments are conducted for droplets of different compositions of ethanol-water mixture at different substrate temperatures. The theoretical models are also developed and compared against the experimental results. Due to the competition between the evaporation rates of the individual components in an ethanol-water binary mixture, we observed contrasting dynamics at the room and elevated temperatures. Thus the presentation of the experimental and theoretical results are classified into three sections. 

First, the results are presented at $T_s=25^\circ$C (which is close to the ambient temperature). The evaporation dynamics of a sessile droplet of ethanol-water binary mixtures at room temperature has been studied by several authors (see, for instance \cite{sefiane2003experimental,saenz2017dynamics,sterlyagov2018}), as also discussed in the introduction. While for pure water and pure ethanol droplets, we observe pinned-stage evaporation, in case of a sessile droplet of (E 50\% + W 50 \%) mixture, two distinct stages, namely, the early pinned-stage and the later receding-stage are observed. It is also found that the height, the contact angles and the volume of the droplet decrease monotonically for droplets of pure fluids. However, in case of a droplet of (E 50\% + W 50 \%) binary mixture, the more volatile ethanol, evaporates faster leading to a nonlinear trend in the evaporation process at the early stage. The contact angles of the droplet during the entire evaporation process are found to be less than 90$^\circ$ at $T_s=25^\circ$C. {The theoretical modelling of the droplet evaporation flux is done by combining the diffusion limited vapour mass flux mechanism \citep{sobac2012,carle2016} with the concentration gradient induced convection flux model developed by \cite{kelly2018correlation}. The binary droplet model also incorporates the binary vapour-liquid equilibrium diagram data \citep{gmehling1993modified} to obtain accurate mass flux calculations. The theoretical evaporation flux model is found to provide an excellent match with the experimental results for the droplet volume evolution with time for both the pure fluid and the binary mixtures.}

Secondly, we study the behaviour of sessile droplets of ethanol-water binary mixture of different compositions at an elevated substrate temperature ($T_s=60^\circ$C). The evaporation dynamics observed in this case is more interesting. At $T_s=60^\circ$C, the lifetime of the droplet, $t_e$ shows a non-monotonic trend with the increase in ethanol concentration in the binary mixture, which can be clearly divided into two regions (E $< 50$\% and E $\ge 50$\%). For E $< 50$\%, $t_e$ decreases linearly, but for E $\ge 50$\%, $t_e$ does not change much, which we believe is due to the non-ideal vapour pressure phase behaviour of the water-ethanol binary mixtures \citep{bejan2016advanced}. At this elevated substrate temperature, except for the (E 50\% + W 50 \%) binary mixture, droplets of other compositions remain pinned at the early stage, followed by a receding stage at the later times. In the case of the droplet of (E 50\% + W 50 \%) mixture, however, we observe an early spreading stage, an intermediate pinned stage and a late receding stage. Unlike $T_s=25^\circ$C, at $T_s=60^\circ$C, the contact angle of the droplet of pure water at the early times is greater than 90$^\circ$ (hydrophobic), but for other compositions the contact angle is less than 90$^\circ$ (hydrophilic) during the entire evaporation process. We observed a self-similar nature in the variations of the normalised volume $(V/V_0)$ against the normalised time $(t/t_e)$ curves for different compositions at $T_s=60^\circ$C. We have also shown that the theoretical mass flux models developed by combining the effects of diffusion, concentration gradient driven convection, and passive vapour transport with the freely convecting air flow agree well with the experimentally observed trends of $(V/V_0)$ versus $(t/t_e)$. 

Finally, we investigate the dynamics of a droplet of (E 50\% + W 50 \%) mixture at different substrate temperatures. It is observed that the lifetime of the droplet, $t_e$ decreases in a logarithmic scale with the increase in the substrate temperature from $25^\circ$C to $60^\circ$C. As expected, at all substrate temperatures, an increasing ethanol concentration in the ethanol-water binary mixture decreases the lifetime of the droplet. It is found that the spreading and the wetting dynamics of a sessile droplet of (E 50\% + W 50 \%) mixture at $T_s = 40^\circ$C are closer to that observed at $T_s = 25^\circ$C, whereas at $T_s = 50^\circ$C, the behaviour is similar to that observed at $T_s = 60^\circ$C. We found that at high substrate temperature ($T_s=60^\circ$C), the normalised volume of (E 50\% + W 50\%) droplet decreases linearly with the normalised time, but the trend becomes nonlinear at low substrate temperatures, which indicates that the evaporation dynamics of a droplet of binary mixture (E 50\% + W 50\%) at $T_s=60^\circ$C is similar to a droplet of another pure fluid with a higher volatility at room temperature. {While the previous studies \citep{sefiane2003experimental} predict four distinct stages in an evaporating sessile droplet of ethanol-water binary mixture at room temperature, our experiments highlight the strong dependence of the duration and the intensity of these stages with different substrate temperatures and for droplets of different compositions of the binary mixture.}

We observe undulations (interfacial instabilities) at the liquid-vapour interface of the droplets with high ethanol concentrations (80\% ethanol and pure ethanol) at room and elevated temperatures of the substrates. The intensity of these undulations is more at the high substrate temperature ($T_s = 60^\circ$C). While the lower surface tension of a droplet with higher ethanol concentration near the end stages of evaporation may be responsible for enhancing the intensity of the instability waves that sometimes even lead to droplet breakup, according to the authors, this has not been understood clearly yet and can be studied in future. \\

\noindent{\bf Acknowledgement:} {K. C. S. thanks Science \& Engineering Research Board, India for their financial support (Grant number: MTR/2017/000029). }


\end{document}